\documentstyle[epsf,12pt]{article}

  \def\pp{{\mathchoice
          {
              \kern 1pt%
              \raise 1pt
              \vbox{\hrule width5pt height0.4pt depth0pt
                    \kern -2pt
                    \hbox{\kern 2.3pt
                          \vrule width0.4pt height6pt depth0pt
                          }
                    \kern -2pt
                    \hrule width5pt height0.4pt depth0pt}%
                    \kern 1pt
           }
            {
              \kern 1pt%
              \raise 1pt
              \vbox{\hrule width4.3pt height0.4pt depth0pt
                    \kern -1.8pt
                    \hbox{\kern 1.95pt
                          \vrule width0.4pt height5.4pt depth0pt
                          }
                    \kern -1.8pt
                    \hrule width4.3pt height0.4pt depth0pt}%
                    \kern 1pt
            }
            {
              \kern 0.5pt%
              \raise 1pt
              \vbox{\hrule width4.0pt height0.3pt depth0pt
                    \kern -1.9pt  
                    \hbox{\kern 1.85pt
                          \vrule width0.3pt height5.7pt depth0pt
                          }
                    \kern -1.9pt
                    \hrule width4.0pt height0.3pt depth0pt}%
                    \kern 0.5pt
            }
            {
              \kern 0.5pt%
              \raise 1pt
              \vbox{\hrule width3.6pt height0.3pt depth0pt
                    \kern -1.5pt
                    \hbox{\kern 1.65pt
                          \vrule width0.3pt height4.5pt depth0pt
                          }
                    \kern -1.5pt
                    \hrule width3.6pt height0.3pt depth0pt}%
                    \kern 0.5pt
            }
        }}

  \def\mm{{\mathchoice
                       {
                             \kern 1pt
               \raise 1pt    \vbox{\hrule width5pt height0.4pt depth0pt
                                  \kern 2pt
                                  \hrule width5pt height0.4pt depth0pt}
                             \kern 1pt}
                       {
                            \kern 1pt
               \raise 1pt \vbox{\hrule width4.3pt height0.4pt depth0pt
                                  \kern 1.8pt
                                  \hrule width4.3pt height0.4pt depth0pt}
                             \kern 1pt}
                       {
                            \kern 0.5pt
               \raise 1pt
                            \vbox{\hrule width4.0pt height0.3pt depth0pt
                                  \kern 1.9pt
                                  \hrule width4.0pt height0.3pt depth0pt}
                            \kern 1pt}
                       {
                           \kern 0.5pt
             \raise 1pt  \vbox{\hrule width3.6pt height0.3pt depth0pt
                                  \kern 1.5pt
                                  \hrule width3.6pt height0.3pt depth0pt}
                           \kern 0.5pt}
                       }}

\catcode`@=11
\def\un#1{\relax\ifmmode\@@underline#1\else
        $\@@underline{\hbox{#1}}$\relax\fi}
\catcode`@=12

\let\du=\du                     

\def\a{\alpha}
\def\b{\beta}
\def\c{\chi}
\def\d{\delta}
\def\e{\epsilon}
\def\f{\phi}
\def\g{\gamma}
\def\h{\eta}

\def\j{\psi}

\def\l{\lambda}
\def\m{\mu}
\def\n{\nu}
\def\o{\omega}
\def\p{\pi}
\def\q{\theta}

\def\s{\sigma}
\def\t{\tau}

\def\x{\xi}
\def\z{\zeta}
\def\D{\Delta}

\def\G{\Gamma}

\def\L{\Lambda}
\def\O{\Omega}

\def\S{\Sigma}

\def\ve{\varepsilon}
\def\vf{\varphi}

\def\vq{\vartheta}

\def\cf{{\cal F}}

\def\ch{{\cal H}}

\def\ck{{\cal K}}

\def\cm{{\cal M}}

\def\bo{{\raise-.5ex\hbox{\large$\Box$}}}               
\def\pa{\partial}                                       
\def\de{\nabla}                                         
\def\pr{\prod}                                          
\def\TH{{\raise.2ex\hbox{$\displaystyle \bigodot$}\mskip-4.7mu \llap H \;}}
\def\face{{\raise.2ex\hbox{$\displaystyle \bigodot$}\mskip-2.2mu \llap {$\ddot
        \smile$}}}                                      
\def\dg{\sp\dagger}                                     

\def\sp#1{{}^{#1}}                              
\def\Tilde#1{\widetilde{#1}}                    
\def\Bar#1{\overline{#1}}                       
\def\VEV#1{\left\langle #1\right\rangle}        
\def\abs#1{\left| #1\right|}                    
\def\leftrightarrowfill{$\mathsurround=0pt \mathord\leftarrow \mkern-6mu
        \cleaders\hbox{$\mkern-2mu \mathord- \mkern-2mu$}\hfill
        \mkern-6mu \mathord\rightarrow$}
\def\dvec#1{\vbox{\ialign{##\crcr
        \leftrightarrowfill\crcr\noalign{\kern-1pt\nointerlineskip}
        $\hfil\displaystyle{#1}\hfil$\crcr}}}           
\def\dt#1{{\buildrel {\hbox{\LARGE .}} \over {#1}}}     

\def\frac#1#2{{\textstyle{#1\over\vphantom2\smash{\raise.20ex
        \hbox{$\scriptstyle{#2}$}}}}}                   
\def\sfrac#1#2{{\vphantom1\smash{\lower.5ex\hbox{\small$#1$}}\over
        \vphantom1\smash{\raise.4ex\hbox{\small$#2$}}}} 
\def\bfrac#1#2{{\vphantom1\smash{\lower.5ex\hbox{$#1$}}\over
        \vphantom1\smash{\raise.3ex\hbox{$#2$}}}}       
\def\afrac#1#2{{\vphantom1\smash{\lower.5ex\hbox{$#1$}}\over#2}}    

\def\[{\lfloor{\hskip 0.35pt}\!\!\!\lceil}
\def\]{\rfloor{\hskip 0.35pt}\!\!\!\rceil}

\def\du#1#2{_{#1}{}^{#2}}
\def\ud#1#2{^{#1}{}_{#2}}

\def\fracm#1#2{\hbox{\large{${\frac{{#1}}{{#2}}}$}}}
\def\ha{{\fracmm12}}
\def\tr{{\rm tr}}

\def\un{\underline}
\def\fracmm#1#2{{{#1}\over{#2}}}

\def\low#1{{\raise -3pt\hbox{${\hskip 0.75pt}\!_{#1}$}}}

\def\Dot#1{\buildrel{_{_{\hskip 0.01in}\bullet}}\over{#1}}
\def\dt#1{\Dot{#1}}

\def\Tilde#1{{\widetilde{#1}}\hskip 0.015in}

\def\sbar#1{\stackrel{*}{\Bar{#1}}}

\newskip\humongous \humongous=0pt plus 1000pt minus 1000pt
\def\caja{\mathsurround=0pt}
\def\eqalign#1{\,\vcenter{\openup2\jot \caja
        \ialign{\strut \hfil$\displaystyle{##}$&$
        \displaystyle{{}##}$\hfil\crcr#1\crcr}}\,}
\newif\ifdtup

\def\pl#1#2#3{Phys.~Lett.~{\bf {#1}B} (19{#2}) #3}
\def\np#1#2#3{Nucl.~Phys.~{\bf B{#1}} (19{#2}) #3}
\def\prl#1#2#3{Phys.~Rev.~Lett.~{\bf #1} (19{#2}) #3}
\def\pr#1#2#3{Phys.~Rev.~{\bf D{#1}} (19{#2}) #3}
\def\cqg#1#2#3{Class.~and Quantum Grav.~{\bf {#1}} (19{#2}) #3}
\def\cmp#1#2#3{Commun.~Math.~Phys.~{\bf {#1}} (19{#2}) #3}

\def\ijmp#1#2#3{Int.~J.~Mod.~Phys.~{\bf A{#1}} (19{#2}) #3}
\def\mpl#1#2#3{Mod.~Phys.~Lett.~{\bf A{#1}} (19{#2}) #3}

\def\ibid#1#2#3{{\it ibid.}~{\bf {#1}} (19{#2}) #3}

\parskip=\medskipamount                 
\thicklines                         

\begin{document}


\thispagestyle{empty}               

\def\border{                                            
        \setlength{\unitlength}{1mm}
        \newcount\xco
        \newcount\yco
        \xco=-24
        \yco=12
        \begin{picture}(140,0)
        \put(-20,11){\tiny Institut f\"ur Theoretische Physik Universit\"at
Hannover~~ Institut f\"ur Theoretische Physik Universit\"at Hannover~~
Institut f\"ur Theoretische Physik Hannover}
        \put(-20,-241.5){\tiny Institut f\"ur Theoretische Physik Universit\"at
Hannover~~ Institut f\"ur Theoretische Physik Universit\"at Hannover~~
Institut f\"ur Theoretische Physik Hannover}
        \end{picture}
        \par\vskip-8mm}

\def\headpic{                                           
        \indent
        \setlength{\unitlength}{.8mm}
        \thinlines
        \par
        \begin{picture}(29,16)
        \put(75,16){\line(1,0){4}}
        \put(80,16){\line(1,0){4}}
        \put(85,16){\line(1,0){4}}
        \put(92,16){\line(1,0){4}}

        \put(85,0){\line(1,0){4}}
        \put(89,8){\line(1,0){3}}
        \put(92,0){\line(1,0){4}}

        \put(85,0){\line(0,1){16}}
        \put(96,0){\line(0,1){16}}
        \put(92,16){\line(1,0){4}}

        \put(85,0){\line(1,0){4}}
        \put(89,8){\line(1,0){3}}
        \put(92,0){\line(1,0){4}}

        \put(85,0){\line(0,1){16}}
        \put(96,0){\line(0,1){16}}
        \put(79,0){\line(0,1){16}}
        \put(80,0){\line(0,1){16}}
        \put(89,0){\line(0,1){16}}
        \put(92,0){\line(0,1){16}}
        \put(79,16){\oval(8,32)[bl]}
        \put(80,16){\oval(8,32)[br]}

        \end{picture}
        \par\vskip-6.5mm
        \thicklines}

\border\headpic {\hbox to\hsize{
\vbox{\noindent DESY 98 -- 059  \hfill hep-th/9806009\\
ITP--UH--12/98 \hfill                       June 1998            }}}

\noindent
\vskip1.3cm
\begin{center}

{\Large\bf       ANALYTIC TOOLS TO BRANE TECHNOLOGY
\vglue.1in       IN $N=2$ GAUGE THEORIES WITH MATTER~\footnote{Supported in part by the 
`Deutsche Forschungsgemeinschaft' and the NATO grant CQG 930789}}\\
\vglue.3in

Sergei V. Ketov \footnote{
On leave of absence from:
High Current Electronics Institute of the Russian Academy of Sciences,
\newline ${~~~~~}$ Siberian Branch, Akademichesky~4, Tomsk 634055, Russia}

{\it Institut f\"ur Theoretische Physik, Universit\"at Hannover}\\
{\it Appelstra\ss{}e 2, 30167 Hannover, Germany}\\
{\sl ketov@itp.uni-hannover.de}
\end{center}
\vglue.2in
\begin{center}
{\Large\bf Abstract}
\end{center}

Exact solutions to the low-energy effective action (LEEA) of the four-dimensional $N=2$ 
supersymmetric gauge theories are known to be obtained either by quantum field theory methods 
from S-duality in the Seiberg-Witten approach, or by the Type-IIA superstring/M-Theory methods of 
brane technology. After a brief review of the standard field-theoretical results for the $N=2$ 
{\it gauge} (Seiberg-Witten) LEEA, we consider a field-theoretical derivation of the exact 
{\it hypermultiplet} LEEA by using the $N=2$ harmonic superspace methods. We illustrate our 
techniques on a number of explicit examples. Our main purpose, however, is to discuss the existing 
{\it analytical} (calculational) support for the alternative methods of brane technology. We 
summarize known exact solutions to the eleven-dimensional and ten-dimensional type-IIA 
supergravities, which describe classical configurations of intersecting BPS branes with eight
supercharges relevant to the non-perturbative $N=2$ gauge theory with fundamental hypermultiplet 
matter. The crucial role of the M-Theory in providing a classical resolution of singularities in 
the ten-dimensional (Type-IIA superstring) brane picture, as well as the $N=2$ extended 
supersymmetry in four dimensions, are made {\it manifest}. The two approaches to a derivation of
the exact $N=2$ gauge theory LEEA are thus seen to be complementary to each other and mutually 
dependent.

\newpage

\begin{center}
{\bf CONTENT}
\end{center}

\noindent
1. Introduction . . . . . . . . . . . . . . . . . . . . . . . . . . . . . . .
. . . . . . . . . . \hfill{3}

1.1. Basic facts about $N=2$ supersymmetry in $D=4$ . . . . . . . . . . . . . . . \hfill{5}

1.2. Basic facts about $N=2$ supersymmetric LEEA . . . . . . . . . . . . . . . . \hfill{8}

\noindent
2. Gauge LEEA in Coulomb branch . . . . . . . . . . . . . . . . . . . . . . . 
. . . . . \hfill{11}

2.1. On instanton calculations . . . . . . . . . . . . . . . . . . . . . . . 
. . . . . . \hfill{12}

2.2. Seiberg-Witten curve . . . . . . . . . . . . . . . . . . . . . . . . . . 
. . . . . . \hfill{13}

\noindent
3. Hypermultiplet LEEA in Coulomb branch  . . . . . . . . . . . . . . . . . . 
. . . . \hfill{15}

3.1. $N=2$ harmonic superspace  . . . . . . . . . . . . . . . . . . . . . . 
. . . . . . \hfill{16}

3.2. Induced Taub-NUT metric or KK-monopole  . . . . . . . . . . . . . . . . . .
\hfill{21}

3.3. Duality transformation and $N=2$ tensor multiplet . . . . . . . . . . . . . \hfill{25}

3.4. Induced multicentre Taub-NUT metrics . . . . . . . . . . . . . . . . . . . . \hfill{28}

\noindent 
4. Brane technology . . . . . . . . . . . . . . . . . . . . . . . . . . . . . 
. . . . . . . . \hfill{31}

4.1. D=11 supergravity and its BPS solutions . . . . . . . . . . . . . . . . . . . 
\hfill{31}

4.2. NS and D branes in D=10 dimensions .  . . . . . . . . . . . . . . . . . . . . \hfill{35}

4.3. Intersecting branes . . . . . . . . . . . . . . . . . . . . . . . . . . . . . . . . .\hfill{40}

4.4. Effective field theory in worldvolume of type-IIA branes . . . . . . . . . . . \hfill{41}

4.5. M-theory resolution . . . . . . . . . . . . . . . . . . . . . . . . . .
. . . . . . . \hfill{45}

4.6. SW solution from classical M-5-brane dynamics . . . . . . . . . . . . . . . . \hfill{47}

4.7. Relation to the HSS results and S-duality . . . . . . . . . . . . . . . . . . . \hfill{52}

\noindent
5. Next-to-leading-order correction to the Seiberg-Witten LEEA . . . . . . . . . . \hfill{54}

\noindent
6. Hypermultiplet LEEA in baryonic Higgs branch  . . . . . . . . . . . . . . . . . . \hfill{58}

\noindent
7. Conclusion . . . . . . . . . . . . . . . . . . . . . . . . . . . . . . . . 
. . . . . . . . . \hfill{61}

\noindent
Appendix A: Taub-NUT hypermultipet self-interaction in components \\
from harmonic superspace . . . . . . . . . . . . . . . . . . . . . . . . . . . . . . . . .\hfill{65}

\noindent
Appendix B: Eguchi-Hanson hypermultipet self-interaction \\
in components from harmonic superspace . . . . . . . . . . . . . . . . . . . . . . . . \hfill{69}

\noindent
References  . . . . . . . . . . . . . . . . . . . . . . . . . . . . . . . . . 
. . . . . . . . . . \hfill{74}

\newpage

\section{Introduction}

Quantum Field Theory (QFT) is the current theoretical foundation of the elementary
particles physics, including the Standard Model (SM) and its (minimal) supersymmetric extension
(MSSM). An experimental success of the SM and a theoretical attractiveness of the MSSM imply
some general lessons to the field theory practicioners. Among them are: 
(i) not just an arbitrary QFT is of particular relevance for physical applications, but a
renormalizable, unitary and asymptotically-free gauge theory, (ii) gauge symmetry and internal 
symmetry play the crucial role in maintaining consistency of a `good' QFT at the quantum level, 
(iii) a `good' QFT should be supersymmetric at sufficiently high energies, while the supersymmetry 
must be broken at lower energies.

The standard textbook description of quantum gauge theories is often limited to 
{\it perturbative} considerations, whereas many physical phenomena (e.g., confinement) are 
essentially {\it non}-perturbative. It is usually straightforward (although, it may be quite 
non-trivial~!) to develop a quantum perturbation theory with all the fundamental symmetries to 
be {\it manifestly} (i.e. linearly) realized. Unfortunately, a perturbative expansion usually does
 not make sense when the field coupling becomes strong. A path integral representing the quantum 
generating functional of QFT has to be defined in practical terms which, generally speaking, can be
(and, in fact, has been) done in {\it many} ways beyond the perturbation theory (e.g., lattice 
regularization, instantons, duality). Because of this reasoning, it was, until recently, common to 
believe among many field theorists that a non-perturbative gauge QFT is not well-defined 
enough, in order to make definitive predictions and non-perturbative calculations from the first 
principles ({\it cf.} the current status of QCD). 

Nowadays, since the discovery of {\it exact} non-perturbative QFT solutions to the 
low-energy effective action (LEEA) in certain $N=2$ supersymmetric four-dimensional $(D=4)$ gauge 
field theories, pioneered by Seiberg and Witten in ref.~\cite{sw}, and subsequent advances in 
non-perturbative M-Theory `formerly known as the theory of superstrings', initiated in another 
Witten's paper \cite{wm},~\footnote{See e.g., refs.~\cite{gen1,gen2} for an introductory review.} 
the conventional wisdom briefly outlined above may have to be revised. Though non-trivial exact
solutions were only found in a certain class of $N=2$ {\it supersymmetric} gauge QFTs having no 
immediate phenomenological applications, they are, nevertheless, of great theoretical value. Indeed,
unlike the SM or MSSM, the $N=2$ extended supersymmetric gauge field theories in four dimensions 
cannot directly serve for phenomenological applications at (low) energies of order 100 GeV, 
essentially because an $N=2$ supersymmetric matter can only be defined in {\it real}~ 
representations of the gauge group and, hence, parity is conserved. The $N=2$ gauge theories may,
nevertheless, appear as a kind of effective QFT description at some intermediate energies provided
that the ultimate (yet unknown at the microscopic level) unified theory of Nature at Planckian
energies (e.g. M-Theory~!) lives in higher spacetime dimensions and has even more supersymmetries.
A solvable (in the LEEA sense) gauge QFT may be a good starting point for further symmetry breaking
towards getting phenomenologically applicable QFT models at lower energies, which are supposed to 
resemble SM or MSSM, while maintaining the integrability properties of the initial supersymmetric 
field theory and thus keeping its non-perturbative LEEA under control.

Exact solutions to the LEEA of the four-dimensional $N=2$ supersymmetric gauge field theories can 
be obtained either by the conventional QFT methods after taking into account S-duality in 
the Seiberg-Witten approach~\cite{sw}, or by the alternative Type-IIA superstring/M-Theory methods 
of {\it brane technology}~\cite{five,hw,witten}.~\footnote{See ref.~\cite{gku} for a recent review.}
Unlike ref.~\cite{gku}, where the brane technology in various dimensions and with various amount of
supersymmetry was considered, we restrict ourselves to the case of four (uncompactified) spacetime
dimensions with the $N=2$ extended supersymmetry there. Our interest to this situation is partially
motivated by some general lessons out of the recent results about the non-perturbative behaviour of
the four-dimensional gauge QFTs, namely,
\begin{itemize}
\item in order to be solvable in the low-energy approximation, a $D=4$ gauge QFT has to be a `good'
one, i.e. it should have $N=2$ extended supersymmetry or, equivalenntly, eight conserved 
supercharges,
\item exact internal symmetries within the $N=2$ extended supersymmetry severely restrict the form 
of allowed non-perturbative solutions to the LEEA in $N=2$ supersymmetric gauge theories, with the 
Seiberg-Witten solution~\cite{sw} being an example,
\item the non-abelian gauge symmetry is always broken in the full (non-perturbative) $N=2$ 
supersymmetric QFT to an abelian subgroup, while the $N=2$ supersymmetry is unbroken.
\end{itemize}

The paper is organized as follows. After a brief review of the standard field-theoretical results 
for the $N=2$ {\it gauge} (Seiberg-Witten) LEEA in sect.~2, we consider a field-theoretical 
derivation of the exact {\it hypermultiplet} LEEA in sect.~3 by using the $N=2$ harmonic superspace
methods. We illustrate our techniques on a number of explicit examples. The brane technology with
eight conserved supercharges is discussed in Sect.~4. The calculational support to the brane
technology is provided by exact BPS-brane solutions to the maximally extended supergravities in 
eleven and ten dimensions, and the harmonic superspace. M-Theory plays the crucial role in the 
brane technology, by providing a classical resolution of singularities in the ten-dimensional 
(Type-IIA superstring) brane picture. The extended supersymmetry with eight supercharges is made 
manifest, which results in many simplifications, transparency and an analytic control (Fig.~1).

\begin{figure}
\vglue.1in
\makebox{
\epsfxsize=4in
\epsfbox{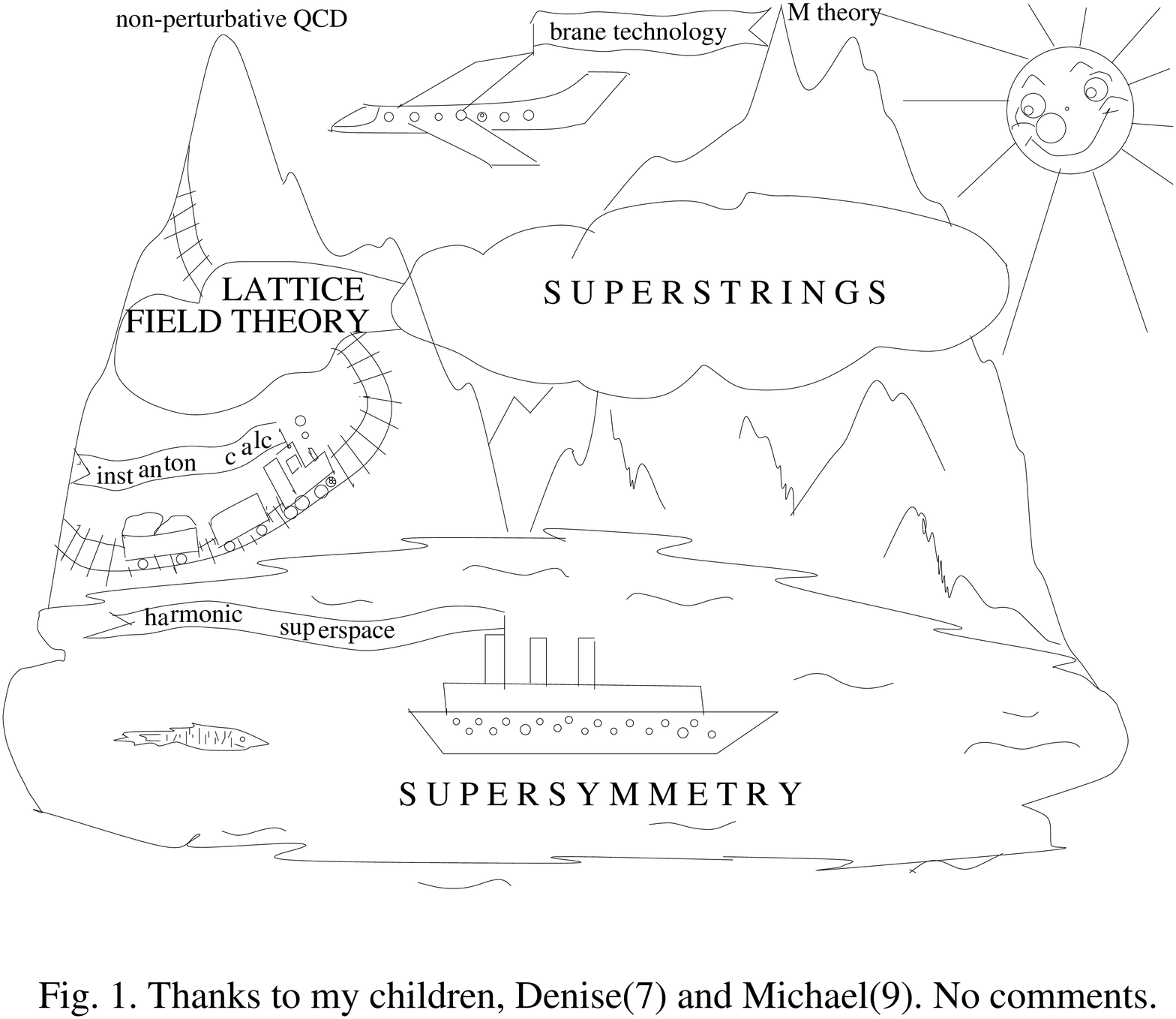}
}
\end{figure}

\subsection{Basic facts about $N=2$ supersymmetry in $D=4$}

Four-dimensional ($D=4$), $N=2$ supersymmetric gauge field theories are not 
integrable, either classically or quantum-mechanically.~\footnote{It is the
{\it self-dual} sector of their Euclidean versions that is integrable in the
classical sense~\cite{mw,kgn}.}  The full quantum effective action $\G$ in 
these theories is highly non-local and intractable. Nevertheless, it can be 
decomposed into a sum of local terms in powers of space-time derivatives or 
momenta divided by some dynamically generated scale $\L$ (in components).
The leading kinetic terms of this expansion are called the {\it low-energy 
effective action} (LEEA). Determining the exact LEEA is a great achievement 
because it provides the information about a non-perturbative spectrum and exact 
static couplings in the full quantum theory at energies which are well below 
$\L$. Since we are only interested in the $D=4$, $N=2$ gauge theories with 
spontaneously broken gauge symmetry via the Higgs mechanism, the effective 
low-energy field theory may include only {\it abelian} massless vector 
particles. All the massive fields (like the charged $W$-bosons) are supposed 
to be integrated out. This very general concept of LEEA is sometimes called 
the {\it Wilsonian} LEEA since it is familiar from statistical mechanics. There
is a difference between the quantum effective action $\G$, usually defined as 
the quantum generating functional of the one-particle-irreducible (1PI) Green's 
functions, and the Wilsonian effective action, as far as the gauge theories 
with massless particles are concerned.~\footnote{See e.g., ref.~\cite{mysw} and
references therein.} 

The $N=2$ supersymmetry severely restricts the form of the LEEA. The very presence
of $N=2$ supersymmetry in the full non-perturbatively defined quantum $N=2$ 
gauge theory follows from the fact that its Witten index~\cite{wind} does not 
vanish, $\D_W=\tr(-1)^F\neq 0$. It just means that $N=2$ supersymmetry cannot 
be dynamically broken. Alternatively, one may try to consistently formulate the whole 
theory in a manifestly $N=2$ supersymmetric way, e.g., in $N=2$ superspace.
 
There are only two basic supermultiplets (modulo classical duality
transformations) in the rigid $N=2$ supersymmetry: an $N=2$ {\it vector}
multiplet and a {\it hypermultiplet}. The $N=2$ vector multiplet components
(in a WZ-gauge) are 
$$\{~ a~,\quad \l^i_{\a}~,\quad V_{\m}~,\quad D^{(ij)} ~\}~,\eqno(1.1)$$ 
where $a$ is a complex Higgs scalar, $\l^i$ is a chiral spinor (`gaugino') 
$SU(2)_A$ doublet, $V_{\m}$ is a real vector gauge field, and $D^{ij}$ is an 
auxiliary scalar $SU(2)_A$ triplet.~\footnote{The internal symmetry $SU(2)_A$ 
is the automorphism symmetry of the $N=2$ supersymmetry \newline ${~~~~~}$ 
algebra, which rotates its two spinor supercharges.} Similarly, the 
on-shell physical components of the {\it Fayet-Sohnius} (FS)~\cite{fs} version of a 
hypermultiplet are 
$$ {\rm FS}:\qquad \{~ q^i~,\quad \j_{\a}~,\quad\bar{\j}_{\dt{\a}}~ \}~, 
\eqno(1.2)$$ 
where $q^i$ is a complex scalar $SU(2)_A$ doublet, and $\j$ is a Dirac spinor.
There exists another (dual) {\it Howe-Stelle-Townsend} (HST) version~\cite{hst} of a 
hypermultiplet, whose on-shell physical components are 
$$ {\rm HST}: \qquad \{~ \o~,\quad \o^{(ij)}~,\quad \c^i_{\a} ~\}~,
\eqno(1.3)$$ 
where $\o$ is a real scalar, $\o^{(ij)}$ is a scalar $SU(2)_A$ triplet, and 
$\c^i$ is a chiral spinor $SU(2)_A$ doublet. The hypermultiplet spinors are
sometimes referred to as `quarks', even though $N=2$ supersymmetry implies
(apparently absent in experiment) extra `mirror' particle for each `true' 
quark in the $N=2$ super-QCD.

The manifestly supersymmetric formulation of supersymmetric field theories
is provided by {\it superspace}~\cite{superspace}. Since superfields are 
generically reducible representations of supersymmetry, they have to be 
restricted by imposing certain superspace constraints. The standard constraints 
defining the $N=2$ {\it super-Yang-Mills} (SYM) theory in the ordinary $N=2$ 
superspace~\cite{gs} essentially amount to the existence of a restricted chiral 
$N=2$ superfield strength $W$, whose leading component is the Higgs field, 
$\left.W\right|=a$. The $N=2$ superfield $W$ also contains the usual Yang-Mills 
field strength $F_{\m\n}(V)$ among its bosonic components, as well as the $SU(2)_A$ 
auxiliary triplet $D^{(ij)}$. Since the latter has to be real in the
sence $\Bar{D^{ij}}=\ve_{ik}\ve_{jl}D^{kl}$, this leads to (non-chiral)
$N=2$ superspace constraints on $W$, which are not easy to solve in terms of 
unconstrained $N=2$ superfields in the non-abelian case. The situation is even
more dramatic in the case of the FS hypermultiplet whose universal off-shell 
formulation does not even exist in the ordinary $N=2$ superspace (i.e. with a finite
number of auxiliary fields). Though a particular (dual) form of the HST hypermultiplet 
can be defined off-shell in the ordinary $N=2$ superspace where it is known as an 
$N=2$ {\it tensor} (or linear) multiplet (see subsect.~3.3), its self-couplings are 
very restricted and not universal there. For instance, in order to be coupled to the 
$N=2$ gauge superfields, the $N=2$ tensor multiplet has to be generalized to a reducible 
(relaxed) version (known as the {\it relaxed} HST multiplet) which is highly complicated 
and cumbersome. The most general (universal) off-shell formulation of a hypermultiplet is,
however, necessary in order to write down its most general couplings, which may appear 
e.g. in the LEEA, in a model-independent way. 

The universal off-shell solution to all $N=2$ supersymmetric field theories in $D=4$ was 
proposed in 1984 by Galperin, Ivanov, Kalitzin, Ogievetsky and Sokatchev~\cite{gikos}. 
They introduced the so-called $N=2$ {\it harmonic superspace} 
(HSS) by adding extra bosonic variables (=harmonics), parameterizing 
a sphere $S^2=SU(2)/U(1)$, to the ordinary $N=2$ superspace coordinates. It 
amounts to the introduction of infinitely 
many auxiliary fields in terms of the ordinary $N=2$ superfields. By the use of
harmonics, one can rewrite the standard $N=2$ superspace constraints to 
another form (that may be called a 'zero-curvature representation') in which the
hidden {\it analytical} structure of the constraints becomes manifest. In the HSS
approach, the harmonics play the role of twistors or spectral parameters
which are well-known in the theory of integrable models. As a result~\cite{gikos}, 
all the $N=2$ supersymmetric field theories can be naturally formulated in terms of
{\it unconstrained} and {\it analytic} $N=2$ superfields, i.e. fully 
off-shell.~\footnote{See subsect.~3.1. for an introduction to $N=2$ HSS.} In 
particular, an off-shell FS hypermultiplet is naturally described by an analytic 
superfield $q^+$ of the $U(1)$ charge $(+1)$, whereas the analytic (sub)superspace
measure has the $U(1)$ charge $(-4)$. A generic hypermultiplet Lagrangian in the
analytic HSS has to be an analytic function of $q^+$ and $\o$, and it may also depend 
upon harmonics $u_i^{\pm}$. 

In the next subsect.~1.2. we are going to discuss the most general form of the LEEA, 
which is dictated by $N=2$ supersymmetry alone. The rest of the paper will be devoted 
to the question how to fix the $N=2$ supersymmetric Ansatz for the vector {\it and} 
hypermultiplet LEEA completely, by using all available methods of calculation (Fig.~1).

\subsection{Basic facts about $N=2$ supersymmetric LEEA} 

We are now already in a position to formulate the general {\it Ansatz} for the
$N=2$ supersymmetric LEEA. As regards the $N=2$ vector multiplet terms, they
can only be in the form of $N=2$ superspace integrals,
$$ \G_V[W,\bar{W}]=\int_{\rm chiral} \cf(W) +{\rm h.c.} + \int_{\rm full}
\ch(W,\bar{W}) + \ldots~,\eqno(1.4)$$
where we have used the fact that the abelian $N=2$ superfield strength $W$ is 
an $N=2$ chiral and gauge-invariant superfield. The leading term in eq.~(1.4)
is given by the chiral $N=2$ superspace integral of a {\it holomorphic} 
function $\cf$ of the gauge superfield strength $W$ which is supposed to be 
valued in the Cartan subalgebra of the gauge group.  The {\it next-to-leading-order} 
term is given by the full $N=2$ superspace integral over the real function $\ch$ of $W$ {\it
and}~ $\bar{W}$. The dots in eq.~(1.4) stand for higher-order terms containing
the derivatives of $W$ and $\bar{W}$.

Similarly, the leading term in the hypermultiplet LEEA is of the form
$$ \G_H[q^+,\sbar{q}{}^+;\o]=\int_{\rm analytic} \ck^{(+4)}(q^+,\sbar{q}{}^+;
\o;u^{\pm}_i)+\ldots~,\eqno(1.5)$$
where $\ck^{(+4)}$ is an analytic function of the FS-type superfields $q^+$, their 
conjugates $\sbar{q}{}^+$, the HST-type superfields $\o$ and, perhaps, harmonics
$u^{\pm}_i$. The action (1.5) is supposed to be added to the standard kinetic 
hypermultiplet action whose analytic Lagrangian is quadratic in $q^+$ and $\o$,
and of $U(1)$-charge $(+4)$ (see sect.~3.1 for details). The function $\ck$ is called 
a {\it hyper-K\"ahler potential}. An arbitrary choice of this function in eq.~(1.5) 
automatically leads to the $N=2$ supersymmetric {\it non-linear sigma-model} 
(NLSM) with a {\it hyper-K\"ahler} metric, because of the $N=2$ extended 
supersymmetry by construction (see Appendices A and B for some explicit examples). 

When being expanded in components, the first term in eq.~(1.4) also leads, 
in particular, to the certain K\"ahler NLSM in the scalar Higgs sector $(a,\bar{a})$.
The corresponding NLSM K\"ahler potential $K_{\cf}(a,\bar{a})$ is dictated by the 
holomorphic function $\cf$ as $K_{\cf}={\rm Im}[\bar{a}\cf'(a)]$, so that the function 
$\cf$ plays the role of a potential for this {\it special} NLSM K\"ahler (but not a 
hyper-K\"ahler) geometry $K_{\cf}(a,\bar{a})$. As regards the hypermultiplet NLSM of 
eq.~(1.5), the relation between a hyper-K\"ahler potential $\ck$ and the corresponding 
K\"ahler potential $K_{\ck}$ of the same NLSM is much more involved. Indeed, 
it is easy to see that the hyper-K\"ahler condition on a K\"ahler potential 
amounts to a non-linear (Monge-Amper\'e) partial differential equation which is 
not easy to solve. It is remarkable that the HSS approach allows one to 
formally get a general 'solution' to the hyper-K\"ahler constraints, in terms of an 
analytic scalar potential $\ck$. Of course, the real problem is now being 
translated into the precise relation between $\ck$ and the corresponding 
K\"ahler potential (or metric) in components, whose determination amounts to 
solving infinitely many linear differential equations altogether, just in 
order to eliminate the infinite number of the auxiliary fields involved (see 
Appendices A and B for examples). Nevertheless, the notion of hyper-K\"ahler potential 
in HSS turns out to be very useful in dealing with the hypermultiplet LEEA (sect.~3).

The LEEA gauge-invariant functions $\cf(W)$ and $\ch(W,\bar{W})$ generically 
receive both perturbative and non-perturbative contributions,
$$ \cf= \cf_{\rm per.} + \cf_{\rm inst.}~,\qquad \ch=\ch_{\rm per.} +\ch_{\rm
non-per.}~,\eqno(1.6)$$
while the non-perturbative corrections to the holomorphic function $\cf$ are 
entirely due to instantons. The last observation is in a sharp contrast with the 
situation in the (bosonic) non-perturbative QCD whose LEEA is dominated by 
instanton-antiinstanton contributions.

Unlike the gauge LEEA, the exact (charged) hypermultiplet LEEA is essentially
perturbative (see sects.~3 and 4), i.e. it does not receive any instanton corrections,
$$ \ck[q^+] =\ck_{\rm per.}[q^+]~. \eqno(1.7)$$

It is quite remarkable that the perturbative contributions to the leading and (in some cases)
even subleading terms in the $N=2$ supersymmetric LEEA entirely come from the {\it one} 
loop only. For example, as regards the leading holomorphic contribution to the gauge LEEA, 
there exists a simple argument: supersymmetry puts the trace of the energy-momentum tensor 
$T\du{\m}{\m}$ and the anomaly $\pa_{\m} j^{\m}_R$ of the abelian $R$-symmetry into a single 
$N=2$ supermultiplet. The trace $T\du{\m}{\m}$ is essentially determined by the perturbative 
renormalization group  $\b$-function~\footnote{Here and in what follows $g$ denotes the gauge 
coupling constant.}, $T\du{\m}{\m}\sim \b(g)FF$, whereas the one-loop contribution to the 
R-anomaly, $\pa\cdot j_R\sim C_{\rm 1-loop}F{}^*F$, is well-known to saturate the exact
solution to the Wess-Zumino consistency condition for the same anomaly (e.g. the one to be 
obtained from the index theorem). Hence, $\b_{\rm per.}(g)=\b_{\rm 1-loop}(g)$ by $N=2$ 
supersymmetry also. Finally, since $\b_{\rm per.}(g)$ is effectively determined by the second 
derivative of $\cf_{\rm per.}$, one concludes that $\cf_{\rm per.}=\cf_{\rm 1-loop}$ too. 

The naive component argument can be extended to a proof \cite{bko} in a manifestly $N=2$ 
supersymmetric way by using the $N=2$ HSS approach, where the whole chiral perturbative 
contribution 
$\int_{\rm chiral} \cf_{\rm per.}(W)$ arises as an anomaly. The non-vanishing central charges of 
the $N=2$ supersymmetry algebra in the $N=2$ gauge field theory under consideration turn out to be 
primarily responsible for the non-vanishing leading holomorphic contribution to the LEEA. A 
perturbative part of the LEEA thus takes the form $\cf_{\rm per.}(W)\sim W^2\log(W^2/M^2)$, 
where $M$ is the renormalization scale, with the coefficient being fixed by the one-loop 
$\b$-function of the renormalization group (see sect.~5). 

The usual strategy in determining the exact LEEA exploits exact symmetries of 
a given $N=2$ quantum gauge theory together with a certain physical input.
As the particularly important example of an $N=2$ supersymmetric gauge theory, one 
can use the $N=2$ supersymmetric QCD with the gauge group $G_c=SU(N_c)$ and
$N=2$ matter described by some number $(N_f)$ of hypermultiplets in the
fundamental representation $\un{N_c}+\un{N_c}^*$ of the gauge group $SU(N_c)$. 
A quantum consistency of the non-abelian gauge theory requires asymptotic freedom which, in
its turn, implies  $N_f<2N_c$ in this case.

All possible $N=2$ supersymmetric vacua can be classified as follows:
\begin{itemize}
\item 
{\it Coulomb branch}: $\VEV{q}=\VEV{\o}=0$, whereas $\VEV{a}\neq 0$; the gauge group 
$G_c$ is broken to its abelian subgroup $U(1)^{{\rm rank}~G_c}$; non-vanishing
'quark' masses are allowed;
\item  {\it Higgs branch}: $\VEV{q}\neq 0$ or $\VEV{\o}\neq 0$ for some hypermultiplets, whereas
$\VEV{a}=0$ and all the `quark' masses vanish; the gauge group $G_c$ is completely broken;
\item {\it mixed (Coulomb-Higgs) branch}: some $\VEV{q}\neq 0$ {\it and}
$\VEV{a}\neq 0$; it requires $N_c>2$, in particular.
\end{itemize}

In the Coulomb branch, one has to specify the both equations (1.6) and (1.7),
whereas in the Higgs branch only eq.~(1.7) really needs to be fixed. In addition, there may be 
less symmetric vacua when e.g., a non-vanishing {\it Fayet-Iliopoulos} (FI) term is present, 
i.e. $\VEV{D^{ij}}=\x^{ij}\neq 0$. Though a FI-term is usually associated with a spontaneous 
or soft breaking of supersymmetry, it does not imply the supersymmetry breaking automatically.
We will only consider a FI term for the fictitious (i.e. non-dynamical) $N=2$ vector multiplet
to be introduced as a Lagrange multiplier $N=2$ superfield (sect.~6). 

\section{Gauge LEEA in Coulomb branch} 

Seiberg and Witten~\cite{sw} gave a full solution to the holomorphic function
$\cf(W)$ by using certain physical assumptions about the global structure of
the quantum moduli space $\cm_{\rm qu}$ of vacua {\it and} electric-magnetic
duality, i.e. not from the first principles. Their main assumption was the
precise value of the Witten index~\footnote{A formal derivation of Witten's
index $\D_W$ from the path integral is plagued with ambiguities.}, $\D_W=2$, 
which implies just two physical singularities in $\cm_{\rm qu}\,$. The electric-magnetic 
duality (also known as {\it S-duality}) was used in ref.~\cite{sw} to connect the weak
and strong coupling regions of $\cm_{\rm qu}\,$.

The Seiberg-Witten solution~\cite{sw} in the simplest
case of the $SU(2)$ gauge group (no fundamental $N=2$ matter) reads
$$ a_D(u) = \fracmm{\sqrt{2}}{\p}\int^u_1\fracmm{dx\sqrt{x-u}}{\sqrt{x^2-1}}~,
\quad
a(u)=\fracmm{\sqrt{2}}{\p}\int^1_{-1}\fracmm{dx\sqrt{x-u}}{\sqrt{x^2-1}}~,
\eqno(2.1)$$
where the renormalization-group independent (Seiberg-Witten) scale $\L^2=1$ 
and, by definition,
$$ a_D=\fracmm{\pa\cf(a)}{da}~.\eqno(2.2)$$
The solution (2.1) is thus written down in the parametric form. Its
holomorphic parameter $u$ can be identified with the second Casimir eivenvalue, 
$u=\VEV{\tr\, a^2}$, that parameterizes $\cm_{\rm qu}\,$. We find convenient to use 
the same lower-case letter $(a)$ to denote the leading component (scalar field) of the
$N=2$ vector multiplet and its expectation value (complex constant) simultaneously. 
The holomorphic function $\cf$ is thus defined over the quantum moduli 
space of vacua, while the S-duality can be identified with the action of the modular 
group $SL(2,{\bf Z})$ there~\cite{sw}. The monodromies of the multi-valued function 
$\cf$ around the singularities are supplied by the perturbative $\b$-functions, whereas the
whole function $\cf$ is a (unique) solution to the corresponding 
Riemann-Hilbert problem.~\footnote{See refs.~\cite{swrev, mysw} for a review.}

In order to make contact with our general discussion in sect.~1, let's consider
an expansion of the SW solution in the semiclassical region, i.e. when 
$\abs{W}\gg\L$,
$$\cf(W)=\fracmm{i}{2\p}W^2\log\fracmm{W^2}{\L^2} +\fracmm{1}{4\p i}W^2
\sum_{m=1}^{\infty} c_m\left(\fracmm{\L^2}{W^2}\right)^{2m}~,\eqno(2.3)$$
where we have restored the $\L$-dependence and written down the interacting 
terms only. It is now obvious that the first term in eq.~(2.3) represents the 
perturbative (one-loop) contribution whereas the rest is just the sum over the
non-perturbative instanton contributions (see subsect.~2.1). It is 
straightforward to calculate the numerical coefficients $\{c_m\}$ from the
explicit solution (2.1) and (2.2)~\cite{klemm}:
$$ \begin{array}{c|cccccc} m & 1 & 2 & 3 & 4 & 5 & \ldots \\
\hline
c_m & 1/2^5 & 5/2^{14} & 3/2^{18} & 1469/2^{31} & 4471/5\cdot 2^{34} & \ldots
\end{array} \eqno(2.4)$$
From the technical point of view, the SW solution is nothing but eq.~(2.4).
It is a challenge for field theorists to reproduce this solution from the
first principles.  

\subsection{On instanton calculations}

The SW solution predicts that the non-perturbative holomorphic contributions 
to the $N=2$ vector gauge LEEA are entirely due to instantons. It is therefore
quite natural to try to reproduce them 'from the first principles', e.g. from 
the path integral approach. The $N=2$ supersymmetric instantons are solutions 
of the classical self-duality equations
$$ F={}^*F~,\qquad i\g^{\m}{D}_{\m}\l=0~,\qquad D^{\m}D_{\m}a
=\[\bar{\l},\l\]~,\eqno(2.5)$$
whose Higgs scalar $a$ approaches a non-vanishing constant at the spacial
infinity so that the whole configuration has a non-vanishing topological charge
$m\in {\bf Z}$. 

{} From the path-integral point of view, the sum over instantons should be of the
form
$$ \cf_{\rm inst.} =\sum^{\infty}_{m=1}\cf_m~,\quad {\rm where}\quad \cf_m=
\int d \m^{(m)}_{\rm inst.}\,\exp\left[-S_{(m)-{\rm inst.}}\right]~.
\eqno(2.6)$$
Each term $\cf_m$ in this sum can be interpreted as the partition function in 
the multi($m$)-instanton background. The non-trivial measure $d\m^{(m)}_{\rm 
inst.}$ in eq.~(2.6) appears as the result of changing variables from the 
original fields to the collective instanton coordinates in the path integral, 
whereas the action $S_{(m)-{\rm inst.}}$ is just the Euclidean action of an $N=2$ 
superinstanton configuration of charge $(m)$. More details about the instanton
calculus can be found e.g., in ref.~\cite{dkm}. One usually assumes that the 
scalar surface term $(\sim \tr\int dS^{\m}a^{\dg}D_{\m}a$) is the only 
relevant term in the action $S_{(m)-{\rm inst.}}$ that contributes. 
In particular, the bosonic and fermionic determinants, which always appear in 
the saddle-point expansion and describe small fluctuations of the fields,
actually cancel in a supersymmetric self-dual gauge background~\cite{adv}. 
Supersymmetry is thus also in charge for the absence of infra-red divergences
present in the determinants.

The functional dependence $\cf_m(a)$ easily follows from the integrated
renormalization group (RG) equation for the one-loop $\b$-function,
$$ \exp\left( -\,\fracmm{8\p^2m}{g^2}\right)
=\left(\fracmm{\L}{a}\right)^{4m}~, \eqno(2.7)$$
and dimensional reasons as follows:
$$ \cf_m(a)=\fracmm{a^2}{4\p i}\left( \fracmm{\L}{a}\right)^{4m}c_m~,
\eqno(2.8)$$
as it should have been expected, up to a numerical coefficient $c_m$. It is
therefore the exact values of the coefficients $\{c_m\}$ that is the issue 
here, as was already noticed above. Their evaluation can thus be reduced 
to the problem of calculating the finite-dimensional multi-instanton measure 
$\{d\m^{(m)}_{\rm inst.}\}$.

A straightforward computation of the measure naively amounts to an explicit
solution of the $N=2$ supersymmetric self-duality equations in terms of the 
collective $N=2$ instanton coordinates for any positive integer instanton 
charge. As is well known, the Yang-Mills self-duality differential equations 
of motion (as well as their supersymmetric counterparts) can be reduced to the
purely {\it algebraic} (though highly non-trivial) set of equations when using
the standard ADHM construction~\cite{adhm}. Unfortunately, an explicit 
solution to the algebraic ADHM equations is known for only $m=1$~\cite{hinst} 
and $m=2$~\cite{osborn}, but it is unknown for $m>2$. Nevertheless, as was 
recently demonstrated by Dorey, Khoze and Mattis~\cite{inmea}, the correct 
multi-instanton measure for any instanton number can be fixed indirectly, by 
imposing $N=2$ supersymmetry and the cluster decomposition requirements together,
and without using the electric-magnetic duality~! This is closed enough to a 
derivation 'from the first principles'. In particular, in the Seiberg-Witten 
model with the $SU(2)$ gauge group considered above, there exists an instanton
solution for $\{c_m\}$ in quadratures~\cite{inmea}. It was demonstrated in 
refs.~\cite{iche} that the leading instanton corrections $(m=1,2)$ agree with the 
exact Seiberg-Witten solution of eq.~(2.4).

\subsection{Seiberg-Witten curve}

{} From the mathematical point of view, the Seiberg-Witten exact solution (2.1) 
in the case of two colors ($N_c=2$) or the $SU(2)$ gauge group is a solution 
to the standard Riemann-Hilbert problem of fixing a holomorphic 
multi-valued function $\cf$ by its given monodromy and singularities. The 
number (and nature) of the singularities is the physical input: they are 
identified with the appearance of massless non-perturbative BPS-like physical 
states (dyons) like the famous t'Hooft-Polyakov magnetic monopole. The
monodromies are supplied by perturbative beta-functions and S-duality 
(see ref.~\cite{mysw} for an introduction).

The $SU(2)$ solution (2.1) can be encoded in terms of the auxiliary 
(Seiberg-Witten) {\it elliptic curve} (or torus) $\S_{\rm SW}$ defined by the algebraic
equation~\cite{sw}:
$$ \S_{\rm SW}~:\qquad  y^2=(v^2-u)^2-\L^4~.\eqno(2.9)$$
The multi-valued functions $a_D(u)$ and $a(u)$ then appear by integration of a
certain abelian (Seiberg-Witten) differential $\l_{\rm SW}$ (of the 3rd kind) over the torus 
periods $A$ and $B$ of $\S_{\rm SW}$:
$$ a(u)=\oint_{A} \l_{\rm SW}~,\qquad a_D(u)=\oint_{B}\l_{\rm SW}~, \eqno(2.10)$$
while the SW differential $\l_{\rm SW}$ itself is simply related to a unique holomorphic 1-form 
$\o$ on the torus $\S_{\rm SW}$,
$$\fracmm{\pa\l_{\rm SW}}{\pa u}=\o~,\qquad \o\equiv \fracmm{dv}{y(v,u)}~~.\eqno(2.11)$$
In the case of eq.~(2.9) one easily finds that $\l_{\rm SW}=v^2dv/y(v,u)$ up to a total derivative.

The fundamental relation to the theory of Riemann surfaces can be generalized
further to more general simply-laced gauge groups and $N=2$ super-QCD 
as well~\cite{swgen1,swgen2}. For instance, a solution to the LEEA of the 
purely $N=2$ gauge theory with the gauge group $SU(N_c)$ is encoded in terms of 
a {\it hyperelliptic curve} of genus $(N_c-1)$, whose algebraic equation
reads~\cite{swgen1}
$$ \S_{\rm SW}~:\qquad y^2=W^2_{A_{N_c-1}}(v,\vec{u})-\L^{2N_c}~.\eqno(2.12)$$
The polynomial $W_{A_{N_c-1}}(v,\vec{u})$ is known in mathematics~\cite{agvbook} as the 
{\it simple singularity} associated with $A_{N_c-1}\sim SU(N_c)$. In two-dimensional 
conformal field theory, the same polynomial is known as the $N=2$ supersymmetric 
{\it Landau-Ginzburg} potential~\cite{mybook}.  Its explicit form is given by
$$ W_{A_{N_c-1}}(v,\vec{u})=\sum^{N_c}_{l=1}\left( v - \vec{\l}_l\cdot \vec{a}
\right)=v^{N_c}-\sum^{N_c-2}_{l=0}u_{l+2}(\vec{a})v^{N_c-2-l}~,\eqno(2.13)$$
where $\vec{\l}_l$ are the weights of $SU(N_c)$ in the fundamental 
representation, and $\vec{u}$ are the Casimir eigenvalues, i.e. the Weyl 
group-invariant polynomials in $\vec{a}$ to be constructed by a classical 
Miura transformation~\cite{mysw}. The simple singularity seems to be the only 
remnant of the fundamental non-abelian gauge symmetry in the Coulomb branch.

Adding fundamental $N=2$ matter does not pose a problem in calculating 
the corresponding Seiberg-Witten curve. The result reads~\cite{swgen2}
$$ \S_{\rm SW}~:\qquad y^2=W^2_{A_{N_c-1}}(v,\vec{u})-\L^{2N_c-N_f}
\prod^{N_f}_{j=1}(v-m_j)~,\eqno(2.14)$$ 
where $\{m_j\}$ are the bare hypermultiplet masses of $N_f$ hypermultiplets
($N_f<N_c$), in the fundamental representation of the gauge group $SU(N_c)$.

In the canonical first homology basis $(A_{\a},B^{\b})$, $\a,\b=1,\ldots,g$, of the 
general genus-$g$ Riemann surface $\S_{\rm SW}$, the multi-valued sections $a(u)$ and 
$a_{\rm D}(u)$ are determined by the equations 
$$ \fracmm{\pa a_{\a}}{\pa u_{\b}}=\oint_{A_{\a}}\,\o^{\b}~,\qquad
\fracmm{\pa a^{\a}_{\rm D}}{\pa u_{\b}}=\oint_{B^{\a}}\,\o^{\b}~,\ \eqno(2.15)$$
in terms of $g$ independent holomorphic 1-forms $\o^{\a}$ on $\S_{\rm SW}$. Eq.~(2.15)
is quite similar to eq.~(2.10) after taking into account that the Seiberg-Witten 
differential $\l_{\rm SW}$ is defined by a relation very similar to that of eq.~(2.11), namely,
$$\fracmm{\pa\l_{\rm SW}}{\pa u_{\a}}=\o^{\a}~.\eqno(2.16)$$
 
The minimal data $(\S_{\rm SW},\l_{\rm SW})$ needed to reproduce the Seiberg-Witten
exact solution to the {\it four-dimensional} LEEA can be associated with a certain 
{\it two-dimensional} integrable system~\cite{swint,dow}. In particular, the SW potential 
$\cf$ appears to be a solution to the Dijkgraaf-Verlinde-Verlinde-Witten-type~\cite{dvvw} 
non-linear differential equations known in the two-dimensional (conformal) topological field 
theory~\cite{mmm}:
$$ \cf_i\cf^{-1}_k\cf_j=\cf_j\cf^{-1}_k\cf_i~,\quad{\rm where}\quad
(\cf_i)_{jk}\equiv\fracmm{\pa^3\cf}{\pa a_i\pa a_j \pa a_k}~~.\eqno(2.17)$$
There also exists another non-trivial equation for $\cf$ which is a consequence
of the anomalous (chiral) $N=2$ superconformal Ward identities in $D=4$~\cite{maw}.

Though the mathematical relevance of the Seiberg-Witten curve is quite clear 
from what was already written above, its geometrical origin and physical 
interpretation are still obscure at this point. This issue can be most naturally 
understood in the context of M-Theory-based brane technology (sect.~4).

\section{Hypermultiplet LEEA in the Coulomb branch}

The previous sect.~2 was entirely devoted to the {\it holomorphic} function 
$\cf$ appearing in the gauge LEEA (1.4) in the Coulomb branch. In this sect.~3
we are going to discuss another {\it analytic} function $\ck$ dictating the
hypermultiplet LEEA (1.5). The function $\ck$ is known as a {\it hyper-K\"ahler 
potential},~\footnote{Any $D=4$, globally $N=2$ supersymmetric NLSM with the 
highest physical spin $1/2$ necessarily \newline ${~~~~~}$ has a 
hyper-K\"ahler metric in its kinetic terms~\cite{alfr}.} and its role in the 
hypermultiplet LEEA is quite similar to that of $\cf$ in the vector gauge LEEA. 
Since the very notion of the hyper-K\"ahler potential requires an introduction of the 
{\it harmonic superspace} (HSS), we begin with a brief introduction into the $N=2$ HSS 
in the next subsect.~3.1 (see refs.~\cite{gikos,hsf} for more).

Being intimately related to the hyper-K\"ahler geometry, the use of a hyper-K\"ahler potential is
by no means limited to the $N=2$ extended supersymmetry in four spacetime dimensions. It fact, 
it is necessary to make the hyper-K\"ahler structure manifest in field theory as well as in brane 
technology. A hyper-K\"ahler structure naturally appears in M-Theory (subsect.~4.5), and it is 
also known to be the important ingredient of an integrable dynamical system~\cite{swint,dow}. For
instance, the use of a hyper-K\"ahler potential allows one to make manifest the symmetry 
enhancement in the case of two 6-branes `on top of each other' (sect.~4.7).

\subsection{$N=2$~ harmonic superspace}

The supersymmetric field theories can be formulated in superspace~\cite{superspace}, usually
in terms of constrained superfields. Unfortunately, the constraints defining a (non-abelian) 
$N=2$ vector multiplet or a hypermultiplet in the ordinary $N=2$ superspace in $D=4$ do not 
have a manifestly holomorphic (or analytic) structure. Accordingly, they do not have a simple 
solution in terms of unconstrained $N=2$ superfields which are needed for supersymmetric 
quantization. The situation is even more dramatic for the hypermultiplets whose known off-shell 
formulations in the ordinary $N=2$ superspace are not universal so that their practical meaning 
is limited.  

In the HSS formalism, the standard $N=2$ superspace~\footnote{Since our $D=4$ spacetime is 
supposed to be flat in this section we identify the flat $(m=0,1,2,3)$ \newline ${~~~~~}$ 
and curved $(\m=0,1,2,3)$ spacetime vector indices.} $Z^M
=(x^m,\q^{\a}_i,\bar{\q}^{\dt{\a}i})$, $\a=1,2$, and $i=1,2$, is extended 
by adding the bosonic variables (or `zweibeins') 
$u^{\pm i}$ parameterizing the sphere $S^2\sim SU(2)/U(1)$. By using these
extra variables one can make manifest the hidden analyticity structure of 
all the standard $N=2$ superspace constraints as well as find their
solutions in terms of unconstrained (analytic) superfields. The harmonic
variables have the following fundamental properties:
$$ \left( \begin{array}{c} u^{+i} \\ u^{-i}\end{array}\right) \in SU(2)~,
\quad {\rm so~~that}\quad u^{+i}u^-_i=1~,\quad{\rm and}\quad
 u^{+i}u^+_i=u^{-i}u^-_i=0~.\eqno(3.1)$$
Instead of using an explicit parameterization of the sphere $S^2$, 
it is convenient to deal with functions of zweibeins, that carry a definite 
$U(1)$ charge $q$ to be defined by $q(u^{\pm}_i)=\pm 1$, and use the following 
integration rules~\cite{gikos}:
$$ \int du =1~,\qquad \int du\, u^{+(i_1}\cdots u^{+i_m}u^{-j_1}\cdots
u^{-j_n)}=0~,\quad {\rm when}\quad m+n>0~.\eqno(3.2)$$
It is obvious that any integral over a $U(1)$-charged quantity vanishes.  

The usual complex conjugation does not preserve analyticity (see below).
However, when being combined with another (star) conjugation that only acts on
the $U(1)$ indices as $(u^+_i)^*=u^-_i$ and $(u^-_i)^*=-u^+_i$, it does 
preserve analyticity. One easily finds~\cite{gikos}
$$ \sbar{u^{\pm i}}=-u^{\pm}_i~,\qquad  \sbar{u^{\pm}_i}=u^{\pm i}~.
\eqno(3.3)$$

The covariant derivatives with respect to the zweibeins, which preserve the 
defining conditions (3.1), are given by 
$$ D^{++}=u^{+i}\fracmm{\pa}{\pa u^{-i}}~,\quad 
D^{--}=u^{-i}\fracmm{\pa}{\pa u^{+i}}~,\quad 
D^{0}=u^{+i}\fracmm{\pa}{\pa u^{+i}}-u^{-i}\fracmm{\pa}{\pa u^{-i}}~.
\eqno(3.4)$$
It is easy to check that they satisfy the $SU(2)$ algebra,
$$\[ D^{++},D^{--}\]=D^0~,\quad \[D^0,D^{\pm\pm}\]=\pm 2D^{\pm\pm}~.
\eqno(3.5)$$
 
The key feature of the $N=2$ HSS is the existence of the so-called
 {\it analytic} subspace parameterized by the coordinates
$$ (\z,u)=\left\{ \begin{array}{c}
x^m_{\rm A}=x^m-2i\q^{(i}\s^m\bar{\q}^{j)}u^+_iu^-_j~,~~
\q^+_{\a}=\q^i_{\a}u^+_i~,~~ \bar{\q}^+_{\dt{\a}}=\bar{\q}^i_{\dt{\a}}u^+_i~,~~
u^{\pm}_i \end{array} \right\}~,\eqno(3.6)$$
which is invariant under $N=2$ supersymmetry and closed under the combined 
conjugation of eq.~(3.3)~\cite{gikos}. This allows one to define {\it analytic} 
superfields of any non-negative and integer $U(1)$ charge $q$, by the analyticity 
conditions
$$D^+_{\a}\f^{(q)}=\bar{D}^+_{\dt{\a}}\f^{(q)}=0~,\quad {\rm where}\quad
D^{+}\low{\a}=D^i_{\a}u^+_i \quad {\rm and}\quad
\bar{D}^+_{\dt{\a}}=\bar{D}^i_{\dt{\a}}u^+_i~.\eqno(3.7)$$

An analytic measure is given by $d\z^{(-4)}du\equiv d^4x_{\rm A}
d^2\q^+d^2\bar{\q}^+du$. It carries the $U(1)$ charge $(-4)$, 
whereas the full neutral measure in the $N=2$ HSS takes the form
$$ d^4xd^4\q d^4\bar{\q}du=d\z^{(-4)}du(D^+)^4~,\eqno(3.8)$$
where
$$(D^+)^4=\fracmm{1}{16}(D^+)^2(\bar{D}^+)^2
=\fracmm{1}{16}(D^{+\a}D_{\a}^+)(\bar{D}^{+}_{\dt{\a}}\bar{D}^{+\dt{\a}})~.
\eqno(3.9)$$
In the analytic subspace, the harmonic derivative $D^{++}$ reads 
$$D^{++}_{analytic} = D^{++}-2i\q^+\s^m\bar{\q}^+\pa_m~,\eqno(3.10)$$ 
it preserves analyticity, and it allows one to integrate by parts. Both the 
original (central) basis and the analytic one can be used on equal footing in 
the HSS. In what follows we omit the subscript {\it analytic}  at the 
covariant derivatives in the analytic basis, in order to simplify the notation.

It is the advantage of the analytic $N=2$ HSS compared to the ordinary $N=2$ 
superspace that both an off-shell $N=2$ vector multiplet and an off-shell
hypermultiplet can be introduced there on equal footing. There exist two 
off-shell hypermultiplet versions in HSS, which are dual to each other.
The so-called {\it Fayet-Sohnius-type} (FS) hypermultiplet is defined as an 
unconstrained complex analytic superfield $q^+$ of $U(1)$-charge $(+1)$, 
whereas its dual, called the {\it Howe-Stelle-Townsend-type} (HST) hypermultiplet, 
is a real unconstrained analytic superfield $\o$ with the vanishing 
$U(1)$-charge.~\footnote{It is worth mentioning here that the FS and HST multiplets 
were originally introduced in the \newline ${~~~~~}$ {\it ordinary} $N=2$ 
superspace~\cite{fs,hst}, whereas we use the same names to denote {\it different} $N=2$
\newline ${~~~~~}$ {\it harmonic} superfields, just in order to keep track of their 
on-shell connection.}
The on-shell physical components of the FS hypermultiplet comprise an $SU(2)_A$
doublet of complex scalars and a Dirac spinor which is a singlet w.r.t. the
$SU(2)_A$. The on-shell physical components of the HST hypermultiplet comprise
real singlet and triplet of scalars, and a doublet of chiral spinors. The FS
hypermultiplet is natural for describing a charged $N=2$ matter (e.g. in the 
Coulomb branch), whereas the HST hypermultiplet is natural for describing
a neutral $N=2$ matter in the Higgs branch. 
Similarly, an $N=2$ vector multiplet is described by an unconstrained analytic
superfield $V^{++}$ of the $U(1)$-charge $(+2)$. The $V^{++}$ is real in the 
sense $\Bar{V^{++}}^{\,*}=V^{++}$, and it can be naturally introduced as a 
connection to the harmonic derivative $D^{++}$. 

A free FS hypermultiplet HSS action is given by
$$ S[q]=-\int d\z^{(-4)}du\,\sbar{q}{}^+D^{++}q^+~,\eqno(3.11)$$
whereas its minimal coupling to an $N=2$ gauge superfield reads
$$ S[q,V]= -\tr\int d\z^{(-4)}du \,\sbar{q}{}^+(D^{++}+iV^{++})q^+~,
\eqno(3.12)$$
where the both superfields, $q^+$ and $V^{++}$, are now Lie algebra-valued.

It is not difficult to check that the free FS hypermultiplet equations of 
motion, $D^{++}q^+=0$, imply $q^+=q^i(Z)u^+_i$ as well as the usual (on-shell)
Fayet-Sohnius constraints~\cite{fs} in the ordinary $N=2$ superspace,
$$D_{\a}^{(i}q^{j)}(Z)=D_{\dt{\a}}^{(i}q^{j)}(Z)=0~.\eqno(3.13)$$

Similarly, a free HSS action of the HST hypermultiplet is given by
$$S[\o]=-\frac{1}{2}\int d\z^{(-4)}du \,(D^{++}\o)^2~,\eqno(3.14)$$
and it is equivalent (dual) to the standard $N=2$ tensor (linear)
multiplet action (see subsect.~3.3).

The standard {\it Grimm-Sohnius-Wess} (GSW) constraints~\cite{gs} defining the 
$N=2$ super-Yang-Mills theory in the ordinary $N=2$ superspace imply 
the existence of a (covariantly) chiral~\footnote{A covariantly-chiral
superfield can be transformed into a chiral superfield by field redefinition.}
and gauge-covariant $N=2$ SYM field strength $W$ satisfying, in addition, the 
reality condition (or the Bianchi `identity') 
$$ {\cal D}^{\a}\low{(i}{\cal D}\low{j)\a}W=\bar{\cal D}_{\dt{\a}(i}
\bar{\cal D}^{\dt{\a}}\low{j)}\bar{W}~.\eqno(3.15)$$
Unlike the $N=1$ SYM theory, an $N=2$ supersymmetric solution to the 
non-abelian $N=2$ SYM constraints in the ordinary $N=2$ superspace is not 
known in an analytic form. It is the $N=2$ HSS reformulation of the $N=2$ SYM 
theory that makes it possible~\cite{gikos}. An exact non-abelian relation 
between the constrained, harmonic-independent superfield strength $W$ and the 
unconstrained analytic (harmonic-dependent) superfield $V^{++}$ is given in 
refs.~\cite{gikos,hsf}, and it is highly non-linear. It is merely its abelian 
version that is needed for calculating the perturbative LEEA in the Coulomb 
branch. The abelian relation is given by
$$ W=\fracmm{1}{4} \{ \bar{\cal D}^+_{\dt{\a}},\bar{\cal D}^{-\dt{\a}}\}
=-\fracmm{1}{4}(\bar{D}^+)^2A^{--}~,\eqno(3.16)$$
where the non-analytic harmonic superfield connection $A^{--}(Z,u)$ to the 
derivative $D^{--}$ has been introduced, ${\cal D}^{--}=D^{--}+iA^{--}$.
As a consequence of the $N=2$ HSS abelian constraint 
$\[{\cal D}^{++},{\cal D}^{--}\]={\cal D}^0=D^0$, the connection $A^{--}$ 
satisfies the relation
$$ D^{++}A^{--}=D^{--}V^{++}~,\eqno(3.17)$$
whereas eq.~(3.15) can be rewritten to the form
$$ (D^+)^2W=(\bar{D}^+)^2\bar{W}~.\eqno(3.18)$$

A solution to the $A^{--}$ in terms of the analytic unconstrained superfield
$V^{++}$ easily follows from eq.~(3.17) when using the identity~\cite{hsf}
$$ D^{++}_1(u_1^+u^+_2)^{-2}=D_1^{--}\d^{(2,-2)}(u_1,u_2)~,\eqno(3.19)$$
where we have introduced the harmonic delta-function $\d^{(2,-2)}(u_1,u_2)$ 
and the harmonic distribution $(u_1^+u^+_2)^{-2}$ according to their 
definitions in refs.~\cite{gikos,hsf}, hopefully, in the self-explaining 
notation. One finds~\cite{zupnik} 
$$ A^{--}(z,u)= \int dv \,\fracmm{V^{++}(z,v)}{(u^+v^+)^2}~,\eqno(3.20)$$
and
$$ W(z)=-\fracmm{1}{4}\int du (\bar{D}^-)^2V^{++}(z,u)~,\quad \bar{W}(z)=
-\fracmm{1}{4}\int du (D^-)^2V^{++}(z,u)~,\eqno(3.21)$$
by using the identity 
$$u^+_i=v^+_i(v^-u^+)-v^-_i(u^+v^+)~,\eqno(3.22)$$
which is the obvious consequence of the definitions (3.1). 

The free equations of motion of an $N=2$ vector multiplet are given by the vanishing analytic 
superfield
$$ (D^+)^4A^{--}(Z,u)=0~,\eqno(3.23)$$
while the corresponding action reads~\cite{zupnik}
$$ \eqalign{
S[V]= & \fracmm{1}{4}\int d^4xd^4\theta\, W^2 + {\rm h.c.}=
\fracmm{1}{2}\int d^4xd^4\theta d^4\bar{\theta}du \,V^{++}(Z,u)A^{--}(Z,u)\cr
= & \fracmm{1}{2}\int d^4xd^4\theta d^4\bar{\theta}du_1du_2\,
\fracmm{V^{++}(Z,u_1)V^{++}(Z,u_2)}{(u_1^+u^+_2)^2}~.\cr}\eqno(3.24)$$

In a WZ-like gauge, the abelian analytic HSS prepotential $V^{++}$ amounts 
to the following explicit expression~\cite{gikos}:
$$\eqalign{
 V^{++}(x_{\rm A},\theta^+,\bar{\theta}^+,u)=&
\bar{\theta}^+\bar{\theta}^+a(x_{\rm A})
+ \bar{a}(x_{\rm A})\theta^+\theta^+ 
-2i\theta^+\s^{m}\bar{\theta}^+V_{m}(x_{\rm A}) \cr
& +\bar{\theta}^+\bar{\theta}^+\theta^{\a +}\j^i_{\a}(x_{\rm A})u^-_i
+\theta^+\theta^+\bar{\theta}^+_{\dt{\a}}\bar{\j}^{\dt{\a}i}(x_{\rm A})u^-_i\cr
&+\theta^+\theta^+\bar{\theta}^+\bar{\theta}^+D^{(ij)}(x_{\rm A})u^-_iu^-_j~,
\cr} \eqno(3.25)$$
where $(a,\j^i_{\a},V_{m},D^{ij})$ are the usual $N=2$ vector multiplet 
components~\cite{gs}.

The (BPS) mass of a hypermultiplet can only come from the central charges 
of the 
$N=2$ SUSY algebra since, otherwise, the number of the massive hypermultiplet 
components has to be increased. The most natural way to introduce central 
charges $(Z,\bar{Z})$ is to identify them with spontaneously broken $U(1)$ 
generators of dimensional reduction from six dimensions via the Scherk-Schwarz
mechanism~\cite{ss}. Indeed, after being written down in six dimensions, 
eq.~(3.10) implies an additional `connection' term in the associated 
four-dimensional harmonic derivative,
$$ {\cal D}^{++}=D^{++}+v^{++}~,\quad {\rm where}\quad 
v^{++}=i(\theta^+\theta^+)\bar{Z}+i(\bar{\theta}^+\bar{\theta}^+)Z~.
\eqno(3.26)$$
Comparing eq.~(3.26) with eqs.~(3.12) and (3.21) clearly shows that the $N=2$ 
central charges can be equivalently treated as a non-trivial $N=2$ gauge
background with the covariantly constant chiral superfield strength
$$ \VEV{W}=\VEV{a}=Z~,\eqno(3.27)$$
where eq.~(3.25) has been used too. See refs.~\cite{ke,bbi,bb,ikz} for more 
details.

\subsection{Taub-NUT metric or KK-monopole}

Since the HSS formulation of hypermultiplets has the manifest off-shell $N=2$
supersymmetry, it is perfectly suitable for discussing possible hypermultiplet 
self-interactions which are highly restricted by $N=2$ supersymmetry. 
Moreover, the manifestly $N=2$ supersymmetric Feynman rules can be derived in 
HSS. The latter can be used to actually calculate the perturbative hypermultiplet 
LEEA (see below). 

To illustrate the power of HSS, let's consider a single FS hypermultiplet for
simplicity. Its free action in HSS can be rewritten in the pseudo-real 
notation, $q^+_a=(q^+,\sbar{q}{}^+)$, $q^a=\ve^{ab}q_b$, $a=1,2$, as follows:
$$S[q]= -\ha \int_{\rm analytic} q^{a+}{\cal D}^{++}q^+_a~,\eqno(3.28)$$
where the derivative ${\cal D}^{++}$ (in the analytic basis) includes central 
charges in accordance with eq.~(3.26). It is obvious from eq.~(3.28) that the 
action $S[q]$ has the extended internal symmetry given by
$$  SU(2)_A \otimes SU(2)_{\rm PG}~,\eqno(3.29)$$
where the $SU(2)_A$ is the automorphism symmetry of the $N=2$ supersymmetry
algebra,~\footnote{It is easy to keep track of the ~$SU(2)_A$~ symmetry in the
~$N=2$~ HSS where this symmetry \newline ${~~~~~}$ amounts to the absence of 
an explicit dependence of a HSS lagrangian upon the harmonic \newline 
${~~~~~}$ variables $u^{\pm}_i$.} whereas the additional {\it Pauli-G\"ursey} 
$SU(2)_{\rm PG}$ symmetry acts on the extra indices $(a,b)$ only. Adding a minimal 
interaction with an abelian $N=2$ vector superfield $V^{++}$ in eq.~(3.28)
obviously breaks the internal symmetry (3.29) down to a subgroup
$$ SU(2)_A\otimes U(1)_{\rm PG}~.\eqno(3.30)$$
It is now easy to see that the {\it only} FS hypermultiplet self-interaction consistent with the 
internal symmetry (3.30) is given by the hyper-K\"ahler potential
$$ \ck^{(+4)}=\fracmm{\l}{2}\left(\sbar{q}{}^+q^+\right)^2~,\eqno(3.31)$$
since it is the only admissible term of the $U(1)$-charge $(+4)$ which can be
added to the FS hypermultiplet action (3.28). We thus get the answer for the LEEA
of a single matter hypermultiplet in the Coulomb branch almost for free, up to the 
induced NLSM coupling constant $\l$.

Similarly, the unique FS hypermultiplet self-interaction in the $N=2$ super-QCD
with $N_c=3$ colors and $N_f$ flavors, and vanishing bare hypermultiplet 
masses, which is consistent with the $SU(N_f)\otimes SU(2)_A\otimes U(1)^2$ symmetry, 
is given by
$$ \ck_{\rm QCD}^{(+4)}=\fracmm{\l}{2}\sum^{N_f}_{i,j=1}\left(\sbar{q}{}^{i+}
\cdot q^+_j\right)\left(\sbar{q}{}^{j+}\cdot q^+_i\right)~,\eqno(3.32)$$
where the dots stand for contractions of color indices.

\begin{figure}
\vglue.1in
\makebox{
\epsfxsize=4in
\epsfbox{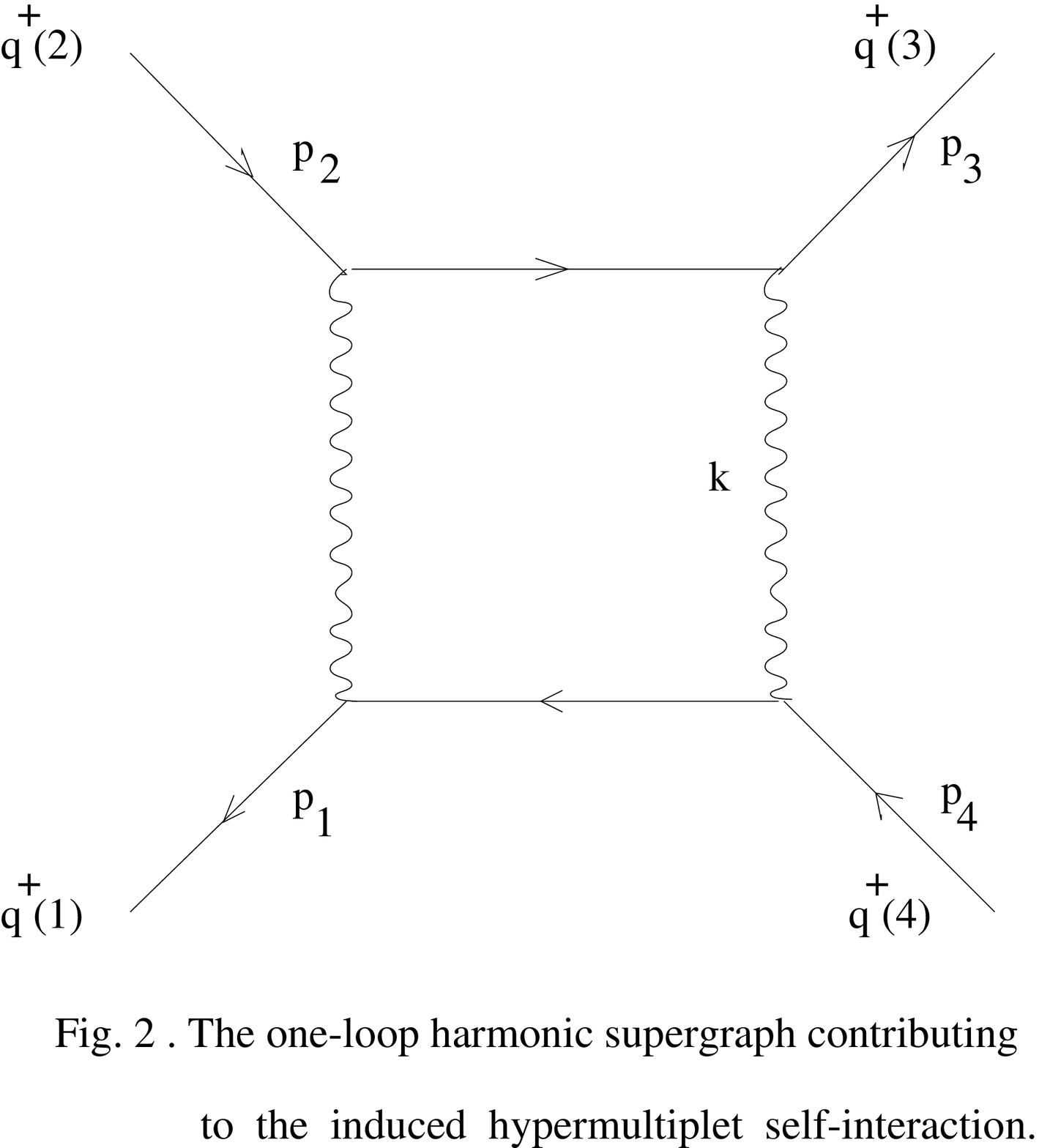}
}
\end{figure}

The induced coupling constant $\l$ in eq.~(3.31) is entirely determined by
the one-loop HSS graph shown in Fig.~2. Since the result vanishes $(\l=0)$
in the absence of central charges,~\footnote{The same conclusion also follows
from the $N=1$ superspace calculations~\cite{gru}.} let's assume that
$Z=\VEV{a}\neq 0$, i.e. we are in the Coulomb branch. The free HSS action of an
$N=2$ vector multiplet and that of a hypermultiplet given above are enough to compute
the corresponding $N=2$ superpropagators. The $N=2$ vector multiplet action
takes the particularly simple form in the $N=2$ super-Feynman gauge (there are
no central charges for the $N=2$ vector multiplet), 
$$ S[V]_{\rm Feynman}=\ha \int_{\rm analytic} V^{++}\Box V^{++}~,\eqno(3.33)$$
so that the corresponding analytic HSS propagator (the wave lines in Fig.~2) 
reads 
$$ i\VEV{ V^{++}(1)V^{++}(2)}=\fracmm{1}{\Box_1}(D_1^+)^4\d^{12}(Z_1-Z_2)
\d^{(-2,2)}(u_1,u_2)~,\eqno(3.34)$$
where the harmonic delta-function $\d^{(-2,2)}(u_1,u_2)$ has been 
introduced~\cite{hsf}. The FS hypermultiplet HSS propagator (solid lines in
Fig.~2) with non-vanishing central charges is more 
complicated~\cite{zupnik,ikz}:
$$ i\VEV{q^+(1)q^+(2)}=\fracmm{-1}{\Box_1+a\bar{a}}
\fracmm{(D^+_1)^4(D^+_2)^4}{(u^+_1u_2^+)^3}e^{\t_3[v(2)-v(1)]}\d^{12}
(Z_1-Z_2)~,\eqno(3.35)$$
where $v$ is the so-called 'bridge' satisfying the equation 
${\cal D}^{++}e^v=0$. One easily finds that
$$ iv =-a(\bar{\theta}^+\bar{\theta}^-)-\bar{a}(\theta^+\theta^-)~.\eqno(3.36)
$$
The rest of the $N=2$ HSS Feynman rules is very similar to that of the 
ordinary $(N=0)$ Quantum Electrodynamics (QED). 

A calculation of the HSS graph in Fig.~2 is now straightforward, while the 
calculational details are given in ref.~\cite{ikz}. One finds the predicted
form of the induced hyper-K\"ahler potential as in eq.~(3.31) indeed, with
the induced NLSM coupling constant given by
$$ \l=\fracmm{g^4}{\p^2}\left[ \fracmm{1}{m^2}\ln\left( 1+\fracmm{m^2}{\L^2}
\right) -\fracmm{1}{\L^2+m^2}\right]~,\eqno(3.37)$$
where $g$ is the gauge coupling constant, $m^2=\abs{a}^2$ is the hypermultiplet
BPS mass, and $\L$ is the IR-cutoff parameter. Note that $\l\to 0$ when the
central charge $a\to 0$.

The only technical problem left is how to decode the HSS result (3.31) in the 
conventional component form. In other words, we have to deduce an explicit
hyper-K\"ahler metric which corresponds to the hyper-K\"ahler potential (3.31).
A general procedure of getting the component form of the bosonic NLSM from a
hypermultiplet self-interaction in HSS consists of the following steps:
\begin{itemize}
\item expand the equations of motion in the Grassmann (anticommuting) 
coordinates, and ignore all the fermionic field components,
\item solve the kinematical linear differential equations for all the 
auxiliary fields, thus eliminating the infinite tower of them in the harmonic 
expansion of the hypermultiplet HSS analytic superfields,
\item substitute the solution back into the HSS hypermultiplet action, and
integrate over all the anitcommuting and harmonic HSS coordinates.
\end{itemize}

Of course, it is not always possible to actually perform this procedure. For
instance, just the second step above would amount to solving infinitely many 
linear differential equations altogether. However, just in the case of 
eq.~(3.31), the explicit solution exists~\cite{giost,ikz}. When using the
parametrization
$$ \left. q^+\right|_{\theta=0}=f^i(x)u^+_i\exp\left[ \l f^{(j}(x)\bar{f}^{k)}
(x)u^+_j u^-_k\right]~,\eqno(3.38)$$
one finds the following bosonic $D=4$ NLSM action (see Appendix A for details of calculation):
$$ S_{\rm NLSM}=\int d^4x\,\left\{ g_{ij}\pa_mf^i\pa^mf^j +\bar{g}^{ij}
\pa_m\bar{f}_i\pa^m\bar{f}_j +h\ud{i}{j}\pa_mf^j\pa^m\bar{f}_i -V(f)
\right\}~,\eqno(3.39)$$
whose metric is given by~\cite{giost}
$$ g_{ij}= \fracmm{ \l(2+\l f\bar{f}}{4(1+\l f \bar{f})}\bar{f}_i\bar{f}_j~,
\quad \bar{g}^{ij}=\fracmm{\l(2+\l f\bar{f}}{4(1+\l f\bar{f})}f^if^j~,$$
$$h\ud{i}{j}=\d\ud{i}{j}(1+\l f\bar{f})-\fracmm{\l(2+\l f\bar{f})}{2(1+
\l f\bar{f}}f^i\bar{f}_j~,\quad f\bar{f}\equiv f^i\bar{f}_i~,\eqno(3.40)$$
and the induced scalar potential reads~\cite{ikz}
$$V(f) =\abs{Z}^2\fracmm{f\bar{f}}{1+\l f\bar{f}}~~~.\eqno(3.41)$$
In the form (3.40) the induced metric is apparently free from singularities.
 
It is usually non-trivial to compare a given metric with any standard hyper-K\"ahler 
metric since the metrics themselves are defined modulo field redefinitions, i.e. 
modulo four-dimensional diffeomorphisms in the case under considerarion. 
Fortunately, it is known how to transform the metric (3.40) to the standard 
{\it Euclidean Taub-NUT} (ETN) form~:
$$ ds^2=\fracmm{r+M}{2(r-M)}dr^2+\ha(r^2-M^2)(d\vartheta^2+\sin^2\vartheta
d\varphi^2)+2M^2\left( \fracmm{r-M}{r+M}\right)
(d\j+\cos\vartheta d\varphi)^2~, \eqno(3.42)$$
by using the following change of variables~\cite{giost}:
$$f^1= \sqrt{2M(r-M)}\cos\fracmm{\vartheta}{2}\exp\fracmm{i}{2}(\j+\varphi)~,
$$
$$f^2= \sqrt{2M(r-M)}\sin\fracmm{\vartheta}{2}\exp\fracmm{i}{2}(\j-\varphi)~,
 \eqno(3.43)$$
$$ f\bar{f}=2M(r-M)~,\qquad r\geq M=\fracmm{1}{2\sqrt{\l}}~,$$
where $M=\ha\l^{-1/2}\sim g^{-2}$ is the instanton mass. The ETN instanton is also known 
in the literature as the KK-instanton~\cite{sork,grpe,townr} (see sect.~4 for more).

We thus showed in this subsection that the induced NLSM metric of the hypermultiplet LEEA 
in the Coulomb branch is generated in the one-loop order of quantum perturbation theory, 
and it is given by the Taub-NUT or its higher-dimensional generalizations. The non-trivial 
scalar potential is also generated by one-loop quantum corrections, with the hypermultiplet 
(BPS) mass being unrenormalized as it should.

\subsection{Duality transformation and $N=2$ tensor multiplet}

There exists an interesting connection between the FS hypermultiplet Taub-NUT
self-interaction in the $N=2$ {\it harmonic} superspace and the $N=2$ tensor
(or linear) multiplet self-interaction in the {\it ordinary} $N=2$ superspace.
Namely, the $N=2$ supersymmetric Taub-NUT NLSM is equivalent to a sum of the
naive (quadratic in the fields and non-conformal) and {\it improved} 
(non-polynomial in the fields and $N=2$ superconformally invariant) actions
for the $N=2$ tensor multiplet in the ordinary $N=2$ superspace~!

The $N=2$ {\it tensor} multiplet in the ordinary $N=2$ superspace is defined by the constraints
$$ D\low{\a}{}^{(i}L^{ik)}(Z)=\bar{D}_{\dt{\a}}{}^{(i}L^{jk)}(Z)=0~,\eqno(3.44)$$
and the reality condition
$$ \Bar{L^{ij}}=\ve_{ik}\ve_{jl}L^{kl}~.\eqno(3.45)$$
Unlike the FS hypermultiplet in the ordinary $N=2$ superspace, the constraints
(3.44) are off-shell, i.e. they do not imply the equations of motion for the
components of the $N=2$ tensor multiplet. The $N=2$ tensor multiplet itself 
can be identified as a restricted HST hypermultiplet (i.e. as an analytic $\o$ 
superfield subject to extra off-shell constraints), while its $N=2$
supersymmetric self-interactions are a subclass of those for $\o$~\cite{gio}.
The $N=2$ tensor multiplet has $8_{\rm B}\oplus 8_{\rm F}$ off-shell 
components:
$$ \vec{L},\quad \z^i_{\a}~,\quad B~,\quad {E'}_{m}=\ha\ve_{mnpq}\pa_nE_{pq}~,
\eqno(3.46)$$
where $\vec{L}$ is the scalar $SU(2)_A$ triplet, $\vec{L}=\tr(\vec{\t}L)$ and
$\vec{\t}$ are Pauli matrices, $\z^i_{\a}$ is a chiral spinor doublet, $B$ is a
complex auxiliary scalar, and $E_{mn}$ is a gauge antisymmetric tensor whose
field strength is ${E'}_m$.

Let's start with our induced hypermultiplet LEEA
$$ S[q^+]_{\rm Taub-NUT} = \int_{\rm analytic} \left[ \sbar{q}{}^+D^{++}q^+
+ \fracmm{\l}{2}(q^+)^2(\sbar{q}{}^+)^2\right]~,\eqno(3.47)$$
and make the following substitution of the HSS superfield variables~\cite{gio}:
$$ \sqrt{\l}q^+=-i\left(2u^+_1+ig^{++}u^-_1\right)e^{-i\tilde{\o}/2}~,
\quad \sqrt{\l}\sbar{q}{}^+=i\left( 2u^+_2-ig^{++}u^-_2\right)e^{i\tilde{\o}
/2}~,\eqno(3.48)$$
where
$$ g^{++}(l,u)\equiv \fracmm{2(l^{++}-2iu_1^+u_2^+)}{1+\sqrt{1-4u^+_1u^+_2
u^-_1u^-_2 -2il^{++}u_1^-u_2^-}}~,\eqno(3.49)$$
and $(l^{++},\o)$ are the new dimensionless analytic superfieds. It is not
difficult to check that eqs.~(3.48) and (3.49) imply, in particular, that
$$ \l \sbar{q}{}^+q^+=2il^{++}~,\eqno(3.50)$$
whereas the action (3.47) takes the form (after the rescaling $l^{++}\equiv
\sqrt{\l}L^{++}$ and $\tilde{\o}=\sqrt{\l}\o$):
$$ S[L^{++};\o]\low{\rm Taub-NUT}=
S\low{\rm free}[L^{++};\o] + S\low{\rm impr.}[L^{++}]~,\eqno(3.51)$$
where 
$$S\low{\rm free}[L^{++};\o]= 
\ha \int_{\rm analytic}\left[ (L^{++})^2+\o D^{++}
L^{++}\right]~,\eqno(3.52)$$
and
$$ S\low{\rm impr.}[L^{++}]= 
\fracmm{1}{2\l}\int_{\rm analytic} \left[g^{++}
(L;u)\right]^2~.\eqno(3.53)$$

The action (3.51) or (3.52) has the superfield $\o$ as a Lagrange multiplier. Hence, on the
one hand side, varying the action (3.51) with respect to $\o$ yields the constraint
$$ D^{++}L^{++}=0~,\eqno(3.54)$$
which, in its turn, implies $L^{++}=u^+_iu^+_jL^{ij}(Z)$ {\it and} 
eq.~(3.44). Therefore, the actions (3.51) and (3.52) describe an $N=2$ tensor 
multiplet in the $N=2$ HSS. On the other hand side, one can vary e.g. the action (3.52) 
with respect to $L^{++}$ first. Then one finds that
$$ L^{++}=D^{++}\o~.\eqno(3.55)$$
Hence, $L^{++}$ can be removed in favor of $\o$. This example is a
manifestation of the classical {\it duality} between the FS 
hypermultiplet $q^+$ and the HST hypermultiplet $\o$ in the $N=2$ HSS. 

The action (3.53) describes the so-called {\it improved} $N=2$ tensor 
multiplet~\cite{impr}. It can be shown that it is fully invariant under the
rigid $N=2$ superconformal symmetry, while the associated hyper-K\"ahler
metric is equivalent to the flat metric up to a $D=4$ diffeomorphism 
\cite{impr}. However, the sum of the actions (3.52) and (3.53) describes an 
interacting theory, and it is just the NLSM with the Taub-NUT instanton (or 
KK-monopole) metric.

Because of this connection between certain $N=2$ supermultiplets and their 
self-interactions in the HSS, it should not be very surprising that the
Taub-NUT self-interaction can also be reformulated in the {\it ordinary} $N=2$
superspace in terms of the $N=2$ tensor multiplet alone, just as the sum of its
naive and improved actions. The most elegant formulation of the latter exists
in the {\it projective} $N=2$ superspace~\cite{klr,myrev} in which the harmonic
variables are replaced by a single complex projective variable $\x\in CP(1)$.
Unlike the $N=2$ HSS, the projective $N=2$ superspace does not have to 
introduce extra auxiliary fields beyond those already present in the off-shell 
$N=2$ tensor multiplet. The starting point are the defining constraints (3.44)
for the $N=2$ tensor multiplet in the ordinary $N=2$ superspace. It is not
difficult to check that they imply (see ref.~\cite{myrev} for more details and
generalizations)
$$ \de_{\a}G\equiv (D^1_{\a}+\x D^2_{\a})G=0~,\qquad \D_{\dt{\a}}G\equiv 
(\bar{D}^1_{\dt{\a}}+\x\bar{D}^2_{\dt{\a}})G=0~,\eqno(3.56)$$
for {\it any} function $G(Q(\x),\x)$ which is a function of
$Q(\x)\equiv \x_i\x_j L^{ij}(Z)$ and $\x_i\equiv (1,\x)$ only.

It follows that we can build an $N=2$ superinvariant just by integrating $G$
over the rest of the $N=2$ superspace coordinates in the directions which are
'orthogonal' to those in eq.~(3.56), namely,
$$ S_{\rm inv.}[L]= \int d^4x\fracmm{1}{2\p i}\oint_C d\x\,\tilde{\de}^2
\tilde{\D}^2G(Q,\x)~,\eqno(3.57)$$
where we have introduced the new derivatives
$$ \tilde{\de}_{\a}=\x D^1_{\a}-D^2_{\a}~,\quad \tilde{\D}_{\dt{\a}}=\x
\bar{D}^1_{\dt{\a}}-\bar{D}^2_{\dt{\a}}~.\eqno(3.58)$$

The choice of the function $G(Q,\x)$ and the contour $C$ in the complex 
$\x$-plane, which yields the Taub-NUT self-interaction in eq.~(3.57), is
given by~\cite{klr,myrev}
$$ S\low{\rm Taub-NUT}[L]= 
\int d^4x\, \tilde{\de}^2\tilde{\D}^2\fracmm{1}{2\p i}
\left\{ \oint_{C_1} d\x\,\fracmm{Q^2}{2\x} +\fracmm{1}{\sqrt{\l}}\oint_{C_2}
d\x\, Q\ln(\sqrt{\l} Q)\right\}~,\eqno(3.59)$$
where the contour $C_1$ goes around the origin, whereas the contour $C_2$
encircles the roots of the quadratic equation $Q(\x)=0$ in the complex
$\x$-plane.

The $N=2$ superfield Feynman rules can also be developed in the $N=2$ projective 
superspace~\cite{gonz}, where they are likely to be deducible from the more general HSS rules.
Partial (abelian) results are also available in the ordinary $N=2$ superspace~\cite{myrev}.

Finally, one may wonder, in which sense an $N=2$ tensor multiplet action
describes a $D=4$, $N=2$ supersymmetric NLSM with the highest physical spin 
$1/2$, because of the apparent presence of the gauge antisymmetric tensor 
$E_{mn}$ among the $N=2$ tensor multiplet components --- see eq.~(3.46). A 
detailed investigation of the component action, which follows from the 
superspace action (3.59), shows that the tensor $E_{mn}$ and its field 
strength ${E'}_m$ enter the action only in the combination 
$$ \left( 1+ \fracmm{1}{\l\vec{L}^2}\right)
({E'}_m)^2+\ha\ve_{mnpq}E_{pq}F_{mn}(L)~,\eqno(3.60)$$
where the tensor
$$ F_{mn}(L)\equiv \left(\pa_m\vec{L}\times\pa_n\vec{L}\right)\cdot
\fracmm{\vec{L}}{\abs{\vec{L}}^3}\eqno(3.61)$$
is formally identical to the electromagnetic field strength of a magnetic 
monopole. Therefore, there exists a vector potential $A_m$ such that 
$F_{mn}(L)=\pa_mA_n -\pa_nA_m$. An explicit magnetic monopole solution for the
locally defined potential $A_m(\vec{L})$ cannot be 'rotationally' invariant 
with respect to the $SO(3)\sim SU(2)_A/Z_2$ symmetry, though it can be written 
down as a function of the $SO(2)$-irreducible $L^{ij}$-components defined by 
$L^{ij}=\d^{ij}S+P^{(ij)}_{\rm traceless}$. After integrating
by parts and introducing a Lagrange multiplier $V$ as
$$ {}*EF={}^*EdA\to -d{}^*EA=-{E'}_mA_m\to -E_mA_m-E_m\pa_mV~,\eqno(3.62)$$
we can integrate out the full vector $E_m$. It results in the bosonic NLSM
action in terms of four real scalars $(S,P^{(ij)}_{\rm traceless},V)$, as it should.

\subsection{Induced multicentre Taub-NUT metrics}

The {\it Euclidean Taub-NUT} (ETN) metric is a unique non-trivial hyper-K\"ahler four-dimensional 
metric having the $U(2)$ isometry whose transformation laws are linear and 
holomorphic~\cite{gipop}. Without the holomorphicity requirement, there exists another 
hyper-K\"ahler solution known as the {\it Eguchi-Hanson} (EH) four-dimensional 
instanton~\cite{eghrev}. The EH metric also possess the $U(2)$ isometry, and its possible 
appearance in the hypermultiplet LEEA will be discussed in sect.~6 (see also Appendix B for 
a derivation of the EH metric from harmonic superspace). 

From the viewpoint of $N=2$ supersymmetry in four spacetime dimensions, the origin of the 
$U(2)=SU(2)\times U(1)$ internal symmetry in the ETN- and EH-type hypermultiplet LEEA is, however,
quite different. In the ETN-case, the $SU(2)$ isometry factor is identified with the $SU(2)_A$ 
automorphisms of $N=2$ supersymmetry, whereas the $U(1)$ isometry factor is identified with the
unbroken subgroup $U(1)_{\rm PG}$ of the Pauli-G\"ursey (PG) symmetry $SU(2)_{\rm PG}$ of the free 
FS-type hypermultiplet action (3.28). In the EH-case, the $SU(2)$ isometry factor has to be 
identified with the $SU(2)_{\rm PG}$ symmetry, whereas the $U(1)$ isometry factor is just the
unbroken abelian part of the $SU(2)_A$ automorphisms. All these identifications are quite obvious 
from the viewpoint of HSS: the ETN-type hypermultiplet self-interaction in HSS maintains $SU(2)_A$ 
but breaks $SU(2)_{\rm PG}$ down to its abelian subgroup $U(1)_{\rm PG}$ (see subsect.~3.2 and 
Appendix A), whereas the EH-type hypermultiplet self-interaction (see sect.~6 and Appendix B) 
breaks  $SU(2)_A$ down to its abelian subgroup $U(1)_A$ but maintains $SU(2)_{\rm PG}\,$.

A natural generalization is provided by the FS-type hypermultiplet LEEA whose induced 
hyper-K\"ahler metric merely has the abelian $U(1)\times U(1)$ isometry, where the first $U(1)$ 
isometry factor is supposed to be identified with $U(1)_A$ whereas the second isometry $U(1)$ 
factor is $U(1)_{\rm PG}$. Since the $SU(2)_A$ internal symmetry is not anomalous in $N=2$
supersymmetric quantum perturbation theory, its breaking can only be caused either 
(i) by $SU(2)_A$ non-invariant terms in the fundamental lagrangian, with the Fayet-Iliopoulos (FI) 
term (see sect.~6) being an example, or 
(ii) non-perturbatively, with the BPS branes intersecting at angles (see sect.~6) being an example.

The most general $U(1)_A\times U(1)_{\rm PG}$-invariant hyper-K\"ahler potential $\ck$ is given by 
a general, of overall charge $(+4)$, analytic function of the product $(\sbar{q}q)^{(+2)}$, i.e.  
$$ S[q]=\int_{\rm analytic}\,\left\{ \sbar{q}{}^+ D^{++}q^+ +\ck^{(+4)}(\sbar{q}q)\right\}~,
\eqno(3.63)$$
with
$$ \ck^{(+4)}(\sbar{q}q)=\sum_{l=0}^{\infty}\x^{(-2l)}\fracmm{(\sbar{q}{}^+q^+)^{l+2}}{l+2}~,
\eqno(3.64)$$
where the harmonic-dependent `coefficients' $\,\x^{(-2l)}(u)$ are defined by 
$$\x^{(-2l)}=\x^{(i_1\cdots i_{2l})}u^-_{i_1}\cdots u^-_{i_{2l}}~, 
\quad l=1,2,\ldots~,\eqno(3.65)$$
and satisfy the reality condition
$$ \sbar{\x}{}^{(-2l)}=(-1)^l\x^{(-2l)}~.\eqno(3.66)$$

As was shown in refs.~\cite{giru,gorv}, in fact, {\it any} four-dimensional hyper-K\"ahler metric 
having the $U(1)_{\rm PG}$ isometry belongs to the class of the multicentre {\it Gibbons-Hawking} 
(GH) metrics~\cite{gha}, which includes the ETN and EH 
metrics as the particular cases. Since the multicentre metrics are of special importance to
the brane technology (see the next sect.~4), where they are usually described in terms of a 
harmonic function H (see eqs.~(3.70)--(3.72) below), it is useful to establish a 
correspondence between the two equivalent descriptions, the one in terms of an {\it analytic} 
hyper-K\"ahler potential $\ck$ and another one in terms of a {\it harmonic} (singular) function 
$H$ satisfying the Laplace equation 
$$ \D H=0 \eqno(3.67)$$
outside the origin $(r=0)$ in three Euclidean dimensions.~\footnote{Because of the $U(1)_{\rm PG}$
global isometry, the function $H$ can be chosen to be independent upon \newline
${~~~~~}$ one coordinate.} This 
correspondence was established in ref.~\cite{gorv}. Given a general solution to eq.~(3.67) in 
spherical coordinates $(r,\vq,\vf)$, which reads
$$ H=\fracmm{1}{2r}+\fracmm{U(\vec{r})}{2}\equiv
\fracmm{1}{2r}+\fracmm{1}{2}\sum^{+\infty}_{l=0}\,\sum_{m=-l}^{m=+l}\,c_{lm}r^lY_l^m(\vq,\vf)~,
\eqno(3.68)$$
in terms of the standard momentum eigenfunctions $Y_l^m(\vq,\vf)$, the one-to-one correspondence 
between the integration constants $c_{lm}$ in eq.~(3.68) and the hyper-K\"ahler potential 
coefficients (3.65) is given by~\cite{gorv}
$$ \x^{i_1=1,\ldots,i_{l-m}=1,i_{l-m+1}=2,\ldots,i_{2l}=2}=\fracmm{c_{lm}}{C}
\fracmm{(2l+1)}{(l+1)}~,\eqno(3.69)$$
which some normalization-dependent constant $C$. 

A multicentre GH metric is usually written down in the form
$$  ds^2=H d\vec{r}\cdot d\vec{r} +H^{-1}(d\varrho +\vec{C}\cdot d\vec{r})^2~,\eqno(3.70)$$
where the vector $\vec{C}(\vec{r})$ satisfies the equation
$$ \vec{\de}H=\vec{\de}\wedge \vec{C}~,\eqno(3.71)$$
while the harmonic function $H$ has the `milticentre' form 
$$ H(\vec{r})=\fracmm{\l}{2}+
\sum^n_{i=1} \fracmm{\abs{k_i}}{2\abs{\vec{r}-\vec{r_i}}}~.\eqno(3.72)$$

The harmonic function (3.72) can always be put into the form (3.68), when choosing the coordinate 
system whose origin coincides with one of $\vec{r}_i$, extracting the leading singular term and 
then expanding the regular rest in the spherical harmonics.

A multicentre metric can therefore be naturally interpreted as the metric describing a static 
multi-monopole configuration whose monopoles have magnetic charges $k_i$ and sit at the space 
points $\vec{r_i}$. The $4n$-parameters $\{\vec{r}_i,k_i\}$ in eq.~(3.72), i.e. the {\it moduli} 
of this hyper-K\"ahler configuration, thus have the very clear physical meaning in terms of the 
harmonic function description, while the harmonic function $H$ itself has singularities at the
positions of the monopoles. In the alternative HSS description of the same multi-monopole 
configuration, though the HSS moduli $\x^{(i_1\cdots i_{2l})}$ have no direct physical 
interpretation, the description itself in terms of the analytic hyper-K\"ahler potential is 
apparently {\it non-singular}. The latter will be useful for the brane technology, when describing 
the internal symmetry enhancement of coinciding branes in non-singular terms (subsect.~4.2).

\section{Brane technology}

It is quite remarkable that one can also get exact LEEA solutions to the $D=4$, $N=2$ {\it quantum}
gauge field theories from the {\it classical} M-theory brane dynamics~\cite{witten} (see 
refs.~\cite{mikh,brit} also). Among a few things one knows about M-Theory, only two general 
statements are going to be used in what follows, namely, that (i) M-theory is the strong coupling 
limit of the type-IIA superstring theory (which is rather non-constructive), and (ii) the 
low-energy limit of M-theory is $D=11$ supergravity. Because of (ii), stable and unique BPS states 
of the $D=11$ supergravity can be exploited to extract non-perturbative
information about M-theory. This information can then be applied to study the effective 
supersymmetric gauge field theories in the BPS brane world-volumes. These `in-M-brane' gauge 
theories are not quite the same as the ones we studied in the previous sections (see the end of
subsect.~4.6 and sect.~5), though they can share the same LEEA under certain circumstances to be 
discussed below. The supersymmetric BPS branes, which are relevant to this {\it brane technology}, 
are introduced in the next subsections, along the lines of the existing reviews~\cite{town,stel,jer} 
(see refs.~\cite{town,stel,jer} and references therein for more details).

\subsection{$D=11$ supergravity and its BPS solutions}

$D=11$ is believed to be the maximum dimension of spacetime (with Lorentzian signature), where 
a consistent interacting supersymmetric field theory exists~\cite{nahm}. This is essentially the 
consequence of the fact that the massless physical particles mediating long-range forces in our 
$D=4$ spacetime can have spin 2 at most, while there is only one type of particles of spin 2 
(i.e. gravitons). A supersymmetry charge (a component of $D=4$ spinor) changes helicity $\l$ of 
a massless particle by a half, so that the maximal non-vanishing product of $N$ supersymmetry 
charges changes $\l$ by $N/2$. Hence, in a massless representation of the $N$-extended $D=4$ 
supersymmetry, the helicity varies from $\l$ to $\l+\ha N$. It immediately implies 
$N\leq 8$ provided that $\abs{\l}\leq 2$. The maximal $N=8$ supersymmetry in $D=4$ 
has $8\times 4=32$ real component charges,
while the maximal spacetime dimension (with Lorentzian signature), where a minimal spinor 
representation also has real 32 components, is just $D=11$. The simplest supersymmetry algebra in 
$D=11$ takes the form
$$ \{ Q_{\a},Q_{\b}\}=(C\G^M)_{\a\b}P_M~,\eqno(4.1)$$
where $Q_{\a}$ is a Majorana spinor supersymmetry charge, $\a=1,2,\ldots,32$, $P_M$ is a $D=11$
spacetime momentum operator, $\G^M$ are $D=11$ real gamma matrices, and $C$ is the charge 
conjugation matrix in $D=11$, while $M=0,1,2,\ldots,10$, and Minkowski flat metric is 
$\h^{MN}={\rm diag}(-,+,\ldots,+)$.

The $D=11$ supergravity~\cite{cjs} is described in terms of three fields: a metric $g_{MN}$, a 
gravitino $\j_{M\a}$ and a 3-form gauge potential $A_{MNP}$ with a gauge transformation 
$\d A_{(3)}=d\L_{(2)}$ and a field strength $F_{(4)}=dA_{(3)}$. The theory has 
$128_{\rm B}+128_{\rm F}$ on-shell physical components. The bosonic part of the $D=11$
supergravity action reads
$$ S_{\rm bosonic}=\frac{1}{2}
\int d^{11}x\left\{ \sqrt{-g}(R-\frac{1}{48}F^2) + \frac{1}{6}F_{(4)}\wedge 
F_{(4)}\wedge A_{(3)}\right\}~,\eqno(4.2)$$ 
where we have taken the gravitational coupling constant to be equal to one. The last (Chern-Simons)
term in eq.~(4.2) is required by $D=11$ supersymmetry of the total action including the 
gravitino-dependent terms.

A quantized $D=11$ supergravity is not expected to be a consistent theory e.g., because of its
apparent non-renormalizability. Instead, it should rather be interpreted as the {\it effective} 
low-energy approximation (LEEA) to the presumably consistent M-Theory. This fact alone leads to 
many far-reaching consequences. Though we are unable to describe underlying (microscopic) dynamics 
of the M-Theory from its LEEA alone, knowing the latter allows us to determine the spectrum of BPS 
states in the M-theory, just by constructing the classical solutions to the $D=11$ supergravity
equations of motion, which preserve some part $(\n)$ of the $D=11$ supersymmetry and, hence, 
some part of the translational invariance as well. Since the 
supersymmetry variations of the bosonic fields of $D=11$ supergravity are all proportional to the 
gravitino field, the latter should vanish in a supersymmetric solution, $\j_{M\a}=0$. The 
vanishing supersymmetry variation of the gravitino field itself,
$$ \left.\d\j_{M}\right|_{\j=0}=\Tilde{D}_M\ve\equiv \left( D_M 
-\frac{1}{288}[\G\du{M}{NPQS}-8\d^N_M\G^{PQS}]F_{NPQS}\right)\ve=0~,\eqno(4.3)$$
where $D_M\ve=(\pa_M +\frac{1}{4}\o\du{M}{BC}\G_{BC})\ve$, then implies the existence of a Majorana
{\it Killing spinor\/} field $\ve$ which satisfies the first-order differential equation (4.3). In 
eq.~(4.3) we denoted by $\G^{M_1\cdots M_p}$ the antisymmetric products of the gamma matrices with 
unit weight. 

Assuming the existence of asymptotic states with a supersymmetric vacuum, and requiring the $D=11$
metric to be asymptotically Minkowskian, it is easy to see that the only BPS states with respect to
the supersymmetry algebra (4.1) are just massless particles, since~\footnote{It is the asymptotical
form of the local $D=11$ supersymmetry algebra that is given by the rigid \newline ${~~~~~}$
superalgebra (4.1). The Killing spinor $\ve(x)$ should also be constant at infinity.}
$$ 0=\det\VEV{\{Q_{\a},Q_{\b} \} }=\VEV{\det(\G\cdot P) }=\VEV{ (P^2)^{16} }~,\eqno(4.4)$$
while the matrix $\{Q_{\a},Q_{\b}\}$ has 16 independent zero eigenvalues $(\n=\ha)$. This
simply means that a massless representation of $D=11$ supersymmetry is $\ha$-shorter than the 
massive one, as is well-known in supersymmetry {\it without} central charges. The corresponding 
asymptotically flat classical BPS solution of $D=11$ supergravity with $P^2=0$ (called M-wave) was
found e.g. in ref.~\cite{hwave}.  

It is one of the lessons of $D=4$ gauge field theory that the massless particles appearing in a 
perturbative spectrum may not be the only BPS states. Non-perturbative (massive) BPS states in 
extended $D=4$ supersymmetry carry electric and magnetic charges saturating the Bogomolnyi bound, 
whereas these charges appear as the central charges on the right-hand-side of the supersymmetry 
algebra. The symmetric matrix on the left-hand-side of eq.~(4.1) belongs to the adjoint 
representation $\un{\bf 528}$ of the Lie algebra of $Sp(32)$, which is decomposed with respect to 
its (Lorentz) subgroup $SO(1,10)$ as 
$$ \un{\bf 528}\to \un{\bf 11}\oplus \un{\bf 55}\oplus\un{\bf 462}~.\eqno(4.5)$$
The $\un{\bf 11}$ is apparently associated with $P_M$ in eq.~(4.1), whereas the rest has to be
associated with some additional `central' charges commuting with supersymmetry charges and monenta,
but not commuting with Lorentz rotations. The $D=11$ Lorentz representations
$\un{\bf 55}$ and $\un{\bf 462}$ are associated with a 2-form $Z_{(2)}$ and a 5-form $Y_{(5)}$, 
respectively, so that the maximal $D=11$ supersymmetry algebra reads~\cite{hopr}
$$ \{ Q_{\a},Q_{\b}\}=(C\G^M)_{\a\b}P_M~+\frac{1}{2}(\G^{MN}C)_{\a\b}Z_{MN}
+\frac{1}{5!}(\G^{MNPQS}C)_{\a\b}Y_{MNPQS}~,\eqno(4.6)$$
where $Z_{MN}$ represent the `electric' charges and $Y_{MNPQS}$ are the `magnetic' ones. The BPS
object carrying non-vanishing electric charges is known as a {\it supermembrane} or an electric
{\it M-2-brane}~\cite{agit}. Associated with the $D=11$ spacetime symmetries broken by the 
supermembrane are the {\it Nambu-Goldstone} (NG) modes. The three-dimensional LEEA action 
describing the dynamics of small fluctuations of the NG fields about the supermembrane in 
a $D=11$ supergravity background was discovered by Bergshoeff, Sezgin and Townsend 
\cite{bst}.~\footnote{An action of NG fields is entirely determined by the broken symmetries 
and, hence, it is unique.}  

The BPS object, which is magnetically {\it dual} to the M-2-brane in eleven dimensions, is a 
magnetically charged 5-brane called {\it M-5-brane}. Indeed, according to Gauss's law, the electric
charge of a particle (i.e. $0$-brane) in some number $(D)$ of spacetime dimensions is measured by 
the dual gauge field strength according to the integral 
$Q_{\rm electric}=\int_{S^{D-2}}{}^*F$ over the sphere $S^{D-2}$ surrounding the particle, where 
$F_{(2)}=dA_{(1)}$ is the abelian field strength of a $U(1)$ gauge field $A_{(1)}$ and 
${}^*F_{(D-2)}$ is the Hodge dual to $F_{(2)}$ in $D$ dimensions. In the case of an `electric' 
p-brane charged with respect to a gauge $(p+1)$-form $A_{(p+1)}$ in $D$ dimensions, the field 
strength is $F_{(p+2)}=dA_{(p+1)}$ and its dual is ${}^*F_{(D-p-2)}$. For {\it magnetically} 
charged objects the roles of $F$ and ${}^*F$ are supposed to be interchanged. For example, the 
object carrying a magnetic charge in $D=4$ is again a 0-brane (i.e. particle or monopole) since the 
dual potential $\tilde{A}$ defined by ${}^*F_{(2)}=d\tilde{A}$ is a 1-form, whereas the charge of 
the D=4 monopole is measured by $F_{(2)}$ as $Q_{\rm magnetic}=\int_{S^2} F$. Similarly, since the 
potential $\tilde{A}$ of the dual field strength $*F_{(D-p-2)}$ is a $(D-p-3)$-form, 
${}^*F_{(D-p-2)}=d\tilde{A}_{(D-p-3)}$, it is a $(D-p-4)$-brane that can support magnetic charges. 
The well-known `golden rule' for an electrically charged p-brane and its dual, magnetically charged 
q-brane thus reads
$$ p+q=D-4~.\eqno(4.7)$$
Given $D=11$ and $p=2$, one has $q=5$. The magnetic charge of an M-5-brane is proportional to the
integral $\int_{S^4}F_{(4)}$ over the sphere $S^4$ surrounding the brane at spacial infinity in 
five directions transverse to its six-dimensional worldvolume. The integral is obviously 
topological (i.e. homotopy invariant) due to the Bianchi identity $dF_{(4)}=0$.

The explicit form of electric (M-2-brane) and magnetic (M-5-brane) BPS solutions to the $D=11$ 
supergravity is known (see, e.g., the reviews~\cite{town,stel,jer} and references therein). For 
our purposes we only need a solitonic 5-brane solution found by G\"uven~\cite{guv}, which reads
$$ 
ds^2=H^{-1/3}(y)dx^{\m}dx^{\n}\h_{\m\n}+ H^{2/3}(y)dy^mdy^n\d_{mn}~,\qquad
F_{(4)}={}^*_5dH~,\eqno(4.8)$$
where the $D=11$ spacetime coordinates have been split into the `worldvolume' coordinates labeled
by $\m,\n=0,1,2,3,4,5$ and the `transverse to the worldvolume' coordinates labeled by 
$m,n=6,7,8,9,10$, according to the spacetime decomposition $R^{1,10}=R^{1,5}\times R^5$. 
In eq.~(4.8), the Hodge dual $({}^*_5)$ in the five transverse dimensions has been
introduced, whereas $H(y)$ is supposed to be a {\it harmonic} function in $R^5$, i.e. 
$\vec{\nabla}{}^2H(y)\equiv \D H(y)=0$. All the other components of $F_{(4)}$ are zero. 
For a single M-5-brane of 
magnetic charge $k$, the harmonic function $H(y)$ is given by
$$ H(y)=1+ \fracmm{\abs{k}}{r^3}~,\quad {\rm where}\quad r^2=y^my_m~.\eqno(4.9)$$
This M-5-brane solution is completely regular (i.e. truly solitonic) and it, in fact, interpolates
between the two maximally supersymmetric $D=11$ `vacua', the one being asymptotically flat in the 
limit $r\to\infty$ while another approaching $(AdS)_7\times S^4$ in the limit 
$r\to 0$~\cite{gtown,dgt}. 
 
When one chooses instead the harmonic function 
$$ H(y) = 1 + \sum _{s=1}^{n} \fracmm{\abs{k_s}}{\abs{\vec{y}-\vec{y}_s}^3}~~, \eqno(4.10)$$
one arrives at the classical configuration of $n$ {\it parallel} and similarly 
oriented M-5-branes of magnetic charges $k_s$, located at $\vec{y}_s$ in the $R^5$-space. This 
{\it multicentre} BPS solution also 
admits 16 Killing spinor fields by construction, so that it preserves $\n=\frac{1}{2}$ of 
supersymmetry in $D=11$. The existence of the multicentre brane solutions can be physically 
interpreted as a result of 
cancellation of gravitational and {\it anti-gravitational} (due to the antisymmetric tensor) 
forces, which is quite similar to the well-known `no force condition' (zero binding energy) in
D=4 physics of monopoles.

\subsection{NS and D branes in $D=10$ dimensions}

In eleven dimensions there are only M-waves, M-2- and M-5-branes as the `elementary' BPS states
preserving exactly a half of the $D=11$ supersymmetry. In order to make contact with the type-IIA
superstring theory in $D=10$, let's now assume that one of the {\it transverse} (to the brane 
worldvolume) dimensions is compactified on a circle $S^1$ of radius $R_{[11]}$. An M-5-brane can 
now be either (i) Kaluza-Klein-like `reduced' to ten dimensions, which results in a solitonic 
{\it NS-5-brane}, or (ii) it can be `wrapped' around the circle $S^1$, which results in a 
{\it D-4-brane}. 

By construction, the NS-5-brane is {\it magnetically} charged with respect to a gauge NS-NS 2-form 
(Kalb-Ramond field) $B^{[10]}_{(2)}$ descending from the gauge 3-form $A^{[11]}_{(3)}$ in eleven 
dimensions. In accordance with  eq.~(4.7), the NS-5-brane is magnetically dual to the `fundamental'
$D=10$ superstring. Since the NS-5-brane still depends upon the compactified (periodic) 
coordinate $\varrho$ of $S^1$, it thus contains all the associated {\it Kaluza-Klein} (KK) physical
modes. In order to become a BPS solution to the Type-IIA $D=10$ supergravity, the NS-5-brane 
solution should therefore be `averaged' over the compactified coordinate $\varrho$, which just
amounts to dropping all the {\it massive} KK modes. Though the latter is fully legitimate for a 
small compactification radius $R_{[11]}$ of $S^1$ (i.e for weakly coupled superstrings --- 
see eq.~(4.15) below), it becomes illegitimate for large $R_{[11]}$ (i.e. for strongly coupled 
superstrings) when some massive KK modes become light. From the viewpoint of the ten-dimensional 
type-IIA superstring theory, all the KK-modes appear as non-perturbative states.  

The wrapped M-5-brane(=D-4-brane) is (RR) charged with respect to the gauge 3-form $A^{[10]}_{(3)}$ 
of the type-IIA supergravity, which is also descending from $A^{[11]}_{(3)}$, so that it is a 
{\it Dirichlet}-4-brane indeed.~\footnote{As is well-known~\cite{polch}, the D-branes have a simple 
interpretation in the perturbative superstring theory as the spacetime topological defects on which
the open type-I superstrings with Dirichlet boundary conditions can end. The charges carried by the
D-branes are known to be the {\it Ramond-Ramond} (RR) charges in superstring theory.}

The KK Ansatz for the bosonic fields of $D=11$ supergravity, which leads to the $D=10$ type-IIA
action (in the so-called string frame), reads \cite{town,stel,jer}
$$ \eqalign{
ds^2_{[11]}= & e^{-\frac{2}{3}\f} ds^2_{[10]} + e^{\frac{4}{3}\f} (d\varrho +C_Mdx^M)^2~,\cr
A^{[11]}_{(3)}=& A^{[10]}_{(3)} + B^{[10]}_{(2)}\wedge d\varrho~,\quad M=0,1,2,\ldots,9~,\cr}
\eqno(4.11)$$
where the $S^1$ coordinate $\varrho$ is supposed to be periodic (with period $2\p$), and the 
$D=10$ {\it dilaton} $\f$ and KK vector $C_M$ have been introduced. 

The $D=10$ bosonic action descending from eq.~(4.2) reads
$$\eqalign{
S_{\rm IIA}=& \frac{1}{2}\int d^{10}x\sqrt{-g}\left\{ e^{-2\f}\left[ R +4\de_M\f\de^M\f-
\frac{1}{12}F_{MNP}F^{MNP}\right] \right.\cr
& \left. -\frac{1}{48}F_{MNPQ}F^{MNPQ}-\frac{1}{4}F_{MN}F^{MN}\right\} 
+({\rm Chern-Simons~~terms})~,\cr}\eqno(4.12)$$
where $F_{(2)}=dC_{(1)}$, $F_{(3)}=dB_{(2)}$ and $F_{(4)}=dA_{(3)}$. The kinetic terms of the 
type-IIA superstring NS-NS fields $(g_{MN},B_{MN},\f)$ in the first line of eq.~(4.12) are thus
uniformly coupled to the dilaton factor $e^{-2\f}$ ({\it cf.} the familiar factor $g^{-2}$ in front
of the Yang-Mills action), whereas the field strengths of the RR-fields $(C_M,A_{MNP})$ in 
the second line of eq.~(4.12) do {\it not} couple to the dilaton at all. Therefore, the superstring 
coupling constant $g_{\rm string}$ is given by the asymptotical value of $e^{\f}$,
$$ g_{\rm string}=\VEV{e^{\f}}~,\eqno(4.13)$$
while the RR-field couplings to the D-branes in the type-IIA supergravity should contain 
non-perturbative information about the type-IIA superstring/M-Theory.

It follows from the KK-Ansatz (4.11) that the compactification radius $R_{[11]}$ is also related to 
the dilaton as
$$ R_{[11]}=\VEV{e^{\frac{2}{3}\f}}~.\eqno(4.14)$$ 
Combining eqs.~(4.13) and (4.14) results in the remarkable relation~\cite{wm}
$$ R_{[11]}=g_{\rm string}{}^{2/3}~.\eqno(4.15)$$
The strong coupling limit of the type-IIA superstring theory is, therefore, 
{\it eleven\/}-dimensional~\cite{wm}~! Since there is only one $D=11$ supersymmetric field theory 
(of the second-order in spacetime derivatives), namely, the $D=11$ supergravity, it should thus be 
interpreted as the LEEA of some eleven-dimensional M-Theory which is supposed to be $D=11$ 
supersymmetric by consistency.

The solitonic (i.e. regular) NS-5-brane BPS solution to the type-IIA, $D=10$ supergravity, which is 
obtained by plain {\it dimensional reduction} of the M-5-brane solution (4.8) down to 
$R^{1,5}\times R^{4}$, is given by~\cite{town,stel}  
$$ ds^2_{[10]}=dx^{\m}dx^{\n}\h_{\m\n}+ H(y)dy^mdy^n\d_{mn}~,\qquad
F_{(3)}={}^*_4dH~,\qquad  e^{2\f}=H~,\eqno(4.16)$$
where $x^{\m}$ $(\m=0,1,2,3,4,5)$ parameterize $R^{1,5}$ and $y^{m}$ $(m=6,7,8,9)$ parameterize 
$R^{4}$, $({}^*_4)$ is the Hodge dual in $R^4$, and $H(y)$ is a harmonic function in $R^4$. A
BPS configuration of $n$ parallel and similarly oriented NS-5-branes is obtained after choosing
the harmonic function $H(y)$ as in eq.~(4.10).

Similarly, the D-4-brane solution to the type-IIA supergravity reads~\cite{town,stel}
$$ ds^2_{[10]}=H^{-\frac{1}{2}}dx^{\m}dx^{\n}\h_{\m\n}+ H^{\frac{1}{2}}dy^mdy^n\d_{mn}~,\qquad  
F_{(3)}=0~,\qquad e^{-4\f}=H~,\eqno(4.17)$$
where $x^{\m}$ $(\m=0,1,2,3,4)$ parameterize $R^{1,4}$, whereas $y^{m}$ $(m=5,6,7,8,9)$ 
parameterize $R^{5}$, and $H(y)$ is a harmonic function in $R^5$. Given the choice of eq.~(4.10) 
for the harmonic $H$-function in eq.~(4.17), one arrives at a BPS configuration of $n$ parallel 
and similarly oriented D-4-branes in static equilibrium. The `no force condition' in this case can 
again be physically interpreted as the result of mutual cancellation of the gravitational (NS-NS) 
and anti-gravitational (R-R) forces between the D-4-branes. However, unlike the similar NS-5-brane 
solution (4.16), the solution (4.17) has isolated singularities at the positions 
$\{\vec{y}_s\}$ of the D-4-branes in $R^5$.

In the case of $n$ parallel and similarly oriented D-4-branes, there are $n$ abelian gauge fields 
in their common worldvolume, which originate as the zero modes of the open superstrings stretched 
between the D-4-branes~\cite{polch}. In the coincidence limit, where all $n$ D-4-branes collapse, 
i.e. when they are `on top of each other', the gauge symmetry $U(1)^n$ enhances to $U(n)$ 
\cite{wkk}. This gauge symmetry enhancement can be understood from the viewpoint of the 
perturbative open (T-dual, or subject to Dirichlet boundary conditions) superstring 
theory~\cite{polch} due to the appearance of extra massless vector bosons in the coincidence limit. 

However, it is not yet the end of the $D=10$ BPS brane story. The KK massive particles associated 
with the compactification circle $S^1$ can be naturally interpreted in $D=10$ as the D-0-branes 
charged with respect to the RR (and KK) gauge field $C_{(1)}$~\cite{wkk}. Indeed, the eleventh 
component of the spacetime momentum in the eleven-dimensional supersymmetry algebra (4.1) plays the
 role of 
an abelian central charge in the compactified theory, whereas this central charge in $D=10$ 
originates from the RR charges of D-0-branes. From the viewpoint of the type-IIA superstring, all 
these BPS states are truly non-perturbative.~\footnote{There are no RR charged states in the 
perturbative superstring spectrum.} 

According to the golden rule (4.7), magnetically dual to the D-0-branes in $D=10$ are 
D-{\it 6-branes}, which are, therefore, also of KK origin in the type-IIA supergravity. 
The corresponding
classical BPS solution of the type-IIA supergravity equations of motion in $D=10$ reads~\cite{ht1}
$$ ds^2_{[10]}=H^{-\frac{1}{2}}dx^{\m}dx^{\n}\h_{\m\n}+ H^{\frac{1}{2}}dy^mdy^n\d_{mn}~,\qquad
F_{(2)}={}^*_3dH~,\qquad  e^{-4\f}=H^3~,\eqno(4.18)$$
where $x^{\m}$ $(\m=0,1,2,3,4,5,6)$ parameterize $R^{1,6}$ and  $y^{m}$ $(m=7,8,9)$ parameterize
$R^{3}$, $({}^*_3)$ is the Hodge dual in $R^3$, and $H(y)$ is a harmonic function in $R^3$. A
BPS configuration of $n$ parallel D-6-branes is described by the harmonic function $H(y)$ similar 
to that of eq.~(4.10) but with the different power (-1) instead of (-3) there. Like a D-4-brane, 
a single D-6-brane is singular in $D=10$ at the position of the D-6-brane in $R^3$. An M-Theory 
resolution of the D-4-brane singularity will be discussed in subsect.~4.5. As regards the D-6-brane 
singularity in $D=10$, its M-Theory resolution in eleven dimensions is provided by the following 
{\it non-singular} $D=11$ supergravity solution~\cite{townr}:
$$ ds^2_{[11]}=dx^{\m}dx^{\n}\h_{\m\n}+ Hd\vec{y}\cdot d\vec{y}+H^{-1}
(d\varrho +\vec{C}\cdot d\vec{y})^2~,\eqno(4.19)$$
where $H=1+\frac{1}{2}r^{-1}$, $r=\abs{\vec{y}}$, and $\vec{\de}\times\vec{C}=\vec{\de}H$
({\it cf.} eq.~(3.70)~!). The $D=11$ spacetime (4.19) is given by the product of the flat space 
$R^{1,6}$ with the Euclidean Taub-NUT (ETN) space (sect.~3). The ETN space can be thought of as a 
non-trivial bundle (Hopf fibration) with the base $R^3$ and the fiber $S^1$. After dimensional
reduction to $D=10$ dimensions, eq.~(4.19) results in a single D-6-brane located at the 
origin of $R^3$. Though the Taub-NUT metric seems to be singular at $r=0$, as is well-known 
\cite{sork,grpe}, it is just a coordinate singularity provided that $\varrho$ is periodic with the 
period $2\p$.~\footnote{The asymptotic ETN metric near the singularity of $H$ is 
diffeomorphism-equivalent to the flat \newline ${~~~~~}$ $R^4$ metric.} 

Therefore, the M-theory interpretation of a D-6-brane is given by the ETN (or KK \cite{sork,grpe}) 
monopole which interpolates between the two maximally 
supersymmetric M-Theory `vacua': the flat $D=11$ spacetime near $r=0$ and the KK spacetime 
$R^{1,9}\times S^1$ near $r\to\infty$ \cite{townr}. It is straightforward to generalize this result
to a system of $n$ parallel and similarly oriented D-6-branes in $D=10$, whose M-theory 
interpretation is given by a Euclidean {\it multi-Taub-NUT} monopole (subsect.~3.4) described 
by the multi-centered harmonic function ({\it cf.} eq.~(3.72)~) 
$$H_{\rm multi-ETN}(\vec{y})=
1 +\sum_{i=1}^{n}\fracmm{ \abs{k_i}}{2\abs{\vec{y}-\vec{y_i}}}~.\eqno(4.20)$$
This solution is non-singular in $D=11$ provided that no two centers coincide, i.e. 
$\vec{y}_i\neq\vec{y}_j$ for all $i\neq j$. In the coincidence limit of parallel and similarly 
oriented D-6-branes with equal RR charges (=1), non-isolated singularities appear. 

To investigate the coincidence limit in some more details, one may have to generalize the 
harmonic function $H$ of eq.~(4.20) to the form (3.72). For instance, in the case of just two 
centers with equal charges, a good choice is given by~\cite{town}
$$ H(\vec{y}) =\fracmm{\l}{2} + \fracmm{1}{2\abs{\vec{y}-\e\vec{y}_0}}
+\fracmm{1}{2\abs{\vec{y}+\e\vec{y}_0}}~,\eqno(4.21)$$
where $\{\l,\e\}$ are some positive constants. The double ETN metric is then obtained by
choosing $\l/2=1$ in eq.~(4.21), whereas the limit $\l\to 0$ results in the Eguchi-Hanson (EH)
metric. In the coincidence limit $\abs{\e}\to 0$, near the singularity of $H$, the value of $\l$ is 
obviously irrelevant, whereas the gauge symmetry in the common worldvolume of the D-6-branes is 
{\it enhanced} from $U(1)\times U(1)$ to $U(2)$ ~\cite{wm,wkk}. From the M-theory perspective, 
the M-2-branes can wrap around the  compactification circle $S^1$, being simultaneously stretched 
between the D-6-branes. It is the massless modes of these M-2-branes that play the role of the 
additional (non-abelian) massless vector particles in the coincidence limit~\cite{town}. In order 
to make this symmetry enhancement manifest, it is useful to employ the non-singular HSS description
of the mixed EH-ETN metric in terms of the harmonic potential (4.21) --- see subsect.~4.7.

The $D=10$ singularity of the single D-6-brane solution can be physically interpreted as the result 
of an illegitimate neglect of the KK particles which become {\it massless} at the D-6-brane core 
and whose inclusion resolves the singularity in D=11~\cite{townr}. This phenomenon is known as the
{\it M-Theory resolution} of short-distance singularities in the $D=10$ type-IIA supergravity by 
relating them (via a strong-weak coupling duality) to the long-distance effects of the massless 
modes of the M-2-brane wrapped around $S^1$~\cite{town}. The M-theory resolution is one of the 
cornerstones of the brane technology in non-perturbative $D=4$, $N=2$ supersymmetric gauge field
theories~\cite{witten} (see subsect.~4.5).

\subsection{Intersecting branes}

Each of the BPS (single or parallel) `elementary' brane solutions to $D=11$ or $D=10$ (type-IIA)
supergravity considered in the previous subsections breaks exactly $1/2$ of the maximal 
supersymmetry having 32 supercharges, and it is governed by a single harmonic function of 
transverse spacial coordinates. More BPS brane solutions preserving some part $\n\leq 1/2$ of the 
maximal supersymmetry can be obtained by a superposition of the (intersecting) `elementary' branes.
A procedure of constructing the corresponding classical supergravity solutions depending upon 
several harmonic functions is outlined in ref.~\cite{tse}, and it is known as the
`harmonic function rule'.~\footnote{See e.g., ref.~\cite{jer} and references therein for details.} 

For our purposes of describing the brane technology towards the gauge field theories with $N=2$ 
supersymmetry in $D=4$ (i.e. having $2\times 4=8$ supercharges), we only need `marginal' (i.e. with
vanishing binding energy) BPS brane configurations which preserve exactly $8/32=1/4$ of the maximal 
supersymmetry and have an (uncompactified) flat spacetime $R^{1,3}$ as their intersection. This 
limits our discussion (e.g. in $D=10$ dimensions) to the {\it orthogonally} intersecting NS-branes, 
D-4-branes and D-6-branes, having $R^{1,3}$ as their common worldvolume, where the effective 
four-dimensional $N=2$ supersymmetric gauge field theory in question lives. From the M-theory 
perspective (i.e. $D=11$ dimensions), we want the NS- and D-4-branes to be represented by a 
{\it single} M-5-brane described by a single harmonic function and, possibly, in the background of 
a (multi-ETN) monopole described by another harmonic function or the corresponding hyper-K\"ahler 
potential. The relevant $1/4$-supersymmetric BPS brane solution is then parameterized by two 
functions, as in the $N=2$ gauge field theory with a hypermultiplet matter. Unlike the $D=10$ 
configuration of the intersecting BPS branes (which is singular), the corresponding M-theory brane 
configuration in $D=11$ is non-singular, and it carries some non-perturbavive information about the
$N=2$, $D=4$ gauge theory in question. Our immediate tasks are, therefore, (i) to establish a
correspondence (i.e. a dictionary) between the (classical) brane- and (quantum) field theory 
quantities, and (ii) to fix the form of the M-5-brane. Both problems were solved by Witten in 
ref.~\cite{witten} (see also refs.~\cite{five,hw} for some earlier work on brane technology).

\subsection{Effective field theory in worldvolume of type-IIA branes}

Exact solutions to the LEEA in $D=4$, $N=2$ supersymmetric gauge field theories (for definiteness,
the $N=2$ super-QCD with $N_c$ colors and $N_f$ flavors) can be interpreted (and, in fact, derived) 
in a nice geometrical way, when considering the effective field theory in the {\it common} 
worldvolume (to be identified with our $D=4$ spacetime $R^{1,3}$) of the magnetically (or RR) 
charged BPS branes of the type IIA superstring and then resolving classical singularities of these 
$D=10$ branes in $D=11$ M-theory~\cite{witten}. 

The relevant $1/4$-supersymmetric configuration of the orthogonally intersecting branes in ten 
dimensions $R^{1,9}$ parameterized by $(x^0,x^1,x^2,x^3,x^4,x^5,x^6,x^7,x^8,x^9)$ is 
schematically depictured in Fig.~3. 
It consists of two parallel (magnetically charged) NS-5-branes, $N_c$ parallel 
Dirichlet-4-branes orthogonally stretching between the NS-5-branes, and $N_f$ Dirichlet-6-branes
which are orthogonal to both NS and D-branes. 

\begin{figure}
\vglue.1in
\makebox{
\epsfxsize=4in
\epsfbox{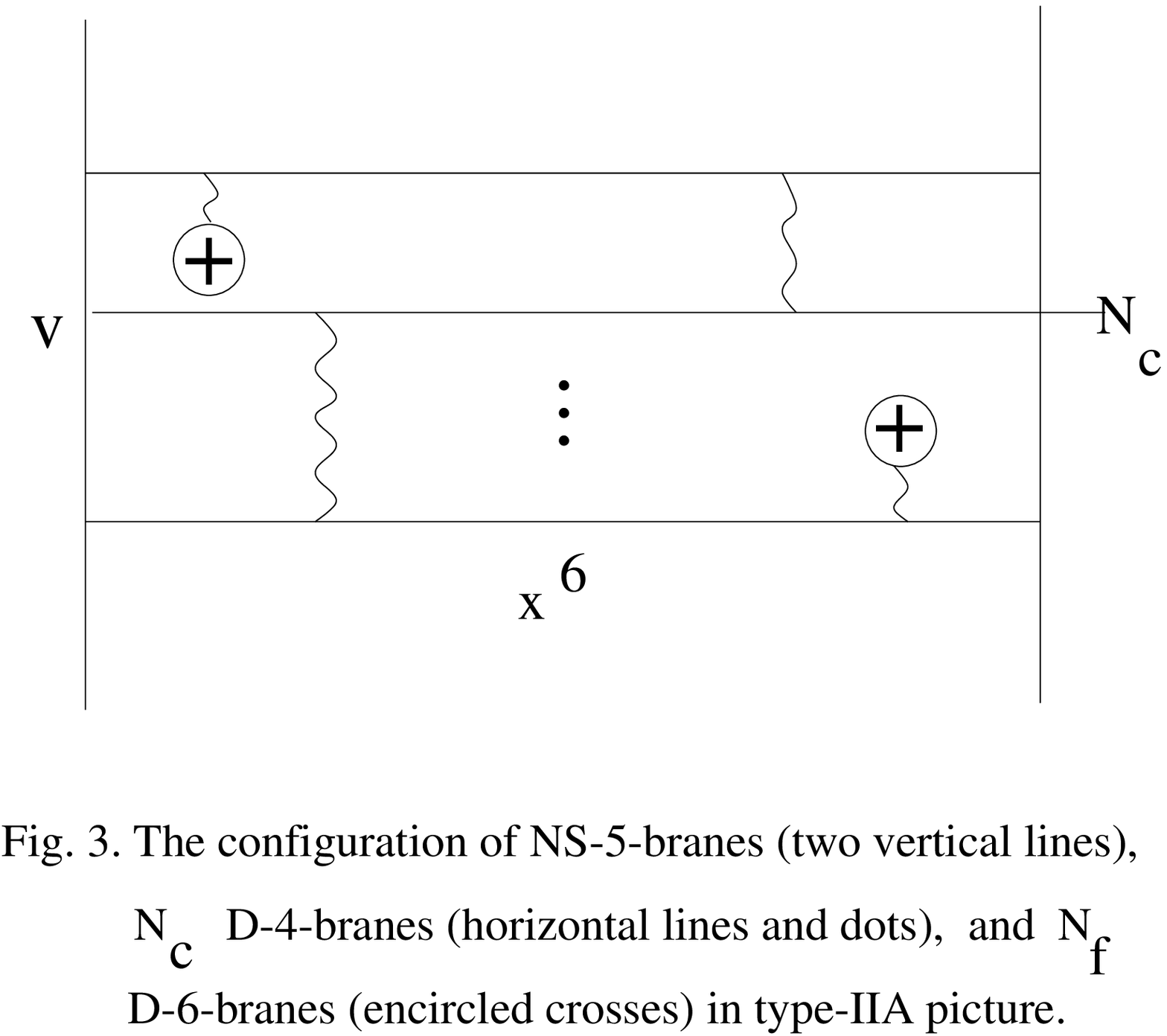}
}
\end{figure}

The two parallel 5-branes are located at $\vec{w}=(x^7,x^8,x^9)=0$ and have classically fixed $x^6$
values. Being parallel to each other but orthogonal to the 5-branes, the 4-branes have their 
worldvolumes to be parameterized by $(x^0,x^1,x^2,x^3)\in R^{1,3}$ and $x^6$. Being orthogonal to
both 5-branes and 4-branes, the 6-branes are located at fixed values of $(x^4,x^5,x^6)$, while their
worldvolumes are parameterized by $(x^0,x^1,x^2,x^3)\in R^{1,3}$ and $\vec{w}\in R^3$.

After `blowing-up' the intersecting NS-5- and D-4-brane configuration depictured in Fig.~3, its 
two-dimensional projection (in the directions orthogonal to the $D=4$ effective spacetime) looks 
like as a {\it hyperelliptic} curve $\S$ of genus $g=N_c-1$, depictured in Fig.~4. Indeed, as is 
well-known in the theory of Riemann surfaces~\cite{fkra}, a hyperelliptic (compact) Riemann surface
 of genus $g$ can be defined by taking two Riemann spheres, cutting each of them 
between $2g+2$ ramification (Weierstrass) points $e_i$, and then identifying the cuts as shown in  
Fig.~4. The corresponding algebraic (complex) equation reads $y^2= \prod^{2g+2}_{i=1}(z-e_i)$, with 
$e_i\neq e_j$ for $i\neq j$. In other words, a two-sheeted cover of the sphere branched over $2g+2$
points is just the (essentially unique) realization of a hyperelliptic surface of genus $g\geq 1$.
Though the surface obtained by the projection of the M-5-brane worldvolume is actually non-compact
(it goes through infinity), nevertheless, we are going to formally apply the theory of compact 
Riemann surfaces to our case (a justification of this procedure is beyond the scope of this paper). 

\begin{figure}
\vglue.1in
\makebox{
\epsfxsize=4in
\epsfbox{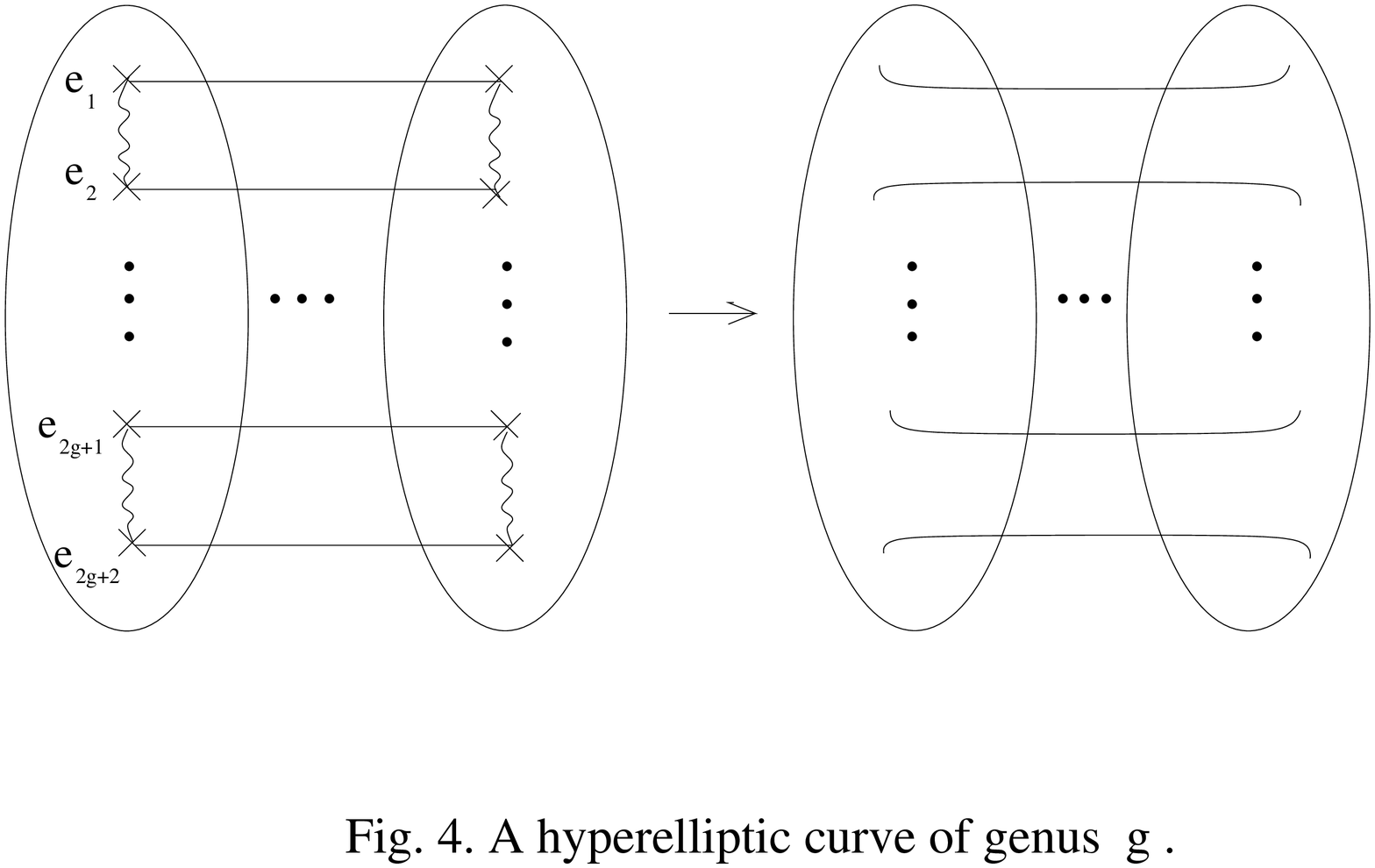}
}
\end{figure}

Back to ten dimensions, the 5-brane worldvolumes are, therefore, given by the {\it local}
product $R^{1,3}\times \S_0$, where $R^{1,3}$ is our $D=4$ 
spacetime parameterized by the coordinates
$(x^0,x^1,x^2,x^3)$, whereas $\S_0$ is the singular, of genus $g=N_c-1$, Riemann surface 
parameterized by the real coordinates $(x^4,x^5)$ or, equivalently, the complex variable 
$v\equiv x^4+ix^5$. 

The type-IIA brane interaction in ten dimensions can be visualized as an exchange of 
open superstrings, even though the ultimate force between some static branes may vanish. 
Associated with the zero modes of such open superstrings are BPS multiplets of a supersymmetric 
field theory. In particular, a gauge $N=2$ vector multiplet in the effective $D=4$ spacetime 
$R^{1,3}$ may be identified with massless modes of an open $(4-4)$ superstring carrying Chan-Paton 
factors at its ends and stretching between two D-4-branes (Fig.~3), whereas the spacetime matter
hypermultiplets are just zero modes of open $(6-4)$ superstrings connecting the D-6-branes to
the D-4-branes. The BPS mass of a hypermultiplet is determined by the distance 
(in $x^{4,5}$ directions) between the corresponding D-6-brane and D-4-brane 
(see also the next subsect.~5).~\footnote{The $(6-6)$ open superstrings decouple from the NS-5- 
and D-4-branes in the LEEA limit.}

On physical reasons, the effective gauge coupling constant $g_{\rm gauge}$ of the N=2 
supersymmetric gauge field theory in our effective $D=4$ spacetime should be proportional to a 
distance between two NS-5-branes~\cite{witten},
$$ \fracmm{1}{g^2_{\rm gauge}}=\fracmm{x^{[6]}_1-x^{[6]}_2}{g_{\rm string}}~,\eqno(4.22)$$
where $g_{\rm string}$ is the type-IIA superstring coupling constant. Indeed, as was explained by
Witten~\cite{witten}, the 5-brane $x^6$-coordinate should be thought of as a function of $v$ by 
minimizing the total (BPS~!) 5-brane worldvolume. The BPS condition for large $v$ is given by a 
two-dimensional Laplace equation on $x^6$, whose solution has a logarithmic dependence upon $v$ for
large values of $v$. Having interpreted $\abs{v}$ as a mass scale in our theory, eq.~(4.22) is  
apparently consistent with the well-known logarithmic behaviour of the four-dimensional effective 
gauge coupling at high energies in an asymptotically free gauge QFT. The presence of the 
superstring coupling $g_{\rm string}$ in eq.~(4.22) can be justified by the way it appears
in the the D-brane effective action induced by open superstrings (ending on a D-brane) in the brane
worldvolume. This effective action (or, at least, its bosonic part) is calculable, e.g. by the use
of the standard sigma-model approach to open string theory~\cite{abo}. As a leading contribution, 
one finds the {\it Born-Infeld}-type effective action~\cite{lei}
$$ S_{\rm BI}=T\int_{\rm worldvolume}e^{-\f}\sqrt{\det(g_{\m\n}+B_{\m\n}+2\p F_{\m\n})}~,
\eqno(4.23)$$
where $T$ is a constant brane tension, $\f$ is the dilaton field, $g_{\m\n}$ is the induced metric 
in the worldvolume, $B_{\m\n}$ and $F_{\m\n}$ are the pull-backs of the 2-form $B$ and the abelian 
field strength $F$, respectively. The factor $\VEV{e^{-\f}}=1/g_{\rm string}$ contributing to the
effective brane tension in eq.~(4.23) is dictated by the disc topology of the relevant open 
superstring tree diagram. Extracting from eq.~(4.23) the term quadratic in $F$ leads to the 
denominator on the right-hand-side of eq.~(4.22).

To this end, let's summarize the fundamental properties of the brane configuration depictured in 
Fig.~3, from the 10-dimensional type-IIA point of view:
\begin{itemize}
\item its common worldvolume is $(1+3)$-dimensional, being infinite in all directions there,
\item it is the BPS (stable and unique) supersymmetric solution to the type-IIA supergravity
equations of motion,
\item it is invariant under $1/2\times 1/2=1/4$ of the maximal supersymmetry, with the first $1/2$ 
factor being due to the parallel NS-5-branes and the second $1/2$ factor due to the parallel 
D-4-branes orthogonal to the NS-5-branes; this results in $32/4=8$ conserved supercharges of the 
$N=2$ extended supersymmetry in the effective $D=4$ spacetime,
\item the D-6-branes, which are orthogonal to both NS-5- and D-4-branes, do not break the 
$N=2$ supersymmetry any more; they correspond to the matter hypermultiplets in $D=4$,
\item the ten-dimensional Lorentz group $SO(1,9)$ is broken down to
$$ SO(1,3)\otimes SU(2)_A\otimes U(1)_{\rm c.c.}~, \eqno(4.24)$$
in accordance to the local decomposition $R^{1,9}= 
R^{1,3}\times R^2_v\times R^1_{x^6}\times R^3$, respectively;
the $SO(1,3)$ factor of eq.~(4.24) can be identified with the Lorentz group of our $D=4$ spacetime 
$R^{1,3}$, the rotational symmetry $SO(3)$ of $R^3$ implies the $N=2$ supersymmetry automorphisms 
$SU(2)_A=SO(3)/Z_2$ in $D=4$, whereas the $U(1)=SO(2)$ factor of eq.~(4.24) can be identified with
the central charge $(v)$ transformations~\cite{ikz}.
\end{itemize}
These are all the fundamental properties that, in fact, uniquely determine the general BPS brane 
configuration of Fig.~3. One gets the effective gauge field theory in the common brane worldvolume 
as the effective spacetime, having the $N=2$ extended supersymmetry, whose LEEA is under control.
In order to accommodate non-perturbative quantum gauge field theory dynamics, the classical brane 
configuration of Fig.~3 should be 'blown-up'. In the type-IIA picture considered above, as we 
already know from the previous subsections, the corresponding BPS solution to the type-IIA
supergravity suffers from singularities. These singularities cannot be described semi-classically in
$D=10$, but they can be resolved in D=11 after reinterpreting the brane configuration of Fig.~3 in 
M-Theory (see the next subsection).

\subsection{M-Theory resolution}

It follows from eq.~(4.22) that one can keep the effective $D=4$ gauge coupling constant 
$g_{\rm gauge}$ fixed while {\it increasing} the distance $L=x^6_1-x^6_2$ between the 
two NS-5-branes and, simultaneously, the type-IIA superstring coupling constant 
$g_{\rm string}$ accordingly. As we already know, at strong coupling the type-IIA superstring 
should be replaced by M-theory. This means that the additional compact dimension $(x^{10})$ 
represented by 
the circle $S^1$ of radius $R\sim g_{\rm string}^{2/3}$, can no longer be ignored. 
Associated with the $S^1$-rotations is a non-perturbative $U(1)_M$ gauge symmetry.

The low-energy classical description of M-theory and its PBS branes turns out to be sufficient 
(see the next subsection) for a purely geometrical derivation of exact solutions to 
the four-dimensional LEEA of the effective $N=2$ supersymmetric gauge field theories in the 
M-theory 5-brane worldvolume, just because {\it all} the relevant distances in 
the non-perturbative eleven-dimensional BPS brane configuration become {\it large} while no 
singularity appears, unlike that in the type-IIA picture considered in the previous subsection. 
In particular, the D-4-branes and NS-5-branes in the type-IIA picture are replaced in M-Theory 
by a {\it single} and {\it smooth} M-5-brane whose worldvolume is given by the {\it local} product 
$R^{1,3}\times\S$, where $\S$ is the genus $g=N_c-1$ hyperelliptic Riemann surface 
{\it holomorphically} embedded into a four-dimensional hyper-K\"ahler manifold $Q$ given by the
local product of $R^3$ and $S^1$.~\footnote{The holomorphicity of the embedding as well as the 
hyper-K\"ahler nature of the manifold $Q$ are \newline ${~~~~~}$ both required by $N=2$ 
supersymmetry of the effective field theory in the $R^{1,3}$ spacetime.}  
The manifold $Q$ is thus topologically a bundle $Q\sim R^3\times S^1$ parameterized by the 
coordinates $(x^4,x^5,x^6)$ and $\varrho$, whose base $R^3$ can be interpreted as a part of 
the D-6-brane worldvolume in the type-IIA picture and whose fiber $S^1$ is the eleventh 
dimension of M-Theory~\cite{townr}. 

The hyperelliptic curve $\S$ lies at a single point in $R^3$. After unifying the real coordinates 
$x^6$ and $\varrho$ $(R=1)$ into a single complex coordinate $s=x^6+i\varrho$ as in 
ref.~\cite{witten}, the analytic equation defining the Riemann surface $\S$ should be of the form
$$ F(s,v)=0\eqno(4.25)$$ 
with a holomorphic function $F$. Given a finite number of branes, the function $F$ has to be a
polynomial in $v$ and $t=e^{-s}$~\cite{witten}. This polynomial can be fixed in terms of the 
$N=2$ gauge field theory data $(SU(N_c),N_f,m_i)$ by using standard techniques of the 
singularity theory (see, e.g. ref.~\cite{agvbook}). Eq.~(4.25) then takes the form of the 
Seiberg-Witten curve $\S_{\rm SW}$ described in subsect.~2.2~\cite{witten}.

Without hypermultiplet matter, the hyper-K\"ahler manifold $Q$ is flat. In the presence of 
(magnetically charged) hypermultiplets, the manifold $Q$ is given by a multi-ETN monopole. The BPS
bound for any Riemann surface $\S$ embedded into a hyper-K\"ahler manifold $Q$ is given 
by~\cite{mikh}
$$ Area_{\S}\geq \abs{ \int_{\S}\,\O}~,\eqno(4.26)$$
where the K\"ahler form $\O$ of $Q$ \cite{bes} has been introduced. The bound (4.26) becomes 
saturated if and only if $\S$ is holomorphically embedded into $Q$, i.e. if the holomorphic 
description (4.25) of $\S$ is valid.

The origin of the abelian $N=2$ vector multiplets in the Coulomb branch of the effective 
$D=4$ gauge field theory also becomes more transparent from the M-theory point of view~\cite{five}. 
The effective field theory in the six-dimensional worldvolume of an M-5-brane should have chiral 
six-dimensional $N=2$ supersymmetry \cite{kap}, while the only admissible $N=2$ chiral 
supermultiplet in six dimensions is given by the {\it tensor} supermultiplet having a two-form 
$B_{(2)}$ with the {\it self-dual} field strength $T_{(3)}$ (see, e.g. ref.~\cite{kkk}). Indeed, 
being invariant under $16$ linearly realized supersymmetries and having $11-6=5$ scalar fields 
describing transverse fluctuations to the 5-brane, the right supermultiplet must have 
$\ha\cdot 16-5=3$ additional bosonic on-shell degrees of freedom, which can only be delivered by a 
bosonic gauge 2-form with self-dual field strength, belonging to an $N=2$ chiral tensor multiplet 
in six dimensions. Since, in our case,  the M-5-brane is wrapped around the Riemann surface, we can
decompose the self-dual 3-form $T_{(3)}$ as
$$ T_{(3)} = F_{(2)}\wedge \o_{(1)} + {}^*_4F_{(2)}\wedge {}^*_2\o_{(1)}~,\eqno(4.27)$$
where $F_{(2)}$ is the two-form in $R^{1,3}$, whereas $\o_{(1)}$ is the one-form on the Riemann
surface $\S_{N_c-1}$ of genus $N_c-1$. The equations of motion $dT=0$ then imply 
$$ dF=d{}^*F=0~,\eqno(4.28)$$
and
$$ d\o=d{}^*\o=0~.\eqno(4.29)$$
Eq.~(4.29) means that the one-form $\o$ is harmonic on $\S_{N_c-1}$. Since the
number of the independent harmonic one-forms on a Riemann surface exactly equals
to its genus~\cite{fkra}, one also has $(N_c-1)$ two-forms $F$, while each
of them satisfies eq.~(4.28). Since eq.~(4.28) is nothing but the Maxwell equations for 
an electro-magnetic field strength $F$, this explains the origin of the abelian gauge group 
$U(1)^{N_c-1}$ in the Coulomb branch of the effective $D=4$ gauge field theory. 

To summarize the above-mentioned in this subsection, we conclude that the geometrical M-Theory 
interpretation of the $N=2$ {\it gauge} LEEA in the Coulomb branch is given by the 
identification~\cite{five}
$$ \S_{N_c-1} = \S_{SW}~.\eqno(4.30)$$
To understand the {\it hypermultiplet} LEEA (sect.~3) in the similar way, one first notices 
that the D-6-branes are {\it magnetically} charged with respect to the non-perturbative 
$U(1)_M$ gauge symmetry. Hence, the fiber $S^1$ of $Q$ has to be non-trivial (i.e. of 
non-vanishing magnetic charge $m\neq 0$). After taking into account the $U(1)_{\rm PG}$ 
isometry of the internal hypermultiplet NLSM target space $Q_{(m)}$ in the Coulomb branch, 
one concludes that $Q_{(m)}$ has to be the Euclidean multicentre Taub-NUT space or a 
multi-KK monopole whose metric was described in subsects.~(3.4) and (4.2). In particular, 
when the magnetic charge $m=1$, the isometry of $Q_{(1)}$ is enhanced to $U(2)$ and one 
arrives at the Taub-NUT space whose metric was described in subsect.~3.2 (see Appendix A also).

One may wonder about the M-Theory reinterpretation of the BPS multiplets of the effective 
four-dimensional $N=2$ gauge field theory from the $D=11$ viewpoint, since there are no 
strings in eleven dimensions, whereas the open superstring zero modes (in $D=10)$ were identified
with the $N=2$ vector and hyper-multiplets in the previous subsection. In $D=11$ dimensions,
there are, however, M-2-branes which are also BPS and can end on an M-5-brane. When considering 
these M-2-branes as the BPS (of minimal area) deformations of the M-5-brane, one can identify the 
BPS states in the effective $N=2$ supersymmetric four-dimensional field theory with zero modes of 
these M-2-branes. The type of an $N=2$ supermultiplet is determined by a static M-2-brane topology:
a cylinder corresponds to an $N=2$ vector multiplet, whereas a disc corresponds to a 
hypermultiplet~\cite{mikh}.  

\subsection{SW solution from classical M-5-brane dynamics}

We are now in a position to ask an {\it educated} question towards a derivation of the 
Seiberg-Witten exact solution from the brane technology.~\footnote{Essentially the same reasoning 
was first suggested in ref.~\cite{brit}. See also refs.~\cite{mikh,berk}.}  We should consider a 
single M-5-brane in eleven dimensions, whose worldvolume is given by the local product of a flat
spacetime $R^{1,3}$ with the hyperelliptic curve $\S$ of genus $g=N_c-1$ (see subsect.~4.4), where 
$\S$ is supposed to vary in the effective $D=4$ spacetime $R^{1,3}$. Moreover, in order to be 
consistent 
with the rigid  $N=2$ supersymmertry in the $D=4$ spacetime, the Riemann surface $\S$ should be {\it
holomorphically} embedded into a {\it hyper-K\"ahler} four-dimensional manifold $Q$ (say, 
a milti-ETN), which is a part of the whole eleven-dimensional spacetime given by the local-product 
$R^{1,6}\times Q$. Since the Nambu-Goldstone-type effective action of the M-5-brane is {\it 
uniquely} determined by its symmetries, it is, in principle, straightforward to calculate it 
explicitly  (see refs.~\cite{blnpst,apps} for a fully supersymmetric and covariant form of the 
action). The KK reduction of the six-dimensional M-5-brane action on the complex curve $\S$ then 
gives rise to the effective $N=2$ supersymmetric gauge field theory action which is nothing but the
Seiberg-Witten effective action~! This approach to a derivation of the SW result can be 
considered as the particular application of the theory of integrable systems \cite{swint,dow}.
To the end of this subsection, we discuss its simplest technical realization (without 
hypermultiplets, i.e. with a flat hyper-K\"ahler manifold $Q$).

The fully covariant and supersymmetric M-5-brane action of refs.~\cite{blnpst,apps} is not really 
needed for this purpose. Even its bosonic part in a flat $D=11$ supergravity background is too 
complicated, because of the need to accommodate off-shell the self-duality condition for the field 
strength  $T_{(3)}\equiv dB_{(2)}$ of the 2-form  $B_{(2)}$ present in the six-dimensional 
chiral $N=2$ tensor multiplet. Rather superficially, the bosonic part of the M-5-brane action has 
the structure
$$ S_{M-5,~{\rm bosonic}}=S_{\rm N-G} + S_{\rm self-dual} + S_{\rm WZ}~,\eqno(4.31)$$
where $S_{\rm N-G}$ is the standard 5-brane {\it Nambu-Goto} (N-G) action, $S_{\rm self-dual}$ 
stands for the naive worldvolume integral over $T^2$ subject to an (implicit) self-duality 
constraint, whereas $S_{\rm WZ}$ is a higher-derivative {\it Wess-Zumino} (WZ) term which can be 
ignored anyway if we are only interested in the leading contribution to the M-5-brane LEEA with two 
spacetime derivatives at most. Since we are going to restrict ourselves to a calculation of the 
{\it scalar} sector of the effective $N=2$ supersymmetric LEEA in $D=4$ (the rest of the field
theory LEEA depending upon the vector and fermionic components of $N=2$ vector supermultiplets is 
entirely determined by the special geometry of the scalar field components due to the $N=2$ 
extended supersymmetry), we are allowed to ignore even the second term in the bosonic M-5-brane 
effective action (4.31).~\footnote{If $\S$ has a component $\S_f$ of 
{\it finite} volume, the 2-form $B$ may be proportional to the $\S_f$ volume form, thus leading to 
an extra scalar. This scalar is, however, compact~\cite{berk}, so that is does not afftect the 
spacetime part of the effective action that we are interested in here.}  

The N-G action in eq.~(4.31) reads
$$ S_{\rm N-G}=T\int \;{\rm Vol}(g)=T\int d^6\x\,\sqrt{-\det(g_{\m\n})}~, \eqno(4.32)$$
in terms of the 5-brane tension $T$ and the induced metric
$$ g_{\m\n}=\h_{MN}\pa_{\m}x^M\pa_{\n}x^N \eqno(4.33)$$
in the M-5-brane worldvolume parameterized by the coordinates $\x^{\m}$, where the functions 
$x^M(\x)$ describe the embedding of the six-dimensional M-5-brane worldvolume into a flat 
eleven-dimensional spacetime, $M=0,1,\ldots,10$, and $\m=0,1,\ldots,5$.

To simplify the form of the N-G action (4.32) before its KK reduction down to four dimensions, we 
make use of (i) the reparametrizational invariance of this action in six worldvolume dimensions,
and (ii) the geometrical information about the M-5-brane configuration collected in subsects.~4.4 
and 4.5. The local symmetry (i) allows us to choose a static gauge 
$$ x^{\un{\m}}=\x^{\un{\m}}~,\eqno(4.34)$$
where we have introduced the notation
$$ \m=\{\un{\m},4,5\}~,\quad {\rm with}\quad \un{\m}=0,1,2,3~.\eqno(4.35)$$
We remind the reader that our M-5-brane has $\vec{w}=(x^7,x^8,x^9)=\vec{0}$, the Riemann surface 
$\S$ is parameterized by the two coordinates $(v,\bar{v})$, whereas the flat (hyper-K\"ahler) 
manifold $Q$ is parameterized by four coordinates $(s,v,\bar{s},\bar{v})$.

Since we are interested in the M-Theory limit, where the supergravity decouples and the central
charge $v=x^4+ix^5$ of the $D=4$, $N=2$ supersymmetry algebra is constant, we are going to assume
in what follows that $v$ is $D=4$ spacetime independent. Moreover, since $\S$ is supposed to be 
holomorphically embedded into $Q$, while $\S$ is also holomorphically dependent upon its complex 
moduli $u_{\a}$, $\a=1,\ldots,g$, an actual dependence of the single remaining non-trivial function
$s$ among the M-5-brane embedding functions $x^M(\x)$ should be {\it holomorphic}, i.e. of the form 
$$ s=s(v,u_{\a}(x^{\un{\m}}))~,\eqno(4.36)$$
where we have taken into account that $u_{\a}=u_{\a}(x^{\un{\m}})$. From the viewpoint of the 
effective $D=4$ gauge theory, the complex moduli $u_{\a}(x)$ of the Riemann surface $\S(x)$  are 
just the scalar field components of $N=2$ vector multiplets in $D=4$ spacetime.

The induced metric (4.33) now takes the form 
$$ g_{\m\n}=\h_{\m\n} +\frac{1}{2}\left(\pa_{\m}s\pa_{\n}\bar{s}+\pa_{\n}s\pa_{\m}\bar{s}\right)~,
\eqno(4.37)$$
as in ref.~\cite{brit}, so that its direct consequences for the N-G action of eq.~(4.32) can be
simply `borrowed' from ref.~\cite{brit}. When keeping only the terms of the second order 
in the derivatives $\pa_{\un{\m}}$ after a substitution of eq.~(4.37) into eq.~(4.32), one gets the 
action
$$ S[s]=\fracm{T}{2}\int d^6\x\,\h^{\m\n}\pa_{\m}s\pa_{\n}\bar{s} ~\to~ 
S[u]=\fracm{T}{2}\int d^6\x\,\pa_{\un{\m}}s\pa^{\un{\m}}\bar{s}~.\eqno(4.38)$$

The KK-reduction of the action (4.38) on $\S$ gives rise to the following four-dimensional
NLSM action:
$$ S[u]=\fracm{T}{4i}\int d^4x\,\pa_{\un{\m}}u_{\a}\pa^{\un{\m}}\Bar{u}_{\b}\,\int_{\S}
\,\o^{\a}\wedge \Bar{\o}{}^{\b}~,\eqno(4.39)$$
where the holomorphic 1-forms $\o_{\a}$ on $\S$ have been introduced as
$$ \o^{\a}=\fracmm{\pa s}{\pa u_{\a}}dv~.\eqno(4.40)$$

The NLSM metric in eq.~(4.39) can be put into another equivalent form, by using the 
Riemann bilinear identity~\cite{fkra}
$$ \int_{\S} \,\o^{\a}\wedge \Bar{\o}{}^{\b}=\sum^g_{\g=1}\left( \int_{A_{\g}}\,\o^{\a}
\int_{B^{\g}}\,\Bar{\o}{}^{\b}-\int_{A_{\g}}\,\Bar{\o}{}^{\b}\int_{B^{\g}}\,\o^{\a}\right)~,
\eqno(4.41)$$
where a canonical (symplectic) basis $(A_{\a},B^{\b})$ of the first homology class has been 
introduced on the Riemann surface $\S_g$ of genus $g$, $\a,\b=1,\ldots,g$, and $g=N_c-1$. 
Substituting eq.~(4.41) into eq.~(4.39) and using the definitions (2.15) of the multi-valued 
functions $a_{\a}(u)$ and their duals $a_{{\rm D}\a}(u)$ yields~\cite{mikh,brit,berk}
$$\eqalign{
 S[u]~=~ & \fracm{T}{4i}\int d^4x\,\sum^{N_c-1}_{\a=1}\left( \pa_{\un{\m}}a_{\a}\pa^{\un{\m}}
\Bar{a}_{\rm D}^{\a}- \pa_{\un{\m}}\Bar{a}_{\a}\pa^{\un{\m}}a_{\rm D}^{\a}\right)=\cr
~ & = -\fracm{T}{2}\,{\rm Im}\left[ \int d^4x\,\pa_{\un{\m}}\Bar{a}_{\a}\pa^{\un{\m}}
a_{\b}\,\t^{\a\b}\right]~,\cr}\eqno(4.42)$$
where the period matrix $\hat{\t}$ of $\S$ has been introduced,
$$ \t^{\a\b}=\fracmm{\pa a_{\rm D}^{\a}}{\pa a_{\b}}=\fracmm{\pa^2\cf}{\pa a_{\a}\pa a_{\b}}~~.
\eqno(4.43)$$

Eq.~(4.42) precisely gives the scalar part of the full SW effective action in the $N=2$ super-QCD, 
as derived in ref.~\cite{sw}. Because of the NLSM special geometry described
by eq.~(4.43), a unique $N=2$ supersymmetric extension of eq.~(4.42), including all fermionic- and 
vector-dependent terms, is just given by the superfield function $\cf(W)$ integrated over the $N=2$
chiral superspace (see sect.~1). 

It is straightforward to generalize this derivation of the exact, purely gauge, Seiberg-Witten LEEA 
to a more general case with hypermultiplet matter by replacing 
the flat background space $Q$ above with a non-flat hyper-K\"ahler manifold $Q$ described by a 
multicentre ETN metric~\cite{berk}.

Being applied to a derivation of the {\it hypermultiplet} LEEA in $D=4$, the brane technology 
suggests to {\it dimensionally reduce} the effective action of a D-6-brane (see subsects.~4.2 
and 4.4) down to four spacetime dimensions. We already know that in M-Theory the D-6-brane is just 
a KK monopole described by the non-singular eleven-dimensional metric (4.19). Hence, the induced 
metric in the 6-brane worldvolume (in a static gauge) is given by
$$ g_{\m\n}=\h_{\m\n} + G_{ij}(y)\pa_{\m}y^i\pa_{\n}y^j~,\eqno(4.44)$$
where $G_{ij}$ is the four-dimensional ETN-metric, $\m,\n=0,1,\ldots,6$ and $i,j=1,2,3,4$. 
Substituting eq.~(4.44) into the LEEA (Nambu-Goto) part of the 6-brane effective action,
$$ S[y]=\int d^7\x\,\sqrt{-\det(g_{\m\n})}~,\eqno(4.45)$$
expanding it up to the second-order in derivatives,
$$ \sqrt{-\det(g_{\m\n})}={\rm const}. - \frac{1}{2}\h^{\m\n}g_{\m\n}+\ldots~,\eqno(4.46)$$
where the dots stand for higher-derivative terms, and performing plain dimensional reduction 
from seven to four dimensions (in fact, our fields do not depend upon three irrelevant coordinates 
already) results in the NLSM action with the Taub-NUT metric, 
$$ S[y]=-\fracm{1}{2}\int d^4x\, G_{ij}(y)\pa_{\un{\m}}y^i\pa^{\un{\m}}y^j~,\eqno(4.47)$$
in full agreement with the QFT results obtained in the $D=4$, $N=2$ harmonic superspace 
(subsect.~3.2).

Therefore, as regards the {\it leading} (i.e. the second-order in spacetime derivatives, in 
components) contributions to the $N=2$  supersymmetric LEEA, the brane technology offers very 
simple classical tools for their derivation, when compared to the conventional and considerably
more complicated QFT methods. Being either {\it holomorphic} in the case of $N=2$ vector multiplets
or {\it analytic} in the case of hypermultiplets, the leading contributions are always given by 
integrals over a half of the $N=2$ superspace anticommuting coordinates on dimensional reasons, 
while they are {\it protected} due to their `anomalous' nature against higher loop quantum 
corrections and mixed instanton/anti-instanton contributions. All the higher-order terms 
(in the spacetime derivatives) in the $N=2$ supersymmetric LEEA are given by integrals over the 
whole $N=2$ superspace, and they are, therefore, `unprotected' which casts some doubts on the 
applicability of the brane technology to their derivation. In sect.~5 we discuss the 
next-to-leading-order correction to the gauge (SW) $N=2$ LEEA as an example.  

\newpage

\subsection{Relation to the HSS results and S duality}

The relation between the HSS results in subsect.~3.2 and that of brane technology
in subsect.~4.6 towards the hypermultiplet LEEA (in fact, their equivalence)
is provided by the {\it S-duality} in field theory (Fig.~5).

\begin{figure}
\vglue.1in
\makebox{
\epsfxsize=4in
\epsfbox{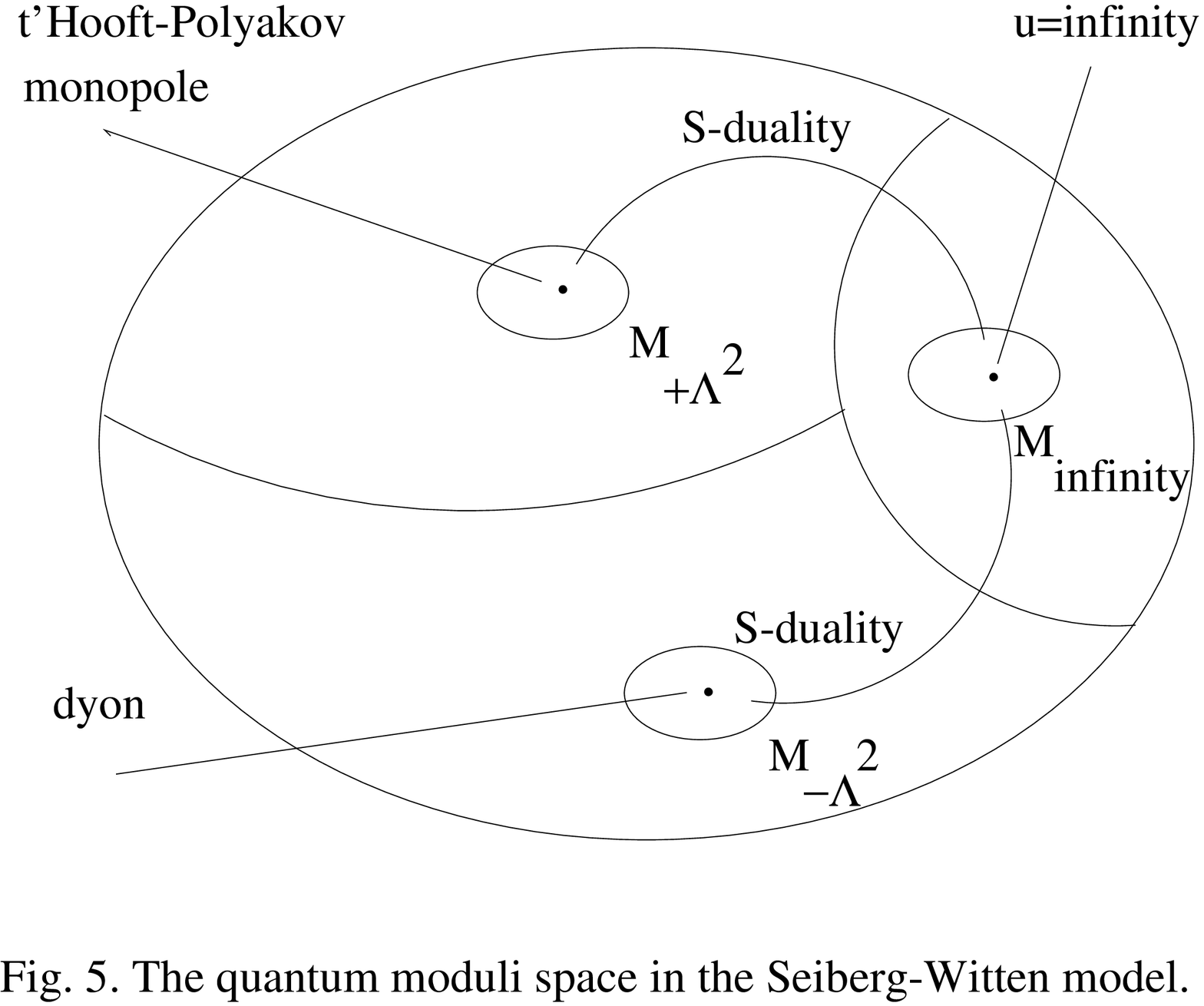}
}
\end{figure}

Consider, for simplicity, the famous Seiberg-Witten model~\cite{sw} whose 
fundamental action describes the purely gauge $N=2$ super-Yang-Mills theory
with the $SU(2)$ gauge group spontaneously broken to its $U(1)_{\rm e}$ subgroup.

On the one hand, in the strong coupling region of the Coulomb branch, 
e.g. near a singularity in the quantum moduli space where a BPS-like 
(t'Hooft-Polyakov) monopole becomes massless, the Seiberg-Witten (SW) theory is 
just described by the S-dual $N=2$ supersymmetric QED. 
In particular, the t'Hooft-Polyakov (HP) monopole belongs to a
{\it magnetically} charged hypermultiplet $q^+_{\rm HP}$ which represents the
non-perturbative degrees of freedom in the strongly coupled (Coulomb) branch of the 
SW theory under consideration (Fig.~5). The HSS results of subsect.~3.2 imply that the 
HP hypermultiplet self-interaction in the vicinity of the HP-monopole singularity is 
regular in terms of the magnetically dual variables,
$$ \ck^{(+4)}\low{\rm Taub-NUT}({q^+}\low{\rm HP}) 
=\fracmm{\l_{\rm dual}}{2}\left(
\sbar{q^+}\low{\rm HP}{q^+}\low{\rm HP}\right)^2~,\eqno(4.48)$$
i.e. it is given by the NLSM with the Taub-NUT metric.

On the other hand, from the type-IIA superstring (or M-theory) point of view,
the HP-hypermultiplet is just the zero mode of the open superstring stretching
between a magnetically charged D-6-brane and a D-4-brane. Therefore, the magnetically 
charged (HP) hypermultiplet is the only $N=2$ matter that survives in the effective
four-dimensional $N=2$ gauge theory (given by the $N=2$ super-QED) near the HP singularity 
after taking the proper LEEA limit of the brane configuration where the supergravity decouples. 
According to the preceeding subsections, the NLSM target space geometry governing the induced HP 
hypermultiplet self-interaction has to be that of Taub-NUT (or KK-monopole) again~!

To conclude this section, we would like to return to the well-known and important phenomenon of 
the {\it symmetry enhancement}, by using the particular case of two coinciding D-6-branes 
`on top of each other' as an example. Their M-Theory resolution was described in subsect.~4.2 
in terms 
of the hyper-K\"ahler metric governed by the harmonic potential (4.21) and describing two nearly 
coinciding ETN monopoles with equal charges put on a line going through the origin. The associated 
{\it double} (i.e. two-centered) ETN metric is, in fact, equivalent to the {\it mixed} EH-ETN 
metric defined by the following NLSM action~\cite{giot}:  
$$\eqalign{
S_{\rm mixed}[q_A,V_{\rm L}] = \int_{\rm analytic}\,& 
\left\{  q^{a+}_A D^{++} q^+_{aA}+V^{++}_{\rm L} \left( \frac{1}{2}
\ve^{AB}q^{a+}_A q^+_{aB} +\x^{++}\right) \right.\cr
~& \left.  +\frac{1}{8} \l \left( q^{a+}_A q^+_{aA}\right)^2
\right\}~,\cr}\eqno(4.49)$$
written down in terms of a gauged $O(2)$ analytic superfield $q^+_A$, $A=1,2$, and the associated 
$O(2)$ vector gauge analytic superfield (Lagrange multiplier) $V^{++}_{\rm L}$ having no kinetic 
term. The hyper-K\"ahler metric in components, corresponding to the HSS action (4.49) leads to the 
harmonic potential (4.21) of the double ETN monopole, as was shown by an explicit calculation in 
ref.~\cite{gorv}. Moreover, the action (4.49) obviously respects the $U(1)_A\times U(1)_{\rm PG}$ 
symmetry. In the limit $\l=0$, eq.~(4.49) leads to the EH metric (see sect.~6), whereas in another 
limit $\x^{++}\equiv \x^{ij}u^+_iu^+_j=0$ it results in the ETN metric up to a superfield 
redefinition~\cite{gorv}. Stated
differently, the HSS action (4.49) interpolates between the ETN action (see Appendix A for details)
and the EH action (see Appendix B for details), as well as between the corresponding hyper-K\"ahler 
metrics. 

Unlike the description in terms of the harmonic potential (4.21) which is not manifestly 
$N=2$ supersymmetric and looks singular, the equivalent description in terms of the
action (4.49) is fully non-singular and manifestly $N=2$ supersymmetric. In particular, the 
coincidence limit $\e\to 0$ in eq.~(4.21) corresponds to the limit $\x\to 0$ in eq.~(4.49) where 
it obviously gives rise to the enhanced $U(2)$ isometry since any explicit dependence of the HSS
lagrangian of eq.~(4.49) upon harmonics drops in this limit.

\newpage

\section{Next-to-leading-order correction to the Seiberg-Witten LEEA}

The next-to-leading-order correction to the $N=2$ gauge (Seiberg-Witten) LEEA in 
the Coulomb branch is determined by a real function $\ch$ of $W$ {\it and} $\Bar{W}$, which is
supposed to be integrated over the whole $N=2$ superspace. Since the full $N=2$ superspace
measure $d^4xd^8\q$ is dimensionless, the function $\ch(W,\Bar{W})$ should also be 
dimensionless, i.e. without any $N=2$ superspace derivatives~\cite{wgr}. Moreover,
the exact function 
$$ \ch(W,\Bar{W})=\ch_{\rm per.}(W,\Bar{W})+\ch_{\rm non-per.}(W,\Bar{W})
\eqno(5.1)$$
has to be S-duality invariant~\cite{henn}. 

It is also worth mentioning that the non-holomorphic function $\ch$ is only defined 
modulo K\"ahler gauge transformations 
$$\ch(W,\Bar{W})\to \ch(W,\Bar{W})+f(W) +\bar{f}(\Bar{W})~,
\eqno(5.2)$$
with an arbitrary holomorphic function $f(W)$ as a gauge parameter.

\begin{figure}
\vglue.1in
\makebox{
\epsfxsize=4in
\epsfbox{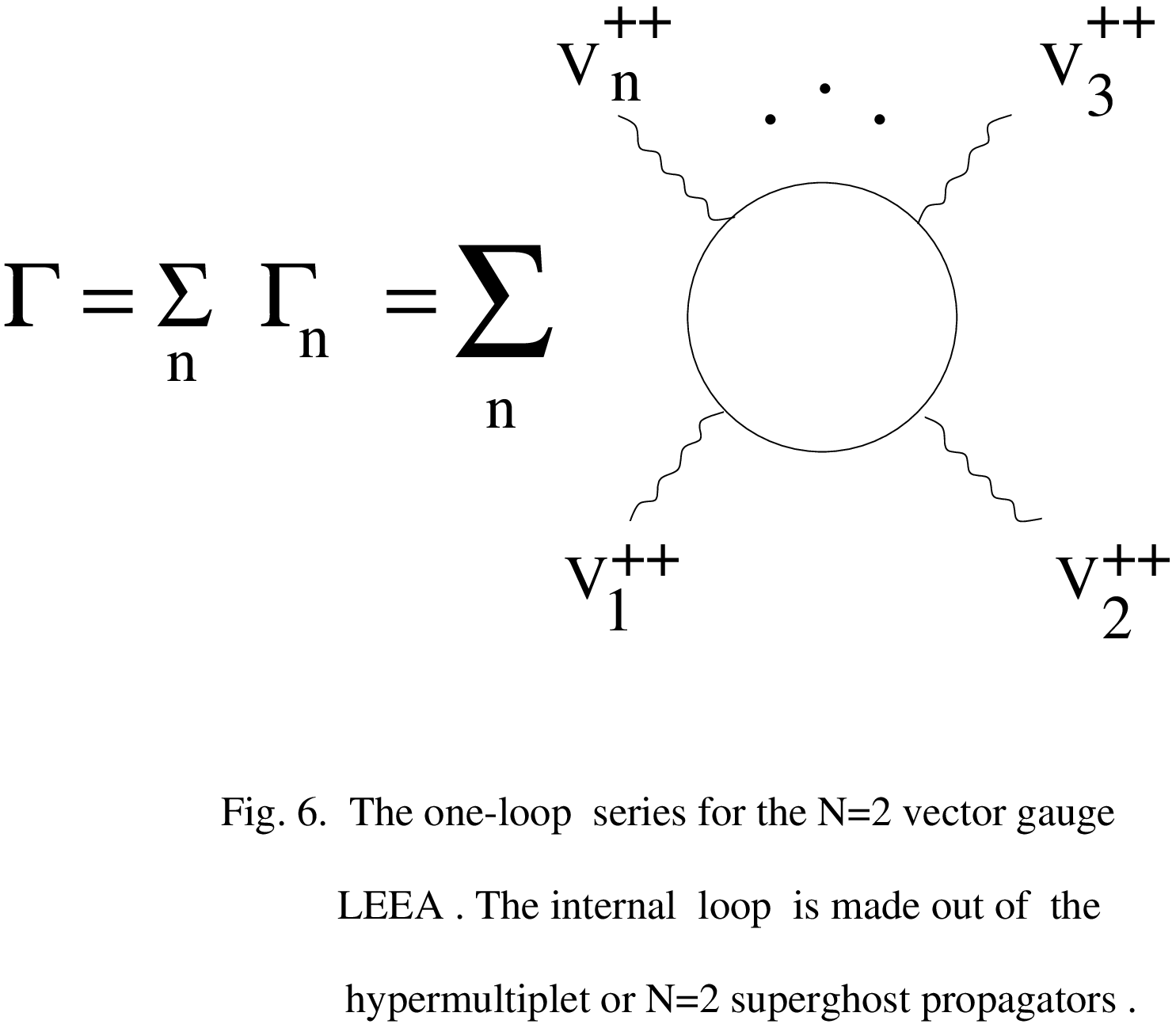}
}
\end{figure}

Within the manifestly $N=2$ supersymmetric background-field approach in $N=2$ HSS, 
the one-loop contribution to $\ch_{\rm per.}$ is given by a sum of the $N=2$ 
HSS graphs schematically pictured in Fig.~6. The sum goes over the external
$V^{++}$-legs, whereas the loop consists of the $N=2$ matter (and $N=2$ ghost) 
superpropagators (sect.~3). $N=2$ ghosts contribute in very much the same
way as $N=2$ matter does, since the $N=2$ ghosts are also described in 
terms of the FS- and HST-type hypermultiplets (with the opposite statistics of 
components) in the $N=2$ HSS~\cite{bb}. Because of the (abelian) gauge 
invariance, the result can only depend upon the abelian $N=2$ superfield 
strength $W$ and its conjugate $\Bar{W}$ via eq.~(3.21). In fact, Fig.~6 also 
determines the one-loop perturbative contribution to the leading holomorphic 
LEEA, which appears as the {\it anomaly} associated with the non-vanishing 
central charges~\cite{bko}. The self-energy $N=2$ HSS supergraph with only two external 
legs in Fig.~6 is the only one which is UV-divergent there. The IR-divergences 
of all the HSS graphs in Fig.~6  can be regularized by introducing the IR-cutoff $\L$, 
which is proportional to the Seiberg-Witten scale $\L_{\rm SW}$ defined in sect.~2,
with the relative coefficient being dependent upon the actual renormalization scheme used
(see e.g., ref.~\cite{svkh} for more details). In the case of
a single $q^+$-type (FS) matter hypermultiplet in an abelian  $N=2$ gauge theory, an 
evaluation of the infinite HSS series depictured in Fig.~6 
yields~\cite{bbi,bb,svkh,bkuz}~\footnote{Our overall normalization differs from that in
refs.~\cite{bbi,bb,bkuz} by some factors of $2$.}
$$ \cf_q(W)=-\fracmm{1}{32\p^2}W^2\ln\fracmm{W^2}{M^2}~,\eqno(5.3)$$
where the renormalization scale $M$ has been fixed by the condition $\cf_q(M)=0$, and
$$ \ch_q(W,\Bar{W})=\fracmm{1}{(16\p)^2} \sum^{\infty}_{k=1}\fracmm{(-1)^{k+1}}{k^2}\left(
\fracmm{W\Bar{W}}{\L^2}\right)^k=\fracmm{1}{(16\p)^2}\int_0^{W\Bar{W}/\L^2}\,\fracmm{d\x}{\x}
\ln(1+\x)~,\eqno(5.4)$$
where we have used the standard integral representation for the dilogarithm function. It is not 
difficult to verify that the asymptotical perturbation series (5.4) can be rewritten as 
$$ \ch_q(W,\Bar{W})=\fracmm{1}{(16\p)^2}\ln\left(\fracmm{W}{\L}\right)
\ln\left(\fracmm{\Bar{W}}{\L}\right)~,\eqno(5.5)$$
or, equivalently, as
$$\ch_q(W,\Bar{W})=\fracmm{1}{2(16\p)^2}\ln^2\left(\fracmm{W\Bar{W}}{\L^2}\right)~,\eqno(5.6)$$
modulo irrelevant K\"ahler gauge terms, see eq.~(5.2). The non-holomorphic contribution of 
eq.~(5.5) or (5.6) does not really depend upon the scale $\L$, again due to the K\"ahler 
invariance (5.2)~\cite{ds}.

The HSS result (5.3) for the perturbative part of the SW $N=2$ gauge LEEA agrees with the well-known
Seiberg argument \cite{sei} based on the perturbative $U(1)_R$ symmetry and an integration of the 
associated chiral anomaly. As is obvious from eq.~(5.5), the next-to-leading-order non-holomorphic 
contribution to the SW gauge LEEA satisfies a simple differential equation~\cite{svkh}
$$ W\Bar{W}\pa_W\pa_{\Bar{W}}\ch_q(W,\Bar{W})=const. \eqno(5.7)$$
which can be considered as the direct consequence of scale and $U(1)_R$ invariances ({\it cf.}
ref.~\cite{ds}).

In a more general case of $N_f$ FS-type hypermultiplets in the fundamental representation of the 
gauge group $SU(N_c)$, i.e. the $N=2$ super-QCD, the extra coefficient in front of the holomorphic
contribution $\cf$ is proportional to the one-loop RG beta-function $(N_f-2N_c)$, whereas the 
extra coefficient in front of the non-holomorphic contribution $\ch$ is proportional to 
$(2N_f-N_c)$, in the $N=2$ super-Feynman gauge~\cite{svkh}. In another interesting case of the 
$N=4$ super-Yang-Mills theory, whose $N=2$ matter content is given by a single HST-type 
hypermultiplet in the {\it adjoint} representation of the gauge group, the numerical coefficient in
front of the holomorphic function $\cf$ vanishes together with the RG beta-function, whereas the 
numerical coeffient in front of the non-holomorphic contribution $\ch$ always appears to be 
positive, in agreement with the earlier calculations in terms of $N=1$ superfields (see page 390 in
the second reference of \cite{superspace}) and some recent $N=2$ supersymmetric calculations by 
different methods~\cite{bkuz,stony}. In the case of finite and $N=2$ 
{\it superconformally} invariant gauge field theories $(N_f=2N_c)$, the leading non-holomorphic 
contribution to the LEEA is given by eq.~(5.5) multiplied by $3N_c$, so that it never vanishes too.

It is also straightforward to check in the HSS approach that there are no {\it two-loop} 
contributions to $\cf_{\rm per.}$ {\it and\/} $\ch_{\rm per.}$, since all the relevant HSS graphs 
shown in Fig.~7 do not actually contribute in the local limit. This conclusion is in agreement with 
calculations in terms of the $N=1$ superfields~\cite{ggruz}, and it is also consistent with the 
general perturbative structure of the $N=2$ supersymmetric gauge field theories within the 
manifestly $N=2$ supersymmetric framework of the background-field method in $N=2$ HSS \cite{bko}. 
It seems, therefore, to be conceivable that {\it all} higher-loop (perturbative) contributions to 
$\ch_{\rm per.}(W,\bar{W})$ may be absent too. It should definitely be the case in the $N=2$ 
superconformally invariant gauge field theories and the $N=4$ SYM theory as well, since any higher 
loop perturbative contributions would break the scale and $U(1)_R$ invariances of these theories. As
regards the non-perturbative (instanton) contributions to $\ch$, they are also expected to vanish in
the four-dimensional gauge field theories with extended {\it superconformal} invariance by the same 
reasoning. Possible non-perturbative instanton corrections in the finite $N=2$ supersymmetric gauge 
field theories were studied in refs.~\cite{npva,ital}, where their absence was argued within the 
standard instanton approach. Stated differently, the one-loop non-holomorphic result above 
represents the exact leading contribution to the LEEA in the $N=2$ superconformally invariant gauge 
field theories in $D=4$.
  
\begin{figure}
\vglue.1in
\makebox{
\epsfxsize=3in
\epsfbox{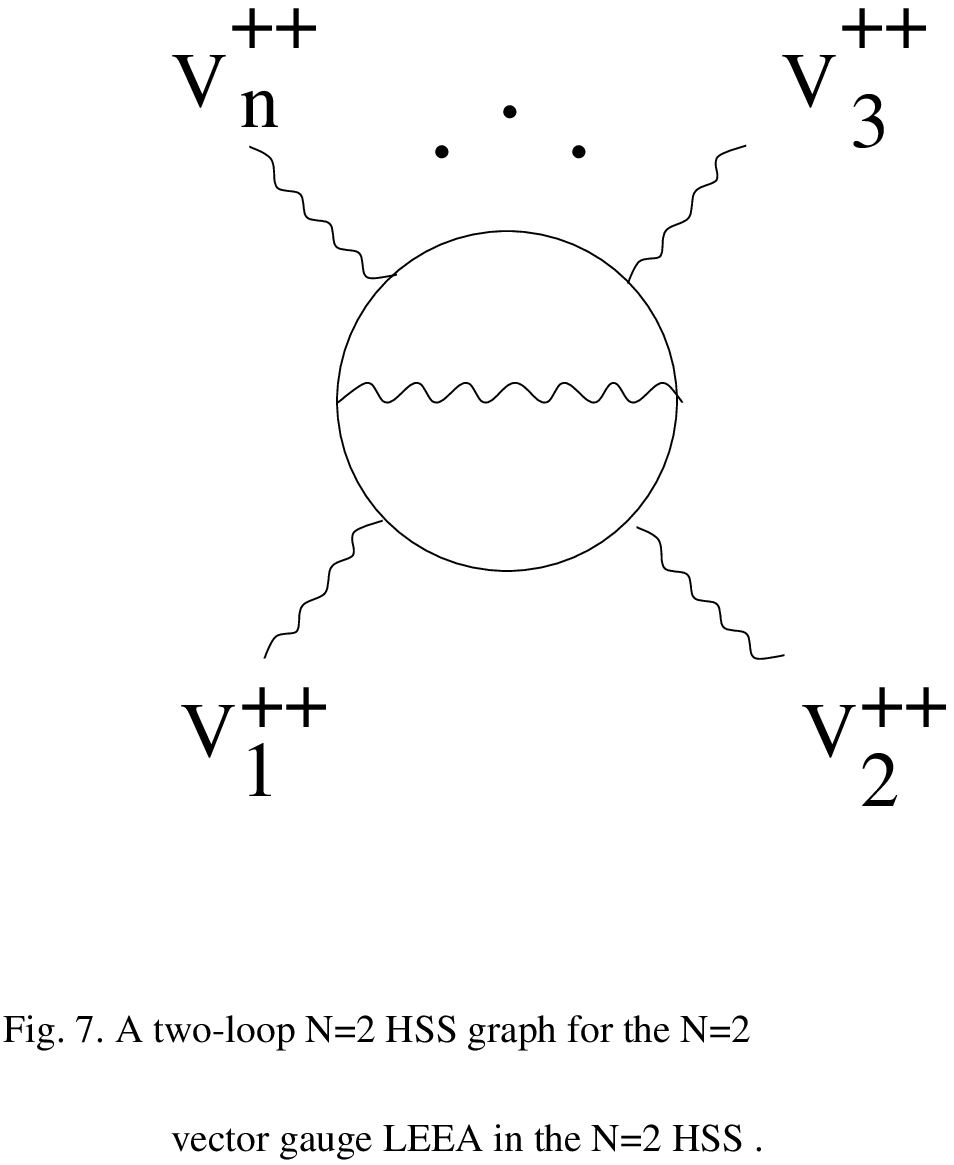}
}
\end{figure}

An exact result for the real function $\ch(W,\Bar{W})$ in a non-superconformally invariant $N=2$
gauge field theory is still unknown. There is, however, an interesting proposal due to Matone 
\cite{matone} that the exact function $\ch(W,\Bar{W})$ may satisfy a non-linear differential 
equation
$$ \pa\low{\Bar{W}}\pa\low{W}\ln\left[\ch\pa\low{W}\pa\low{\Bar{W}}\ln\ch
\right]=0~,\eqno(5.8)$$
which can be interpreted as the fully non-perturbative non-chiral superconformal
`Ward identity'.~\footnote{An explicit solution to eq.~(5.8) was also proposed
in ref.~\cite{matone}.} The leading one-instanton contribution 
in the pure $N=2$ gauge (SW) theory was already calculated in ref.~\cite{yung1}, 
and it does not vanish. The full non-perturbative contribution is unlikely 
to be given by the sum over instanton contributions alone, since it can also 
include (multi)anti-instanton and mixed (instanton-anti-instanton) 
contributions, which are allowed in a non-conformal theory.

The brane technology of sect.~4 might offer the alternative way of calculating the exact 
next-to-leading-order non-holomorphic correction, e.g. by using the covariant 
M-5-brane action describing the classical M-theory 5-brane dynamics, simply by expanding
it further in powers of the spacetime derivatives, up to the fourth-order terms in components. 
Such an investigation was done by de Boer, Hori, Ooguri and Oz in ref.~\cite{berk}, who showed 
that the actual results obtained from the brane technology considerably differ from those of the 
$N=2$ quantum gauge field theory, as regards the non-holomorphic terms in the LEEA. There is only 
one dynamically generated dimensionful scale $\L$ in the four-dimensional LEEA of the QFT under 
consideration, whereas there are at least two dimensionful parameters in M-Theory, namely, the  
radius $R$ of the compactified 11-th dimension and the typical scale $L$ of the brane configuration.
As was explained in ref.~\cite{berk}, the non-holomorphic contributions derived from the M-5-brane 
effective action are dependent upon the radius $R$ in a highly non-trivial way, while this 
dependence does not decouple in any simple limit. From the physical viewpoint, this implies the
absence of a natural field theory limit where the supergravity and the associated KK modes decouple
from the bulk.

Yet another possible approach to an explicit calculation of non-perturbative non-holomorphic terms 
in the $N=2$ field theory LEEA may be based on the use of $N=2$ HSS in instanton-type calculations,
which is, however, yet to be developed.

\section{Hypermultiplet LEEA in baryonic Higgs branch}

The Higgs branch can be naturally divided into two phases called `{\it baryonic}' and 
`{\it non-baryonic}' according to ref.~\cite{apsh}. In the baryonic phase, Fayet-Iliopoulos (FI)
terms are present so that the color symmetry is completely broken. In the non-baryonic phase, all 
the FI terms vanish so that a non-trivial gauge subgroup of the color symmetry survives. In this 
section we are going to discuss the baryonic phase only. It intersects with the Coulomb branch at 
a {\it single} point in the quantum moduli space of the Coulomb branch~\cite{apsh}.

As was already mentioned in sect.~3, the most natural and manifestly $N=2$
supersymmetric description of hypermultiplets in the Higgs branch is provided 
by HSS in terms of the HST-type analytic superfield $\o$ of vanishing
$U(1)$ charge. The $N=2$ HSS is also the quite natural framework to address 
possible symmetry breakings.

The free action of a single $\o$ superfield reads
$$ S[\o]=-\ha\int_{\rm analytic} \left(D_A^{++}\o\right)^2~.\eqno(6.1)$$
Similarly to the free action (3.11) for a $q^+$-type analytic superfield, the
action (6.1) also possesses the extended internal symmetry
$$SU(2)_A\otimes SU(2)_{\rm PG}~,\eqno(6.2)$$
where $SU(2)_A$ is well-known automorphism symmetry of the $N=2$ supersymmetry algebra
(sometimes also called the $SU(2)_R$ symmetry). The additional $SU(2)_{\rm PG}$ symmetry of
eq.~(6.1) is less obvious~\cite{ikz}~:
$$ \d\o=c^{--}D_A^{++}\o-c^{+-}\o~,\eqno(6.3)$$
where $c^{--}=c^{(ij)}u^-_iu^-_j$ and $c^{+-}=c^{(ij)}u^+_iu^-_j$, and
$c^{(ij)}$ are the infinitesimal parameters of $SU(2)_{\rm PG}\,$.

It is quite clear now that it is not possible to construct any non-trivial 
self-interaction in terms of the $U(1)$-chargeless superfield $\o$ alone, without
breaking the $SU(2)_A$ symmetry, simply because a hyper-K\"ahler potential has to
have the non-vanishing $U(1)$ charge $(+4)$. Hence, when $N=2$ supersymmetry and the 
$SU(2)_A$ internal symmetry are not broken, one gets the well-known result~\cite{sw}:
$$\ck^{(+4)}\low{\rm Higgs}(\o)=0~,\eqno(6.4)$$
i.e. the induced hyper-K\"ahler metric in the fully $N=2$ supersymmetric Higgs
branch is flat and, in particular, it does not receive quantum corrections.

It is, however, possible to break the internal symmetry (6.2) down to 
$$ U(1)_A\otimes SU(2)_{\rm PG}~,\eqno(6.5)$$
by introducing the FI term
$$\VEV{D^{ij}}=\x^{ij}={\it const.}\neq 0~,\eqno(6.6)$$
which may softly break the $N=2$ supersymmetry too~\cite{amz}. This way of symmetry breaking 
still allows one to maintain control over the $N=2$ supersymmetric hypermultipet LEEA due to the 
presence of the non-abelian internal symmetry (6.5). The only non-trivial hyper-K\"ahler potential, 
which is invariant with respect to the symmetry (6.5), is given by~\cite{ikz}
$$\ck^{(+4)}\low{\rm EH}(\o)=-\,\fracmm{(\x^{++})^2}{\o^2}~,\eqno(6.7)$$
where $\x^{++}=\x^{ij}u^+_iu^+_j$. It is straightforward to deduce the
corresponding hyper-K\"ahler metric from eq.~(6.7) by using the procedure
already described in subsect.~3.2. One finds that the metric is
equivalent to the standard {\it Eguchi-Hanson} (EH) instanton
metric in four dimensions~\cite{giot,kup} (see Appendix B also).

In order to understand better the origin of the EH-metric from HSS,
it is quite useful to employ the {\it gauging} procedure of generating 
hyper-K\"ahler metrics (see ref.~\cite{hits} for a review). In the $N=2$ HSS,
the additional resources for generating new hyper-K\"ahler potentials
are given by (i) gauging isometries and (ii) adding (electric) FI terms. For instance, 
given two FS-type hypermultiplets $q^+_A\in \un{\bf 2}$ of  
$SU(2)_f$, one can gauge  a $U(1)$ subgroup of $SU(2)_f$ and simultaneously add a FI
term as follows~\cite{giot}:
$$ S_{\rm EH}=\int_{\rm analytic}\,\left\{ q^{a+}_A D^{++} q^+_{aA}+V^{++}_{\rm L}\left( \frac{1}{2}
\ve^{AB}q^{a+}_A q^+_{aB} +\x^{++}\right)\right\}~,\eqno(6.8)$$
where $V^{++}_{\rm L}$ is the corresponding $N=2$ vector gauge potential 
without a kinetic term (i.e. the Lagrange multiplier),
$\x^{++}=\x^{(ij)}u_i^+u^+_j$, and $A,B=1,2$. The action (6.8) has the manifest PG symmetry 
$SU(2)_{\rm PG}$, which rotates the lower-case Latin indices only. The $SU(2)_A$ symmetry is
explicitly broken down to its $U(1)_A$ subgroup by a non-vanishing FI-term $\x^{++}$. After some
algebra, the Lagrange multiplier $N=2$ superfield $V^{++}_{\rm L}$ and one of the hypermultiplet
superfields can be eliminated, while the resulting action equivalent to eq.~(6.8) 
takes the form~\cite{giot}
$$ S_{\rm EH}=\int_{\rm analytic}\,\left\{ q^{a+} D^{++} q^+_{a} -
\fracmm{(\x^{++})^2}{(q^{a+}u^-_a)^2}\right\}~.\eqno(6.9)$$
Changing the variables $q^+_a=u^+_a\o +u^-_af^{++}$, in terms of the dual $\o$-type hypermultiplet
and yet another Lagrange multiplier $f^{++}$, gives us the HSS action with the hyper-K\"ahler 
potential (6.7) after eliminating $f^{++}$ according to its algebraic HSS superfield equations of 
motion. It should be noticed that the hyper-K\"ahler potential (6.7) already implies that 
$\VEV{\o}\neq 0$, so that we are in the Higgs branch indeed. 

Let's now try to understand how a FI-term could be non-perturbatively generated. First, we can 
slightly generalize this problem by allowing non-vanishing vacuum 
expectation values for all bosonic components of the abelian $N=2$ superfield strength $W$,
$$ \VEV{W}=\left\{~ \VEV{a}=Z~,\quad \VEV{F_{\m\n}}=n_{\m\n}~,\quad
\VEV{\vec{D}}=\vec{\x} ~~\right\}~,\eqno(6.10)$$
where all the parameters $(Z,n_{\m\n},\vec{\x})$ are constants. Generally
speaking, it also implies {\it soft} $N=2$ supersymmetry breaking~\cite{amz}.
We already know about the physical meaning of $Z$ --- it is just the complex 
central charge in the $N=2$ supersymmetry algebra. The related gauge-invariant 
quantity $u\sim\VEV{\tr\,a^2}$ parameterizes the quantum moduli space of vacua in
the Coulomb branch. The central charge $Z$ can be naturally generated via the standard 
Scherk-Schwarz mechanism of dimensional reduction from six dimensions~\cite{ikz}. 
Similarly, $n_{\m\n}\neq 0$ can be interpreted as a {\it toron} background after replacing 
the effective spacetime $R^{1+3}$ by a hypertorus $T^{1+3}$ and imposing t'Hooft's twisted 
boundary conditions~\cite{toot}. The $\vec{\x}\neq 0$ is just a FI term.

The brane technology can help us to address the question of {\it dynamical} 
generation of both $n_{\m\n}$ and $\vec{\x}$ in a very geometrical way~\cite{ikz}. Namely,
let's deform the brane configuration of Fig.~3 by allowing some of the NS-5- or D-4-branes
to {\it intersect at angles} instead of being parallel~! Indeed, the vector
$\vec{w}=(x^7,x^8,x^9)$ is the same in Fig.~3 for both solitonic 5-branes.
Its non-vanishing value 
$$\vec{\x}=\vec{w}_1-\vec{w}_2\neq 0 \eqno(6.11)$$
effectively generates the FI term. Similarly, when allowing the D-4-branes to
intersect at angles, some non-trivial values of $\VEV{F_{\m\n}}=n_{\m\n}\neq 0$
are generated~\cite{dtoron}.

Since the spacetime LEEA of BPS branes is governed by a gauge field theory, it does not seem 
to be very surprising that torons can also be understood as the BPS bound states
of certain D-branes in the field theory limit $M_{\rm Planck}\to\infty$ \cite{dtoron}. 
Moreover, torons are known to generate a gluino condensate~\cite{zhit}
$$ \VEV{\l^i\l^j}=\L^3(\x^2)^{ij}~,\quad \x^{ij}\sim \d^{ij}\exp\left(-\,
\fracmm{2\p^2}{g^2}\right)~,\eqno(6.12)$$
where $\vec{\x}\sim \{\x^{ij}\}$ have to be constant~\cite{itepi}, so that they can
be identified with the FI term by $N=2$ supersymmetry.

Finally, it can be useful to understand the origin of the 
hypermultiplet EH-type self-interaction in the Higgs branch from the viewpoint
of brane technology. It is worth mentioning here that the D-4-branes can also
end on the D-6-branes (in the type-IIA picture), while these  D-4-branes
actually support hypermultiplets, not $N=2$ vector multiplets~\cite{witten}. It
results in the hyper-K\"ahler manifold $Q$ with has the different (versus ETN) topology 
$\sim S^3/Z_2$ in its spacial infinity. It now suffices to mention that the
EH-instanton is the only hyper-K\"ahler manifold having this topology among
the four-dimensional ALF spaces~!

\section{Conclusion}

Though being very different, the three approaches depictured in Fig.~1, namely, 
\begin{itemize}
\item[(i)] instanton calculus, 
\item[(ii)] Seiberg-Witten approach and M theory (=brane technology),
\item[(iii)] harmonic superspace,
\end{itemize}
lead to the {\it consistent} results, as regards the {\it leading} terms in the LEEA
of the four-dimensional $N=2$ supersymmetric gauge field theories. It is the second (ii) and 
third (iii) approaches that were extensively discussed in this paper, whereas the instanton calculus
(i) was merely mentioned. Apparently, there is no single universal method to handle the whole
range of problems associated with non-perturbative $D=4$ supersymmetric gauge field theories in a
natural and easy way. Instead, each approach has its own advantages and disadvantages. For example, 
in the Seiberg-Witten approach, the physical information is encoded in terms of functions defined 
over the quantum moduli space whose modular group is identified with the S-duality group. The SW
approach is based on knowing exact perturbative limits of the non-abelian $N=2$ gauge field theory 
under consideration, whereas the HSS approach is the most efficient one in quantum perturbative
calculations. At the same time, the HSS approach up to now cannot be directly applied to address 
truly non-perturbative phenomena yet. In this paper, the HSS description of some non-perturbative
features related to hypermultiplets was only used in combination with the strong-weak couling 
duality (=S-duality). In its turn, the instanton calculus is very much dependent upon applicability 
of its own basic assumptions, while it is not manifestly supersymmetric at any rate if it is 
supersymmetric at all. Moreover, the known instanton methods sometimes need an additional input. 
Being geometrically very transparent, the M-Theory brane technology still has a limited analytic 
support, whereas its successful applications were limited so far to only those terms in the LEEA 
which are protected by anomalous symmetries, i.e. which are either holomorphic or analytic. In any 
case, some care should be exercised to play safely with the brane technology. I believe, it is 
merely a clever combination of all the methods available that has the strongest potential for 
further progress, while it can simultaneously teach us how to proceed with each particular approach.

I would like to conclude with a few comments about $N=2$ supersymmetry breaking
and confinement, in order to indicate on a possible importance of the exact
hypermultiplet low-energy effective action towards a solution to these 
problems. Indeed, it seems to be quite natural to take advantage of the 
existence of exact solutions to the low-energy effective action in $N=2$ 
supersymmetric gauge field theories, and apply them to the old problem of 
color confinement in QCD. In fact, it was one of the main motivations in the 
original work of Seiberg and Witten~\cite{sw}. The most attractive mechanism 
for color confinement is known to be the dual Meissner effect or the dual 
(Type II) superconductivity \cite{mant}. It takes three major steps to connect
an ordinary BCS superconductor to the simplest Seiberg-Witten model in quantum
field theory: first, one defines a {\it relativistic} version of the 
superconductor, known as the (abelian) Higgs model in field theory, second, one
introduces a {\it non-abelian} version of the Higgs model, known as the 
Georgi-Glashow model, and, third, one {\it N=2 supersymmetrizes} the 
Georgi-Glashow model in order to get the Seiberg-Witten model~\cite{sw}. Since
the t'Hooft-Polyakov monopole of the Georgi-Glashow model belongs to a (HP)
hypermultiplet in its $N=2$ supersymmetric (Seiberg-Witten) generalization,
it is quite natural to explain confinement as the result of a monopole
condensation (= the dual Meissner effect as the consequence of the dual 
Higgs effect), due to a non-vanishing vacuum expectation value for the 
magnetically charged (dual Higgs) scalar belonging to the HP hypermultiplet. 
Of course, it is only possible after a judicious $N=2$ supersymmetry breaking.

It is worth emphasizing that our whole discussion in this paper was essentially based 
on having eight conserved supercharges or, equivalently, the $N=2$ extended supersymmetry 
in the $D=4$ spacetime. Accordingly, most of the results are not applicable to a more 
interesting case of $D=4$ gauge field theories with only $N=1$ supersymmetry 
(see e.g., ref.~\cite{peskin} for a review of the known field theory results, and ref.~\cite{gku} 
for a review of brane technology with less than eight supercharges). An exact solution to the 
low-energy effective action of a $D=4$ quantum gauge field theory is generically available only 
with $N=2$ supersymmetry which is directly connected to integrable systems, and neither with merely 
$N=1$ supersymmetry nor in the bosonic QCD. Therefore, on the one hand side, it is the $N=2$ 
supersymmetry that crucially simplifies an evaluation of the $D=4$ QFT low-energy effective action. 
However, on the other hand side, it is the same $N=2$ supersymmetry in $D=4$ that is obviously 
incompatible with phenomenology e.g., because of equal masses of bosons and fermions inside 
$N=2$ supermultiplets (this is applicable, in fact, to any $N\geq 1$ supersymmetry), and the 
non-chiral nature of the $N=2$ supersymmetry (e.g. `quarks' can only appear in {\it real} 
representations of the gauge group). Therefore, if one believes in the fundamental role of $N=2$ 
supersymmetry in high energy physics, one has to find a way of judicious $N=2$ supersymmetry 
breaking. The associated with supersymmetry breaking, dual Higgs mechanism may then be responsible 
for the chiral symmetry breaking and the appearance of the pion effective Lagrangian too, provided
that the dual Higgs field has flavor charges~\cite{sw}. In fact, Seiberg and Witten~\cite{sw} used 
a mass term for the $N=1$ chiral multiplet which is a part of the $N=2$ vector multiplet, in 
order to {\it softly} break $N=2$ supersymmetry to $N=1$ supersymmetry `by hand'. As a result, they 
found a non-trivial vacuum solution with a monopole condensation and, hence, a confinement. The weak
point of their approach is an {\it ad hoc} assumption about the existence of the mass gap, i.e. the 
mass term itself. 

The $N=2$ supersymmetry can be broken either softly or spontaneously, if one
wants to preserve the benefits of its presence (e.g. maintaining the full control
over the low-energy effective action) at high energies. A detailed investigation
of {\it soft} $N=2$ supersymmetry breaking in the $N=2$ supersymmetric QCD was done by 
Alvarez-Gaum\'e, Mari\~no and Zamora~\cite{amz}.~\footnote{See e.g., ref.~\cite{dms} for 
a similar analysis in $N=1$ supersymmetric gauge field theories.} The soft $N=2$ supersymmetry  
breaking is most naturally done by the use of FI-terms~\cite{amz}. Though being pragmatic, 
the soft $N=2$ supersymmetry breaking has a limited predictive power because of many 
parameters, whose number, however, is significatly less than that in the $N=1$ case. It may be
more reasonable to search for {\it spontaneous} $N=2$ supersymmetry breaking, where the 
non-vanishing FI-terms would appear as stationary solutions to the dynamically generated 
scalar potential. This would imply the existence of a non-supersymmetric
vacuum solution for the exact $N=2$ supersymmetric scalar potential which is quite difficult to
get. Since the $N=2$ supersymmetry remains unbroken for any exact Seiberg-Witten solution in the
gauge sector, we may have to consider the induced scalar potentials in the hypermultiplet sector 
of an $N=2$ gauge theory in $D=4$. Indeed, given non-trivial kinetic terms in the hypermultiplet
low-energy effective action to be represented by a hyper-K\"ahler NLSM, in the presence of 
non-vanishing central charges they also imply  a non-trivial hypermultiplet {\it scalar} potential 
whose form is entirely determined by the hyper-K\"ahler metric of the kinetic terms and $N=2$ 
supersymmetry (see Appendices A and B for two explicit examples). Though it may not be easy to 
search for the most general solution with spontaneously broken $N=2$ supersymetry, partly because 
of complications associated with general hyper-K\"ahler geometries having no isometries, our toy 
examples presented in the Appendices are enough to demonstrate a richness of opportunities there.

\section*{Acknowledgements}

I would like to thank the Physics Department of the University of Maryland at 
College Park, the Physics Department of Pennsylvania State University, the
Lyman Laboratory of Physics in Harvard University, the ICTP in Trieste and the 
Theory Division of CERN for a kind hospitality extended to me during the 
completion of this work. I am also grateful to Luis Alvarez-Gaum\'e, Jonathan 
Bagger, Diego Bellisai, Friedemann Brandt, Joseph Buchbinder, Norbert Dragon, Jim Gates, 
Marc Grisaru, Evgeny Ivanov, Andrei Johansen, Olaf Lechtenfeld, Wolfgang Lerche, Marcus Luty, 
Andrei Mironov, Alexei Morozov, Burt Ovrut, Norisuke Sakai, J\"urgen Schulze, John Schwarz, 
Dima Sorokin, Cumrun Vafa, Alexei Yung and Boris Zupnik for useful discussions.

\newpage

\noindent
{\large\bf Appendix A: Taub-NUT hypermultipet self-interaction \newline 
${~~~~~~~~~~~~~~~~~}$  in components from harmonic superspace} 

In this Appendix we explicitly calculate the induced (effective) metric and the induced 
(effective) scalar potential in components from the $N=2$ harmonic superspace results 
(subsect.~3.2) for a single charged FS-type hypermultiplet in the Coulomb branch. The 
corresponding hypermultiplet LEEA is given by the NLSM, whose action in the analytic subspace 
$\z^M=(x^m_A,{\theta^+}\low{\a},\bar{\theta}^+_{\dt{\a}})$ of the $N=2$ HSS reads~\footnote{We
use a separate numeration of equations in each Appendix.}
\begin{eqnarray}
S_{\rm ETN}[q]= -\int d\z^{(-4)} du \left\{~ \sbar{q}{}^+D^{++}q^+ 
+\fracmm{\lambda}{2}(q^+)^2(\sbar{q}{}^+)^2 ~\right\}~,
\label{action}
\end{eqnarray}
where the HSS covariant derivative $D^{++}$, in the analytic basis and with 
non-vanishing central charges $Z$ and $\Bar{Z}$, has been introduced,
\begin{eqnarray}
D^{++} =\partial^{++}
-2i\theta^+\sigma^m\bar{\theta}^+\partial_m +i\theta^+\bar{\theta}^+\Bar{Z}
+i\bar{\theta}{}^+\bar{\theta}{}^+Z~,
\label{der}
\end{eqnarray}
whereas $\l$ is the induced (Taub-NUT) NLSM coupling constant. Eq.~(\ref{der})
can be most naturally obtained by (Scherk-Schwarz) dimensional reduction from six
dimensions~\cite{ikz}. For simplicity, we ignore here possible couplings to an
abelian $N=2$ vector superfield. An explicit expression for $\l$ in terms of 
the fundamental gauge coupling and the hypermultiplet BPS mass was calculated in
ref.~\cite{ikz}. Our $q^+$ superfields are of dimension minus one (in units of 
length), while the coupling constant $\l$ is of dimension two. 

Our aim here is to find the component form of the action (\ref{action}). 
Without central charges it was done in ref.~\cite{giost}. The non-vanishing 
central charges were incorporated in ref.~\cite{ikz}. Here we are going to concentrate 
on a derivation of the component metric {\it and\/} the scalar potential originating 
from the $N=2$ HSS action (\ref{action}). The corresponding equations of motion are
\begin{equation}
D^{++}q^+ +\lambda(q^+\sbar{q}{}^+)q^+ = 0 \quad {\rm and}\quad 
D^{++}\sbar{q}{}^+ -\lambda(q^+\sbar{q}{}^+)\sbar{q}{}^+ = 0 ~.
\label{eqmo}
\end{equation}

Since we are only interested in the purely bosonic part of the action 
(\ref{action}), we drop all fermionic fields in the $\theta^+$,
$\bar{\theta^+}$ expansion of $q^+$,
\begin{eqnarray}
q^+(\z,u) &=& F^+(x_A,u) + i\theta^+\sigma^m\bar{\theta}^+A^-_m(x_A,u) 
\nonumber\\
& &+\theta^+\theta^+M^-(x_A,u)+\bar{\theta}^+\bar{\theta}^+N^-(x_A,u)
\nonumber\\
& & +\theta^+\theta^+\bar{\theta}^+\bar{\theta}^+P^{(-3)}(x_A,u) ~.
\label{exp}
\end{eqnarray}
After being substituted into eq.~(\ref{eqmo}), eq.~(\ref{exp}) yields
\begin{eqnarray}
\label{bw1}
\partial^{++}F^+ +\lambda(F^+\sbar{F}{}^+)F^+ &=& 0~, \\
\label{bw2}
\partial^{++}A^-_m - 2\partial_mF^+ +2\lambda A^-_m \sbar{F}{}^+F^+ 
+ \lambda (F^+)^2 \sbar{A}_m{}^- &=& 0~, \\
\label{bw3}
\partial^{++}M^- +2\lambda M^- \sbar{F}{}^+F^+ + \lambda \sbar{N}{}^-(F^+)^2 
+i\Bar{Z}F^+ &=& 0~, \\
\label{bw4}
\partial^{++}N^- +2\lambda N^- \sbar{F}{}^+F^+ + \lambda \sbar{M}{}^-(F^+)^2 
+iZF^+ &=& 0~, \\
\label{bw5}
\partial^{++} P^{(-3)} + \partial^m A^-_m + 2\lambda F^+\sbar{F}{}^+P^{(-3)}
+\lambda(F^+)^2\sbar{P}{}^{(-3)} & & \nonumber \\
-\fracmm{\lambda}{2} A^{-m} A^-_m \sbar{F}{}^+ 
-\lambda A^{-m}\sbar{A}_m{}^- F^+
+2\lambda F^+(M^-\sbar{M}{}^- +N^- \sbar{N}{}^-) & &\nonumber \\
 +2\lambda \sbar{F}{}^+M^-N^- +i\Bar{Z}N^- +iZM^- &=& 0~, 
\end{eqnarray}
as well as their conjugates. Integrating over $\theta^+, \bar{\theta}^+$ in eq.~(\ref{action}) 
results in the bosonic action
\begin{eqnarray}
S_T &=& 
-\int d\z^{(-4)} du \left\{~ 
\sbar{q}{}^+D^{++}q^+ +\fracmm{\lambda}{2}(q^+)^2
(\sbar{q}{}^{+})^2 ~\right\}\nonumber\\
&\to & S= -\frac{1}{2}\int d^4xdu~ \left\{~ 
(\sbar{A}_m{}^-\partial^mF^+ - A^-_m\partial^m\sbar{F}{}^+)\right.\nonumber\\
& &\left. 
+iF^+(\Bar{Z}\sbar{M}{}^- +Z\sbar{N}{}^-)+i \sbar{F}{}^+(ZM^-+\Bar{Z}N^-)
~\right\}~.
\label{baction}
\end{eqnarray}
Since the action (\ref{action}) has the global $U(1)$ invariance
\begin{eqnarray}
q^{+~'}=e^{i\alpha}q^+, \qquad \sbar{q}{}^{+'}=e^{-i\alpha}\sbar{q}{}^+ ~,
\end{eqnarray}
there exists the conserved Noether current $j^{++}$~,
\begin{eqnarray}
D^{++}j^{++}=0, \qquad j^{++}=iq^+\sbar{q}{}^+ ~.
\end{eqnarray}
It implies, in particular, that $\partial^{++}(F^+\sbar{F}{}^+)=0$ and, hence,
\begin{eqnarray}
F^+(x,u)\sbar{F}{}^+(x,u)=C^{(ij)}u^+_{~i}u^+_{~j}~, \\
(F^+\sbar{F}{}^+)^{\frac{*}{}} = -F^+\sbar{F}{}^+ \rightarrow 
\Bar{C^{(ij)}}= -\epsilon_{il}\epsilon_{jn}C^{(ln)}~,
\end{eqnarray}
where the new function $C^{(ij)}(x)$ has been introduced. Changing the 
variables as
\begin{eqnarray}
F^+(x,u)=f^+(x,u)e^{\lambda \varphi}, \qquad 
\varphi(x,u)=-C^{(ij)}(x)u^+_{~i}u^-_{~j}=-\sbar{\varphi}(x,u)~,
\end{eqnarray}
reduces eq.~(\ref{bw1}) to the linear equation
\begin{eqnarray}
\partial^{++}f^+(x,u) = 0 \rightarrow f^+(x,u)= f^i(x)u^+_{~i}~.
\end{eqnarray}
After taking into account that
\begin{eqnarray}
F^+\sbar{F}{}^+=f^+\sbar{f}{}^+ \rightarrow 
C^{(ij)}(x)=-f^{(i}(x)\bar{f}^{j)}(x)~,
\end{eqnarray}
where $\bar{f}^i = \epsilon^{ij}\bar{f}_j$ and $\bar{f}_j\equiv\Bar{(f^j)}$,
we obtain a general solution in the form
\begin{eqnarray}
F^+(x,u)&=&f^{i}u_{~i}^{+} e^{\lambda\varphi} \nonumber \\
&=& f^{i}(x)u^+_{~i}\exp\{ \lambda f^{(j}\bar{f}^{k)}u_{~j}^{+}u_{~k}^{-} \}~.
\end{eqnarray}
The same conclusion appears when using the {\it Ansatz}
\begin{eqnarray}
F^{+}= e^{C}[ f^{i}u_{~i}^{+}+ B^{ijk}u_{i}^{+}u_{~j}^{+}u_{~k}^{-} ]
\label{1ansatz}
\end{eqnarray}
in terms of functions $C$ and $B^{ijk}$ at our disposal. After 
substituting eq.~(\ref{1ansatz}) into the equation of motion (\ref{bw1}),
we find
\begin{equation}
B^{ijk}= 0  \qquad {\rm and} \qquad
C = \lambda f^{(i}\bar{f}^{j)}u_{~i}^{+}u_{~j}^{-}~,
\end{equation}
so that 
\begin{eqnarray}
F^{+}= f^{i}u_{i}^{+}\exp\{ \lambda f^{(j}\bar{f}^{k)}u_{~j}^{+}u_{~k}^{-} \}
\end{eqnarray}
again. To get a similar equation for $A^{-}_{m}$, we use eq.~(\ref{bw2})
and the {\it Ansatz}
\begin{eqnarray}
A^{-}_{m} &=& e^{\lambda \varphi} \left\{ a\lambda f^{i}u_{~i}^{+}
\partial_{m}(f^{(k}\bar{f}^{j)}u^{-}_{~k}u^{-}_{~j}) \right. \nonumber\\
& &\left. +b \partial_{m}f^{i}u_{~i}^{-}
+c\, \fracmm{\lambda f^{i}u_{~i}^{-}}{1+\lambda f\bar{f}}
(f^{j}\partial_{m}\bar{f}_{j}-\bar{f}_{j}\partial_{m}f^{j})
\right\}
\label{2ansatz}
\end{eqnarray}
with some coefficients $(a,b,c)$ to be determined. After substituting 
eq.~(\ref{2ansatz}) into eq.~(\ref{bw2}), we find the relations
\begin{equation}
b = 2~, \quad 2a=2+b~, \quad{\rm and}\quad c= b-a+1 ~,
\end{equation}
so that $a = b = 2$ and $c = 1$. Therefore, we have
\begin{eqnarray}
A^{-}_{m} &=& e^{\lambda \varphi} \left\{ 2\lambda f^{i}u_{~i}^{+}
\partial_{m}(f^{(k}\bar{f}^{j)}u^{-}_{~k}u^{-}_{~j})\right. \\ \nonumber
& & \left. +2 \partial_{m}f^{i}u_{~i}^{-}
+ \fracmm{\lambda f^{i}u_{~i}^{-}}{1+\lambda f\bar{f}}
(f^{j}\partial_{m}\bar{f}_{j}-\bar{f}_{j}\partial_{m}f^{j})
\right\}~.
\end{eqnarray}

To solve the remaining equations of motion (\ref{bw3}) and (\ref{bw4})
for the auxiliary fields $M^{-}$ and $N^{-}$ (the rest of equations of motion 
in eq.~(\ref{bw5}) is irrelevant for our purposes), 
we introduce the {\it Ansatz}
\begin{eqnarray}
\label{m}
M^{-}&=&e^{\lambda \varphi}R^{-}\equiv R e^{\lambda\varphi}f^{i}u^{-}_{~i}~,\\
\label{n}
N^{-}&=&e^{\lambda \varphi}S^{-}\equiv S e^{\lambda \varphi}f^{i}u^{-}_{~i}~,
\end{eqnarray}
with some coefficient functions $R$ and $S$ to be determined.
After substituting eq.~(\ref{m}) into eq.~(\ref{bw3}) we get
\begin{eqnarray*}
\partial^{++}R^{-}
-\lambda f^{(j}\bar{f}^{k)}u^{+}_{~j}u^{+}_{~k}R^{-}
+\lambda f^{i}f^{j}u^{+}_{~i}u^{+}_{~j}\sbar{S}{}^{-}
+i\Bar{Z}f^{i}u^{+}_{~i}&=&0~, \nonumber \\
Rf^{i}u^{+}_{~i}
-R\lambda f^{(m}\bar{f}^{n)}f^{i}u^{+}_{~m}u^{+}_{~n}u^{-}_{~i}
+i\Bar{Z}f^{i}u^{+}_{~i} 
-\lambda\bar{S}f^{m}f^{n}\bar{f}^{i}u^{+}_mu^{+}_{~n}u^{-}_{~i}&=& 0~, 
\nonumber \\
Rf^{i}u^{+}_{~i}
+R\lambda f^{m}\bar{f}_{n}f^{i}(u^{-~n}u^{+}_{~i}+\delta^{n}_{i})u^+_m
+i\Bar{Z}f^{i}u^{+}_{~i}
-\lambda\bar{S}f^{m}f^{n}\bar{f}^{i}u^{+}_mu^{+}_{~n}u^{-}_{~i}&=& 0~, 
\nonumber\\
f^{i}u^{+}_{~i}[R(1+\lambda f^{j}\bar{f}_{j})+i\Bar{Z}]
-\lambda(R+\bar{S})f^{m}\bar{f}^{n}f^{i}u^{+}_{~m}u^{+}_{~i}u^{-}_{~n}&=& 0~.
\nonumber \\
\end{eqnarray*}
It follows
\begin{equation}
R=-\,\fracmm{i\Bar{Z}}{1+\lambda f\bar{f}} \quad {\rm and,~hence,}
\quad
M^{-}=-\,\fracmm{i\Bar{Z}}{1+\lambda f\bar{f}}\,e^{\lambda \varphi}
f^{i}u^{-}_{~i}~.
\end{equation}

Similarly, we find from eqs.~(\ref{bw4}) and (\ref{n}) that
\begin{eqnarray}
N^{-}= -\,\fracmm{iZ}{1+\lambda f\bar{f}}\,e^{\lambda \varphi}f^{i}u^{-}_{~i}~.
\end{eqnarray}
Substituting now the obtained solutions for the auxiliary fields 
$F^{+},A^{-}_{a},M^{-}$ and $N^{-}$ into the action (\ref{baction}) yields the
bosonic NLSM action  
\begin{equation}
S = \frac{1}{2}\int d^{4}x\, \left\{ 
g_{ij}\partial_{m}f^{i}\partial^{m}f^{j}
+\bar{g}^{ij}\partial_{m}\bar{f}_{i}\partial^{m}\bar{f}_{j}
+2h^{i}_{~j}\partial_{m}f^{j}\partial^{m}\bar{f}_{i}) -V(f)\right\}~,
\end{equation}
whose metric takes the form~\cite{giost}
\begin{eqnarray}
g_{ij}&=& \fracmm{\lambda(2+\lambda f\bar{f})}{2(1+\lambda f\bar{f})}
\bar{f}_i\bar{f}_j ~,\quad
\bar{g}^{ij}=  \fracmm{\lambda(2+\lambda f\bar{f})}{2(1+\lambda f\bar{f})}
f^if^j ~,\\
h^i_j&=& \delta^i_j(1+\lambda f\bar{f})  
-\fracmm{\lambda(2+\lambda f\bar{f})}{2(1+\lambda f\bar{f})}f^i\bar{f}_j~.
\nonumber
\label{1nlsm}
\end{eqnarray}
This metric is known to be equivalent to the standard Taub-NUT metric up to a 
field redefinition~\cite{giost}. The scalar potential in eq.~(\ref{1nlsm}) takes
the form~\cite{ikz}
\begin{equation}
V(f)=\fracmm{Z\Bar{Z}}{1+\lambda f\bar{f}}\,f\bar{f}~.
\label{taubpot}
\end{equation}

By construction, the effective scalar potential (\ref{taubpot}) for a single
charged hypermultiplet is generated in the one-loop perturbation theory (subsect.~3.2), 
and it is exact in the Coulomb branch. The vacuum expectation  values for the scalar 
hypermultiplet components, which are to be calculated from this effective potential, all vanish. 
Notably, the BPS mass $m^2_{\rm BPS}=\abs{Z}^2$ is not renormalized, as it should. However, the 
exact effective scalar potential (\ref{taubpot}) is not merely the quadratic (BPS mass) contribution.

\newpage

\noindent
{\large \bf Appendix B: Eguchi-Hanson hypermultipet self-interaction \newline 
${~~~~~~~~~~~~~~~~~}$ in components from harmonic superspace}

\setcounter{equation}{0}

As was argued in sect.~6, a non-trivial hypermultiplet self-interaction can be 
non-perturbatively generated in the Higgs branch, in the presence of the non-vanishing 
constant FI-term $\x^{(ij)}=\ha(\vec{\t}\cdot\vec{\x})^{ij}$, where $\vec{\t}$ are 
Pauli matrices. The FI-term is given by the vacuum expectation value of the $N=2$ 
vector multiplet auxiliary components (in a WZ-like gauge), and it has a nice 
geometrical interpretation in the underlying ten-dimensional type-IIA superstring brane 
picture (sect.~6).

The simplest non-trivial LEEA for a single dimensionless $\o$-hypermultiplet 
in the baryonic Higgs branch reads
\begin{equation}
S_{\rm EH}[\o]=-\,\fracmm{1}{2\kappa^2}\int d\zeta^{(-4)} du \left\{ 
\left(D^{++}\omega\right)^2
-\fracmm{(\x^{++})^2}{\omega^2}\right\} ~,
\label{omega}
\end{equation}
where $\x^{++}=u^+_iu^+_j\x^{(ij)}$ is the FI-term, and $\kappa$ is the
coupling constant of dimension one (in units of length). After changing the
variables to $q^+_a=u^+_a\omega + u^-_af^{++}$ and eliminating the Lagrange 
multiplier $f^{++}$ via its algebraic equation of motion, one can rewrite 
eq.~(\ref{omega}) to the equivalent form, in terms of the dual 
$q^+$-hypermultiplet, as~\cite{giot}
\begin{equation}
S_{EH}[q]=-\,\fracmm{1}{2\kappa^2}
\int d\zeta^{(-4)} du \left\{  q^{a+} D^{++}q^+_{a} -
\fracmm{(\x^{++})^2}{(q^{a+}u^-_{a})^2}\right\} ~,
\label{qu}
\end{equation}
where we have used the notation $q_a^+=(\sbar{q}{}^+,q^+)$ and $q^{a+}=
\varepsilon^{ab}q^+_b$. In its turn, eq.~(\ref{qu}) is classically equivalent
to the following gauge-invariant action in terms of {\it two\/} FS 
hypermultiplets $q^+_{aA}$ $(A=1,2)$ and the auxiliary (acting as Lagrange multiplier) 
real analytic $N=2$ vector gauge superfield $V^{++}$~\cite{giot}:
\begin{equation}
S_{EH}[q,V]=-\,\fracmm{1}{2\kappa^2}
\int d\zeta^{(-4)} du \left\{ q^{a+}_A D^{++}q^+_{aA}+
V^{++}\left(\frac{1}{2}\varepsilon^{AB}q^{a+}_Aq^+_{Ba}+\x^{++}\right)
\right\}~.
\label{qv}
\end{equation}
We now calculate the component form of this hypermultiplet self-interaction
by using eq.~(\ref{qv}) as our starting point. In a bit more explicit form, it 
reads  
\begin{eqnarray}
S& = & -\fracmm{1}{2\kappa^2}
\int d\z^{(-4)} du \left\{ \sbar{q}_{1}{}^{+}D^{++}q_{1}^{+}
+\sbar{q}_{2}{}^{+}D^{++}q_{2}^{+} \right.\nonumber \\
& & \left. V^{++}(\sbar{q}_{1}{}^{+}q_{2}^{+}-\sbar{q}_{2}{}^{+}q_{1}^{+}+\xi^{++})
\right\}~.
\label{ehaction}
\end{eqnarray}

The equations of motion are given by
\begin{eqnarray}
\label{bew1}
D^{++}q_{1}^{+}+V^{++}q_{2}^{+}&=& 0~, \\
\label{bew2}
D^{++}q_{2}^{+}-V^{++}q_{1}^{+}&=& 0~, \\
\label{bew3}
\sbar{q}_{1}{}^{+}q_{1}^{+}-\sbar{q}_{2}{}^{+}q_{1}^{+}+\xi^{++}&=&0~,
\end{eqnarray}
while the last equation is clearly the algebraic constraint on the two FS
hypermultiplets. In what follows, we ignore fermionic contributions and use a
WZ-gauge for the $N=2$ vector superfield $V^{++}$, so that $D^{++}$ and $q^+$
are still given by eqs.~(\ref{der}) and (\ref{exp}), whereas
\begin{eqnarray}
V^{++}&=& -2i\theta^{+}\sigma^{m}\bar{\theta}^{+}V_{m}(x_{A})
+\theta^+\theta^+\bar{a}(x_{A})+\bar{\theta}^+\bar{\theta}^+a(x_{A})\\
& & +\theta^+\theta^+\bar{\theta}^+\bar{\theta}^+ 
D^{(ij)}(x_A)u^{-}_{~i}u^{-}_{~j}~.
\end{eqnarray}
The equation of motion (\ref{bew1}) in components reads
\begin{eqnarray}
\label{eq11}
\partial^{++}F_{1}^{+}&=& 0~, \\
\label{eq12}
-2\partial_{m}F^{+}_{1}+\partial^{++}A^{-}_{1m}-2V_{m}F^{+}_{2} &=& 0~, \\
\label{eq13}
i\Bar{Z}F^{+}_{1}+\partial^{++}M^{-}_{1}+\bar{a}F^{+}_{2}&=&0~, \\
\label{eq14}
iZF^{+}_{1}+\partial^{++}N^{-}_{1}+aF^{+}_{2} &=& 0~, \\
\label{eq15}
\partial^{++}P^{(-3)}_{1}+\partial^{m}A^{-}_{1m}+i\bar{Z}N^{-}_{1}+iZM^{-}_{1}
& & \nonumber \\
+V^{m}A^{-}_{2m}+\bar{a}N^{-}_{2}+aM^{-}_{2}
+D^{(ij)}u^{-}_{~i}u^{-}_{~j}F^{+}_{2} &=&0~,
\end{eqnarray}
whereas eq.~(\ref{bew2}) gives
\begin{eqnarray}
\label{eq21}
\partial^{++}F_{2}^{+}&=& 0~,\\
\label{eq22}
-2\partial_{m}F^{+}_{2}+\partial^{++}A^{-}_{2m}+2V_{m}F^{+}_{1} &=& 0~, \\
\label{eq23}
i\bar{Z}F^{+}_{2}+\partial^{++}M^{-}_{2}-\bar{a}F^{+}_{1}&=&0~, \\
\label{eq24}
iZF^{+}_{2}+\partial^{++}N^{-}_{2}-aF^{+}_{1} &=& 0~, \\
\label{eq25}
\partial^{++}P^{(-3)}_{2}+\partial^{m}A^{-}_{2m}
+i\bar{Z}N^{-}_{2}+iZM^{-}_{2} & & \nonumber \\
-V^{m}A^{-}_{1m}-\bar{a}N^{-}_{1}-aM^{-}_{1}
-D^{(ij)}u^{-}_{~i}u^{-}_{~j}F^{+}_{1}&=&0~. 
\end{eqnarray}
The constraint (\ref{bew3}) in components is given by
\begin{eqnarray}
\label{eq31}
\sbar{F}_1{}^{+}F^{+}_{2}-\sbar{F}_2{}^{+}F^{+}_{1}+\xi^{++}&=& 0~, \\
\label{eq32}
\sbar{A}_{1a}{}^- F^{+}_{2}+\sbar{F}_1{}^{+}A^{-}_{2a}
-\sbar{A}_{2a}{}^- F^{+}_{1}-\sbar{F}_2{}^{+}A^{-}_{1a}&=& 0~, \\
\label{eq33}
\sbar{F}_1{}^{+}M^{-}_{2}-\sbar{F}_2{}^{+}M^{-}_{1}+\sbar{N}_1{}^{-}F^{+}_{2}
-\sbar{N}_2{}^{-}F^{+}_{1} &=& 0~, \\
\label{eq34}
\sbar{F}_1{}^{+} N^{-}_{2}-\sbar{F}_2{}^{+}N^{-}_{1}+\sbar{M}_1{}^{-}F^{+}_{2}
-\sbar{M}_2{}^{-}F^{+}_{1} &=& 0~,  \\
\label{eq35}
-\frac{1}{2}\sbar{A}_1{}^{m-}A^{-}_{2m}+\sbar{M}_1{}^{-}M^{-}_{2}
+\sbar{N}_1{}^{-}N^{-}_{2}+\sbar{P}_1{}^{(-3)} F^{+}_{2}
+\sbar{F}_1{}^{+} P^{(-3)}_{2} & & \nonumber\\
+\frac{1}{2}\sbar{A}_2{}^{m-}A^{-}_{1m}-\sbar{M}_2{}^{-} M^{-}_{1}
-\sbar{N}_2{}^{-} N^{-}_{1}-\sbar{P}_2{}^{(-3)} F^{+}_{1}
-\sbar{F}_2{}^{+} P^{(-3)}_{1} &=& 0~. \nonumber \\
\end{eqnarray}
Substituting the component expressions for $q_{A}^{+}$ and $V^{++}$ into 
the action (\ref{ehaction}) results in the following bosonic action:
\begin{eqnarray}
S &=& -\,\fracmm{1}{2\kappa^2}\int d^{4}x du\left\{
\sbar{F}_1{}^{+}\partial^{m}A^{-}_{1m}+\sbar{F}_2{}^{+}\partial^{m}A^{-}_{2m}
+V^{m}(\sbar{F}_1{}^{+}A^{-}_{2m}-\sbar{F}_2{}^{+}A^{-}_{1m})\right.\nonumber\\
& & +\sbar{a}(\sbar{F}_1{}^{+}N^{-}_{2}-\sbar{F}_2{}^{+} N^{-}_{1})
+a (\sbar{F}_1{}^{+}M^{-}_{2}-\sbar{F}_2{}^{+} M^{-}_{1})\nonumber\\
& & +iD^{(ij)}u^{-}_{~i}u^{-}_{~j}(\x^{++}+\sbar{F}_1{}^+F^+_2
-\sbar{F}_2{}^+ F^+_1)\nonumber\\
& &\left.  +\sbar{F}_1{}^{+}(i\Bar{Z}N^{-}_{1}+iZM^{-}_{1})
+\sbar{F}_2{}^{+}(i\Bar{Z}N^{-}_{2}+iZM^{-}_{2})\right\}~.
\label{egaction}
\end{eqnarray}

The next step in our calculation is to fix the harmonic dependence of the 
fields $F^{+}_{i},A^{-}_{ia},M^{-}_{i}$ and $N^{-}_{i}$. Eqs.~(\ref{eq11}) 
and (\ref{eq21}) imply
\begin{eqnarray}
F^{+}_{1} = f^{i}_{1}u^{+}_{~i} \quad{\rm and}\quad  
F^{+}_{2} = f^{i}_{2}u^{+}_{~i}~,
\label{Ff}
\end{eqnarray}
whereas eq.~(\ref{eq12}) yields
\begin{eqnarray}
-2\partial_{m}F^{+}_{1}+\partial^{++}A^{-}_{1m}-2V_{m}F^{+}_{2} = 0 ~.
\end{eqnarray}
After introducing the {\it Ansatz}
\begin{eqnarray}
A^{-}_{1m}= A_{1m}^{i}u^{-}_{~i}+B_{1m}^{ijk}u^{+}_{~i}u^{-}_{~j}u^{-}_{~k}
\end{eqnarray}
we find that
\begin{eqnarray}
A^{-}_{1m}= (2 \partial_{m}f^{i}_{1}+2V_{m}f^{i}_{2})u^{-}_{~i}~.
\label{A1f}
\end{eqnarray}
Similarly, it follows from  eq.~(\ref{eq22}) that
\begin{eqnarray}
A^{-}_{2m}= (2 \partial_{m}f^{i}_{2}-2V_{m}f^{i}_{1})u^{-}_{~i}~.
\label{A2f}
\end{eqnarray}
Eqs.~(\ref{eq13}), (\ref{eq14}), (\ref{eq23}) and (\ref{eq24}) now imply
\begin{eqnarray}
\label{M1f}
M^{-}_{1} &=& -(\sbar{a}f^{i}_{2}+i\Bar{Z}f^{i}_{1})u^{-}_{~i}~, \\
\label{N1f}
N^{-}_{1} &=& -(af^{i}_{2}+iZf^{i}_{1})u^{-}_{~i}~, \\
\label{M2f}
M^{-}_{2} &=& (\sbar{a}f^{i}_{1}-i\Bar{Z}f^{i}_{2})u^{-}_{~i}~, \\
\label{N2f}
N^{-}_{2} &=& (af^{i}_{1}-iZf^{i}_{2})u^{-}_{~i} ~.
\end{eqnarray}
After substituting all the component solutions into the action 
(\ref{egaction}), we find the (abelian) gauged NLSM action
\begin{eqnarray}
S &=&\fracmm{1}{2\kappa^2}\int d^{4}x\left\{
(\partial_{m}f_{1}^{~i}+V_{m}f_{2}^{~i})
(\partial^{m}\bar{f}_{1i}+V^{m}\bar{f}_{2i}) \right. \nonumber \\
& & +(\partial_{m}f_{2}^{~i}-V_{m}f_{1}^{~i})
(\partial^{m}\bar{f}_{2i}-V^{m}\bar{f}_{1i})  \nonumber\\
& &\left. -\,\fracmm{Z\Bar{Z}}{(f_1\bar{f}_1+f_2\bar{f}_2)}
\left[(f_1^i\bar{f}_{2i}-f_2^i\bar{f}_{1i})^2
+(f_1^i\bar{f}_{1i}+f_2^i\bar{f}_{2i})^2\right]\right\}~, \nonumber\\
\label{wirkung2}
\end{eqnarray}
where the scalar hypermultiplet components $f^i_{1,2}$ are still subject to the
constraint 
\begin{eqnarray}
\x^{(ij)}&=& \bar{f}_{1}^{(i}f^{j)}_2-f^{(i}_1\bar{f}^{j)}_2~. 
\label{xi}
\end{eqnarray}
In calculating the action (\ref{wirkung2}) we have also used the equation
of motion for the $N=2$ vector multiplet auxiliary field $a$, whose solution
reads
\begin{eqnarray}
\label{a}
a&=& -iZ\fracmm{f_1^i\bar{f}_{2i}-f_2^{i}\bar{f}_{1i}}
{f_1^i\bar{f}_{1i}+f_2^i\bar{f}_{2i}}~.
\end{eqnarray}
A solution to the equation of motion for the vector gauge field $V_m$ is given
by
\begin{eqnarray}
2V_m &=&\fracmm{\partial_m\bar{f}_{1j}f_2^j
-\bar{f}_{1j}\partial_mf_2^j-\partial_m\bar{f}_{2j}f_1^j
+\bar{f}_{2j}\partial_m f_1^j}
{\bar{f}_{1j}f_1^j+\bar{f}_{2j}f_2^j}~~.
\label{va}
\end{eqnarray}

In terms of the two complex scalar $SU(2)$ doublets $f^i_{1,2}$ subject to the
three real constraints (\ref{xi}) and one abelian gauge invariance, we have
$2\times 2\times 2 - 3 -1=4$ independent degrees of freedom, as it should. 
After eliminating the auxiliary vector potential $V_m$ via eq.~(\ref{va}) and 
solving the constraint (\ref{xi}), one finds the NLSM with a non-trival 
hyper-K\"ahler metric (by construction, as the consequence of $N=2$ 
supersymmetry) {\it and\/} a non-trivial scalar potential as well,
\begin{equation}
V=\fracmm{Z\Bar{Z}}{(f_1\bar{f}_1+f_2\bar{f}_2)}
\left[ (f_1^i\bar{f}_{2i}-f_2^i\bar{f}_{1i})^2
+(f_1^i\bar{f}_{1i}+f_2^i\bar{f}_{2i})^2\right]~.
\label{pot}
\end{equation}
The kinetic terms in the NLSM (\ref{wirkung2}) are known to represent the 
{\it Eguchi-Hanson} instanton metric up to a field redefinition~\cite{giot}, 
so that to this end we concentrate on the scalar potential (\ref{pot})
only.  Let's introduce the notation
\begin{eqnarray}
\bar{f}_{(1,2)}^{~1} = \stackrel{\ast}{f}_{(1,2)}{}^{2}~~, \qquad 
\bar{f}_{(1,2)}^{~2} = -\stackrel{\ast}{f}_{(1,2)}{}^{1}~,
\label{nota}
\end{eqnarray}
and keep the positions of indices as above. The operator $\ast$ denotes 
the usual complex conjugation. The constraints (\ref{xi}) now take the form
\begin{eqnarray*}
\x^{11}&=& \bar{f}_{1}^{1}f_2^{~1}-f_1^{~1}\bar{f}_2^{~1}=
\stackrel{\ast}{f}_1{}^{2}f_2^{~1}- f_1^{~1}\stackrel{\ast}{f}_2{}^{2}~,
\nonumber \\
\x^{12}&=& \frac{1}{2}(\bar{f}_1^{~1}f_2^{~2}+\bar{f}_1^{~2}f_2^{~1})
-\frac{1}{2}(f_1^{~1}\bar{f}_2^{~2}+f_1^{~2}\bar{f}_2^{~1})~, \nonumber\\ 
\x^{22}&=& \bar{f}_{1}^{2}f^2_2-f^2_1\bar{f}^2_2 ~.\nonumber\\ 
\end{eqnarray*}
After mulitplying these constraints with Pauli matrices 
$(\tau_1, 1,\tau_3)_{ij}$, we get 
\begin{eqnarray}
\x^1&=& \bar{f}_{1}^{1}f_2^{~2}+\bar{f}_1^{~2}f_2^{~1}
-(f_1^{~1}\bar{f}_2^{~2}+f_1^{~2}\bar{f}_2^{~1})~, \\ 
\x^2&=& \bar{f}_{1}^{1}f^1_2-f^1_1\bar{f}^1_2 +
\bar{f}_{1}^{2}f^2_2-f^2_1\bar{f}^2_2~, \\
\x^3&=& \bar{f}_{1}^{1}f^1_2-f^1_1\bar{f}^1_2
-\bar{f}_{1}^{2}f^2_2+f^2_1\bar{f}^2_2~,
\end{eqnarray}
while we have $\vec{\x}^2\equiv (\x^{1})^2 +(\x^{2})^2 + (\x^{3})^2  \neq 0$.
We now choose the direction $\x^{2}= \xi^{3}=0$ and $\xi^{1}=2i$, so that 
our constraints now take the form
\begin{eqnarray}
\bar{f}_{1}^{1}f_2^{~2}+\bar{f}_1^{~2}f_2^{~1}
-(f_1^{~1}\bar{f}_2^{~2}+f_1^{~2}\bar{f}_2^{~1})&=& 2i ~,\\ 
(-(f_1^{~1})^{\ast}f_2^{~1}+f_1^{~1}(f_2^{~1})^{\ast})
+((f_1^{~2})^{\ast}f_2^{~2}-f_1^{~2}(f_2^{~2})^{\ast})&=& 2i~,
\label{stra1}
\end{eqnarray}
and 
\begin{equation}
f_2^{~1}(f_1^{~2})^{\ast}= f_1^{~1}(f_2^{~2})^{\ast}~~,\quad
f_2^{~2}(f_1^{~1})^{\ast}= f_1^{~2}(f_2^{~1})^{\ast} ~.
\label{stra2}
\end{equation}
We thus end up with only two+one real constraints and one gauge invariance
\begin{eqnarray}
\left( \begin{array}{c}f_1 \\f_2 \end{array}\right)'
= \left( \begin{array}{cc} \cos(\alpha) & \sin(\alpha)\\ -\sin(\alpha) &
\cos(\alpha) \end{array} \right)
\left( \begin{array}{c} f_1 \\ f_2\end{array} \right)~.
\end{eqnarray}
In the parameterization
\begin{eqnarray}
f_i^{~j}= m_i^{~j} exp(i \varphi_i^{~j})
\end{eqnarray}
the constraints (\ref{stra1}) and (\ref{stra2}) read
\begin{eqnarray}
m_1^1 m_2^2 = m_1^2m_2^1 
e^{-i\varphi_2^1-i\varphi_2^2+i\varphi_1^1+i\varphi_1^2}~, \\
m_1^1m_2^1\sin(\varphi_1^1-\varphi_2^1)+m_2^2m_1^2 
\sin(\varphi_2^1-\varphi_1^2)=1~.
\end{eqnarray}

We now want to fix the local $U(1)$ symmetry by imposing the gauge condition 
\begin{eqnarray}
\varphi_2^1+\varphi_2^2&=&\varphi_1^1+\varphi_1^2~.
\label{gauge}
\end{eqnarray}
When using
\begin{eqnarray}
\label{indep}
\abs{f_2^1}\equiv  m, \qquad \abs{f_2^2}\equiv n, \qquad \varphi_1^1 \equiv
 \theta~,\qquad \varphi_2^2 \equiv \phi~,
\end{eqnarray}
as the independent fields, our constraints above can be easily solved as
\begin{equation}
\label{varia}
-\varphi_2^1 = \varphi_2^2 = \phi, \qquad \varphi_1^1= -\varphi_1^2= \theta~,
\qquad m_2^1=m, \quad m_2^2 = n~,
\end{equation}
and
\begin{eqnarray}
m_1^1 = \fracmm{m}{(m^2+n^2)\sin(\theta+\phi)}~,\quad 
m_1^2 = \fracmm{n}{(m^2+n^2)\sin(\theta+\phi)}~~.
\end{eqnarray}

It is straightforward to deduce the other fields $F_i^+, A_{im}^-, M_i^-$ and 
$N_i^-$ in terms of the independent components (\ref{indep}). The scalar 
potential (\ref{pot}) in terms of these independent field variables 
takes the form (no indices and constraints any more~!)
\begin{equation}
V= \fracmm{\abs{Z}^2\sin^2(\theta +\phi)}{m^2+n^2}\left[
\fracmm{4(m^2-n^2)^2}{1+(m^2+n^2)^2\sin^2(\theta +\phi)}
+\fracmm{1+(m^2+n^2)^2\sin^2(\theta +\phi)}{\sin^4(\theta +\phi)}\right]~.
\end{equation}
It is clear from this equation that the potential $V$ is positively definite, 
and it is only non-vanishing due to the non-vanishing central charge $\abs{Z}$.
This signals spontaneous breaking of $N=2$ supersymmetry in our model.

\end{document}